\documentclass[preprint,showpacs,preprintnumbers,amsmath,amssymb]{revtex4}
\usepackage[english]{babel}
\usepackage{graphicx}
\usepackage{latexsym}
\usepackage{amsfonts}
\usepackage{epsfig}
\usepackage{ifthen}
\usepackage{textcomp}
\usepackage{color}
\usepackage{boxedminipage}

\newcommand{\bfphi}{\mbox{\boldmath ${\varphi}$}}
\newcommand{\bfrho}{\mbox{\boldmath ${\varrho}$}}

\newcommand{\bfbeta}{\mbox{\boldmath ${\beta}$}}
\newcommand{\bfalpha}{\mbox{\boldmath ${\alpha}$}}
\newcommand{\tensR}{\mbox{\underline{\underline{\mbox{\boldmath ${\cal R}$}}}}}

\newcommand{\bfcalE}{\mbox{\boldmath${\cal E}$}}
\newcommand{\bfcalH}{\mbox{\boldmath${\cal H}$}}

\begin{document}
\title{Focused Fields of given Power with Maximum  Electric Field Components}
\author{H. P. Urbach  and  S. F. Pereira \\
  Optics Research Group \\
Department of Imaging Science and Technology \\
Delft University of Technology, P.O. Box 5046 \\
2600 GA Delft, The Netherlands (h.p.urbach@tudelft.nl)}
\date{\today}
\begin{abstract}
Closed formulas are derived for the
  field in the focal region of a diffraction limited lens, such that
the electric field component in a given direction  at the focal point is larger than that of all other focused fields with the
same power in the entrance pupil of the lens.
Furthermore, closed formulas are  derived for the corresponding optimum field distribution in the lens pupil.
Focused fields with maximum longitudinal or maximum transverse are considered in detail. The latter field is similar,
 but not identical,  to the focused linearly polarized plane wave.
\end{abstract}
  \maketitle
\section{Introduction}

When a linearly polarized plane wave is focused by a
diffraction-limited lens, the intensity distribution in the focal
plane is in the scalar theory  the well-known
 Airy pattern. However, when the lens has high numerical aperture, the
 rotation  of polarization must be accounted for and the
  vector diffraction theory
 of Ignatowsky \cite{ignatowsky}, \cite{ignatowskyb} and
 Richards and Wolf \cite{wolf}, \cite{richards} has to be
 applied to obtain the field distribution in the focal region.
 We then get three electric and three magnetic field components in the focal region.
 When the  beam in the lens aperture is uniformly linearly polarized plane wave,
 the dominant electric field component in the focal region
is   found to be parallel to the polarization direction of the
incident plane wave. But, as the numerical aperture
increases, the maximum value of the longitudinal component of the
electric field in the focal plane becomes quite substantial,
although it vanishes at the focal point itself.

An appropriately shaped focused spot is essential in
many applications such as optical recording, photolithography and microscopy.
Furthermore, a field in focus with maximum electric component in a specific direction is
important for
  manipulating single molecules and particles, and
in materials processing \cite{xie},  \cite{novotny}, \cite{zhan},
\cite{helseth}, \cite{meier}.
The focused wavefront can be tailored
by  setting a proper amplitude, phase and polarization distributions  in
the pupil of the focusing lens.
Nowadays it is possible to realize almost any
complex transmission function in the pupil plane,
using for example liquid crystal-based devices \cite{sanner},
\cite{zande}, \cite{wilson}, \cite{stalder}, \cite{iglesias}.

The optimization  of focused fields has been studied by other authors.

In this paper, we maximize a specified
component of the electric field in the focal point of a diffraction limited lens.
First we consider  fields in free space or in homogeneous matter, without taking into account the way
these fields are realized, in particular without considering the lens.
 We merely suppose that, with respect to a cartesian coordinate system
$(x,y,z)$, the fields considered consist of  plane waves  propagating in the positive
$z$-direction and  have wave vectors
with   angles with the positive $z$-axis that do not exceed a specific maximum angle, i.e.,
the numerical aperture of the plane waves is restricted.
The optimization problem  is then  to find the complex plane wave amplitudes such that,
for a given direction in space and for given  mean flow of electromagnetic
power through a plane $z$=constant, the amplitude at some point (chosen at the origin)
of the electric field component that is parallel to the chosen direction
  is  larger than that
 of any other field for the same numerical aperture and the same
mean total  power flow.
We shall derive closed formulas for the plane wave amplitudes
of the optimum field.

The solutions of the optimization problems for the field propagating in homogeneous space is
 rigorous results  since they are derived from Maxwell's equations without any further assumptions.
Next we will consider the  realization of the optimum field using a diffraction limited lens with the
origin as focal point.
By using the vector diffraction theory of Ignatowksy and Richards and Wolf,
 closed expressions for the optimum pupil  distributions will be derived, which after focussing give the
 optimum field component in the focal point of the lens. In contrast with the solution in terms of
 the plane wave expansion, the formulas for the optimum pupil fields
 are approximate since they are based on the  vector diffraction theory which is an approximate theory
 that is valid
 for lenses of which the focal distance and the pupil radius are many wavelengths.

When one considers the focussing by a lens it is obvious that the plane wave expansion in image space has
 finite numerical aperture of that of the lens: $\mbox{NA} = n \sin \alpha_{\max}$, where $n$ is the refractive
 index in image space and $\alpha_{\max}$ is half the top angle of the cone with top the focal point and
 base the pupil. But also when one would consider only waves in free space without
a focussing lens, there is a good reason to restrict the fields to finite numerical aperture, in that case
 in particular
to $\mbox{NA}=n$.
In fact, when $\mbox{NA} >n$  a part
  of the evanescent waves are taken into account in the expansion.
 One can then construct fields
with a given power which have arbitrary large components. The evanescent waves do not contribute
to the total  power  and hence one can increase their amplitudes by any desired amount to achieve arbitrarily high
local fields. Stated differently, by constructing
suitable time-harmonic source distributions that emit singular fields  with finite power flow,
one can achieve arbitrary large field components by approaching these source
distributions. At small distances to the sources the evanescent waves play of course a major role.

Among the directions for the optimized electric field component,
  two are of particular interest, namely the directions parallel and perpendicular to the optical $z$-axis.
These directions are also called the longitudinal and transverse directions.
The field with maximum longitudinal component has been
 discussed  in
\cite{urbachPRL}, but without derivation. In this paper details of the derivation are provided
  and the optimization problem is generalized to arbitrary
directions of the electric field vector.

As was announced in \cite{urbachPRL}, the pupil field that  when focused gives
maximum longitudinal component, is radially polarized. This means that in all
points of the lens pupil, the electric field is linearly polarized with the electric field
 pointing in the radial direction. Furthermore, the electric fields in all points of the
 pupil are in phase and the electric field amplitudes are rotationally symmetric.
 The amplitude of the electric pupil field  vanishes at the center of the pupil
 and is a monotonically increasing function of the radial coordinate. The shape of this function depends on the
 numerical aperture.

 It was noted by several authors
\cite{dorn}, \cite{quabis}, \cite{quabis_AP},
\cite{sheppard} that when a radially polarized beam is focused,
 the distribution of the
longitudinal component can be considerably narrower than the focussed spot obtained by focusing
 a linearly polarized plane wave.
With the development of a new generation of photoresists \cite{sanchez}, it is
possible to control the photosensitive material in such a way that
it will react to only one of the polarization components of the
electric field.
 Materials with molecules having  fixed absorption dipole moments have been
applied in \cite{novotny} to be able to probe  field components individually.  When this component is
the longitudinal component, a tighter spot can thus be obtained
than with the classical Airy pattern.

Often the amplitude distribution of the radially polarized beam in the pupil
plane is chosen to be a
doughnut shape or a ring mask function \cite{quabis}.  But these distributions do  not
give the maximum possible longitudinal electric field component in focus for given power and its amplitude
as a function of the radial pupil coordinate differs from the optimum function derived in the present paper.

The other case of particular interest is  the optimization of the transverse electric field vector.
 Since the optical system is assumed to be
rotationally symmetric around the optical axis, we may choose this
direction parallel to the $x$-axis.  The solution of the
optimization problem is then the field for which the amplitude of the $x$-component of the electric field
in the focal point is maximum for the given numerical aperture and the given
total power flow.
We will show that the corresponding pupil field is linearly polarized
with direction of polarization predominantly,  although not exactly, parallel to the
$x$-axis. Therefore, the focused optimum field is  similar to the vectorial Airy pattern
of a focused linearly polarized plane wave, although it is
{\em not} identical to it.

 The paper is organized as follows. In Section 2,
 we will formulate the optimization problem and we will prove that the optimization problem has
 one and only one solution. In Section 3, we will  apply the Lagrange multiplier rule to
 obtain closed formulas for the plane wave amplitudes of the optimum field and for the
 optimum field distributions near focus. In Section 4, we will  study the optimum
 fields in the focal region, in particular their mean energy flow.
 Then, in Section 5, we apply the vector diffraction theory of
 Ignatowsky, Richards and Wolf to  derive  the electric field
 distribution in the pupil of the lens that, when focussed,
 yields the maximum field component in the focal region.

\section{Formulation of the optimization problem for arbitrary electric field component}
\label{section.formulation}
We begin with some notations.
Consider a time-harmonic electromagnetic field in a homogeneous unbounded
medium  with real refractive index
$n$ (i.e., the material does not absorb electromagnetic radiation of the given frequency):
\begin{eqnarray}
\bfcalE(\mathbf{r},t) & = & \mbox{Re}\left[ \mathbf{E}(\mathbf{r}) e^{-i \omega t}\right],
\label{eq.calE}\\
\bfcalH(\mathbf{r},t) & = & \mbox{Re}\left[ \mathbf{H}(\mathbf{r}) e^{-i \omega t}\right],
\label{eq.calH}
\end{eqnarray}
where $\omega >0$.
As stated in the Introduction, the lens  is first not considered in the optimization problem.
It is merely assumed that, with respect to the cartesian coordinate system $(x,y,z)$ with
unit vectors $\hat{\mathbf{x}}$, $\hat{\mathbf{y}}$ and $\hat{\mathbf{z}}$,
the electromagnetic field (\ref{eq.calE}), (\ref{eq.calH}) has numerical aperture $\mbox{NA} \leq n$ and that
the plane wave vectors have positive $z$-component:
\begin{eqnarray}
  \mathbf{E}(\mathbf{r}) & = &\frac{1}{4 \pi^2}
  \int\!\!\int_{\sqrt{k_x^2+k_y^2} \leq  k_0 n \sin \alpha_{\max}} \mathbf{A}(k_x, k_y)
  e^{i \mathbf{k}\cdot\mathbf{r}} \, d k_x d k_y, \label{eq.planeE}\\
  \mathbf{H}(\mathbf{r}) & = & \frac{1}{4 \pi^2}\frac{1}{\omega\mu_0}
  \int\!\!\int_{\sqrt{k_x^2+k_y^2} \leq  k_0 n \sin \alpha_{\max}} \mathbf{k} \times \mathbf{A}(k_x, k_y)
  e^{i \mathbf{k}\cdot\mathbf{r}} \, d k_x d k_y, \label{eq.planeH}
  \end{eqnarray}
  where $ \mathbf{k}=(k_x, k_y, k_z)$ with
  \begin{equation}
     k_z=(k_0^2n^2 - k_x^2 -k_y^2)^{1/2},
     \label{eq.mathbfk}
     \end{equation}
     with $k_0=\omega \sqrt{\epsilon_0\mu_0}=2\pi/\lambda_0$ where $\lambda_0$ is the wavelength in vacuum, and where
$\mbox{NA} = n\sin \alpha_{\max}$
with  $\alpha_{\max}$ the maximum angle that the wave vectors
make with the positive $z$-direction.
If $\mbox{NA}=n$ we have
$\alpha_{\max}=\pi/2$ and the plane wave spectrum then consists
of all homogeneous plane waves  that propagate in  the non-negative
$z$-direction (there are no evanescent waves in the expansion). When
$\mbox{NA}< n$, the cone of allowed wave vectors has top angle
$\alpha_{\max}=\arcsin(\mbox{NA}/n) < 90^o$. Because the electric field
is free of divergence we have that
\begin{equation}
\mathbf{A}\cdot \mathbf{k}=0.
\label{eq.div0}
\end{equation}
We shall use spherical coordinates in reciprocal $\mathbf{k}$-space:
\begin{eqnarray}
   \mathbf{\hat{k}} & = &  \sin \alpha \cos \beta \, \hat{\mathbf{x}} +
                           \sin \alpha \sin\beta\,\hat{\mathbf{y}} +
                          \cos \alpha \, \hat{\mathbf{z}},  \label{eq.hatk}\\
   \hat{\bfalpha} & = &  \cos \alpha \cos \beta \, \hat{\mathbf{x}} +
                         \cos \alpha  \sin \beta\,\hat{\mathbf{y}} -
                          \sin \alpha \, \hat{\mathbf{z}},  \label{eq.hatalpha}\\
   \hat{\bfbeta} & = & -\sin \beta  \, \hat{\mathbf{x}} +
                          \cos \beta  \, \hat{\mathbf{y}},  \label{eq.hatbeta}
 \end{eqnarray}
 where $0 \leq\alpha\leq \alpha_{\max}$ and $0\leq\beta< 2\pi$ are the
  polar and azimuthal angles, respectively.
Conversely, we have
 \begin{eqnarray}
 \hat{\mathbf{x}} & = &  \sin \alpha  \cos\beta\,\hat{\mathbf{k}}
                       + \cos \alpha \cos \beta
 \,\hat{\bfalpha} - \sin \beta \, \hat{\bfbeta}, \label{eqhatx}\\
 \hat{\mathbf{y}} & = &  \sin \alpha \sin \beta \,\hat{\mathbf{k}}
 +  \cos \alpha \sin \beta
 \,\hat{\bfalpha} + \cos\beta \, \hat{\bfbeta}, \label{eqhaty}\\
 \hat{\mathbf{z}} & = &  \cos \alpha \,\hat{\mathbf{k}} -  \sin \alpha
 \, \hat{\bfalpha}. \label{eqhatz}
 \end{eqnarray}
Note that $\{\mathbf{\hat{k}}, \hat{\bfalpha}, \hat{\bfbeta}\}$ is a positively oriented
orthonormal basis:
\begin{equation}
\mathbf{\hat{k}} \times \hat{\bfalpha}=\hat{\bfbeta}, \;\;\;
\hat{\bfalpha}\times \hat{\bfbeta} = \mathbf{\hat{k}}, \;\;\;
\hat{\bfbeta}\times \mathbf{\hat{k}}=\hat{\bfalpha}.
\label{eq.vecprbasis}
\end{equation}
 Furthermore,
 $\mathbf{k} = k_0 n \mathbf{\hat{k}}$ and the Jacobian of the
 transformation $(\alpha, \beta) \mapsto (k_x,k_y)$ is:
 \begin{equation}
   \left( \begin{array}{cc}
        \frac{\partial k_x}{\partial \alpha} & \frac{\partial k_x}{\partial \beta} \\
        \frac{\partial k_y}{\partial \alpha} & \frac{\partial k_y}{\partial \beta}
      \end{array}\right) = k_0 n \left( \begin{array}{cc}
                     \cos \alpha \cos\beta & -\sin \alpha \sin\beta \\
           \cos \alpha \sin\beta & \sin \alpha \cos\beta
           \end{array}\right),
           \label{eq.Jacobian}
           \end{equation}
           so that
    \begin{equation}
    d k_x \, d k_y = k_0^2 n^2 \cos\alpha\, \sin \alpha \, d \alpha\, d \beta.
    \label{eq.Jacobian2}
    \end{equation}
    Because of (\ref{eq.div0}) we have:
 \begin{equation}
 \mathbf{A}(\alpha,\beta) = A_\alpha(\alpha,\beta) \hat{\bfalpha} +
A_\beta(\alpha,\beta)   \hat{\bfbeta},
                 \label{eq.Aphitheta}
 \end{equation}
 for some functions $ A_\alpha$ and $A_\beta$.
Then, using (\ref{eq.vecprbasis}):
\begin{eqnarray}
\mathbf{k} \times \mathbf{A}
  & = & k_0 n ( -A_\beta \hat{\bfalpha} + A_\alpha \hat{\bfbeta}).
\label{eq.ktimesA}
\end{eqnarray}
The plane wave expansion can thus be written as
\begin{eqnarray}
\mathbf{E}(\mathbf{r})& = &\frac{ n^2}{\lambda_0^2}
 \int_0^{\alpha_{\max}}\!\int_0^{2\pi}\!
   ( A_\alpha \hat{\bfalpha}+A_\beta \hat{\bfbeta}  ) \cos \alpha \sin\alpha \,
e^{i \mathbf{k}\cdot \mathbf{r}}\,
 d\alpha \, d\beta,\label{eq.bfE}\\
\mathbf{H}(\mathbf{r})& = & \frac{
n^3}{\lambda_0^2}\left(\frac{\epsilon_0}{\mu_0}\right)^{1/2}
\int_0^{\alpha_{\max}}\!\int_0^{2\pi}
   ( -A_\beta \hat{\bfalpha} + A_\alpha \hat{\bfbeta} ) \cos \alpha \sin \alpha \,
e^{i \mathbf{k}\cdot \mathbf{r}}\,
d\alpha \, d\beta. \label{eq.bfH}
\end{eqnarray}
The $\hat{\bfalpha}$-component is parallel to the plane through the wave vector and the
$z$-axis, whereas the $\hat{\bfbeta}$-component is perpendicular  to this plane.

Let $\hat{\mathbf{v}}=v_x \hat{\mathbf{x}}+v_y\hat{\mathbf{y}} + v_z \hat{\mathbf{z}}$ be a real unit vector.
We consider the projection of the electric field at the origin at time $t=0$ on the direction of
$\hat{\mathbf{v}}$:
\begin{equation}
\mathbf{E}(\mathbf{0})\cdot \hat{\mathbf{v}} = \frac{n^2}{\lambda_0^2} \int_0^{\alpha_{\max}}\!\int_0^{2\pi}\!
    \left[ A_\alpha(\alpha,\beta) \, v_\alpha + A_\beta(\alpha,\beta)\, v_\beta\right] \, \cos \alpha \sin \alpha \,
      d\alpha\, d\beta,
    \label{eq.Ez0}
    \end{equation}
    where
    \begin{eqnarray}
    v_\alpha & = & \hat{\mathbf{v}}\cdot \hat{\bfalpha}=
    v_x \cos\alpha \cos \beta + v_y \cos \alpha \sin \beta - v_z \sin \alpha,  \label{eq.palpha}\\
    v_\beta & = & \hat{\mathbf{v}}\cdot \hat{\bfbeta}=-v_x \sin\beta + v_y \cos \beta.  \label{eq.pbeta}
    \end{eqnarray}
    We will consider $\mathbf{E}(\mathbf{0})\cdot \hat{\mathbf{v}}$ as a (linear) functional of
$\mathbf{A}=A_\alpha\hat{\bfalpha}+ A_\beta\hat{\bfbeta}$,
which  for brevity we will denote by $F(\mathbf{A})$. Hence,
\begin{equation}
F(\mathbf{A}) \stackrel{\mbox{def.}}{=}
   \frac{n^2}{\lambda_0^2} \int_0^{\alpha_{\max}}\!\int_0^{2\pi}\!
   \left[  A_\alpha(\alpha,\beta)\, v_\alpha + A_\beta(\alpha,\beta)
    v_\beta\right]\, \cos \alpha\, \sin \alpha \, d\alpha\, d\beta.
    \label{eq.defF}
     \end{equation}

Next, we calculate the total mean flow of power through a
plane $z=$constant. The total mean power flow is obtained by integrating the normal component of
the vector
$(1/2) \mbox{Re} \, \mathbf{S}$  over the plane $z=\mbox{constant}$,
where  $\mathbf{S}=
\mathbf{E}\times \mathbf{H}^*$ is the complex  Poynting vector. By
using Plancherel's formula, the integral of $\mbox{Re} \, \mathbf{S}$ over this
plane  can be written as an
integral over $k_x$ and $k_y$:
\begin{eqnarray}
  \int_{-\infty}^\infty\! \int_{-\infty}^\infty
   \frac{1}{2}\mbox{Re}\, \left[\mathbf{S}(\mathbf{r})\right]
\, d x \,dy  =
 \frac{1}{2}  \mbox{Re} \,
  \int_{-\infty}^\infty\! \int_{-\infty}^\infty  \mathbf{E}(\mathbf{r}) \times
                                                 \mathbf{H}(\mathbf{r})^*
                         \, dx\, dy \nonumber \\
 =  \frac{1}{8 \pi^2}  \mbox{Re} \,
  \int\!\!\int_{\sqrt{k_x^2+k_y^2} \leq k_0 n \sin\alpha_{\max}}
      \mathbf{A}(k_x, k_y) e^{ik_z z}
      \times \left[ \frac{\mathbf{k}}{\omega \mu_0} \times
         \mathbf{A}(k_x,k_y)^*\right] e^{-i k_z z}
     \, d k_x \, d k_y \nonumber \\
 =  \frac{1}{8 \pi^2} \frac{1}{\omega \mu_0}
  \int\!\!\int_{\sqrt{k_x^2+k_y^2} \leq k_0 n \sin \alpha_{\max}}
      | \mathbf{A}(k_x, k_y)|^2 \, \mathbf{k}\,   d k_x \,d k_y,
      \label{eq.totS}
      \end{eqnarray}
      where we used that $\mathbf{k}$ is real and
      $\mathbf{A}(k_x, k_y)\cdot \mathbf{k}=0$.
            The total time-averaged flow of energy in the positive $z$-direction
through the plane $z=\mbox{constant}$ is given by the
      $z$-component of (\ref{eq.totS}):
      \begin{eqnarray}
      \int_{-\infty}^\infty\! \int_{-\infty}^\infty \frac{1}{2}\mbox{Re} \, [S_z(\mathbf{r})]\, d x\, dy
    = \frac{1}{8 \pi^2} \frac{1}{\omega \mu_0}
  \int\!\!\int_{\sqrt{k_x^2+k_y^2} \leq  k_0 n \sin \alpha_{\max}}
      | \mathbf{A}(k_x, k_y)|^2 \,  k_z  d k_x \,d k_y  \nonumber \\
 = \frac{n^3}{2 \lambda_0^2} \left( \frac{\epsilon_0}{\mu_0}\right)^{1/2}
       \,  \int_0^{\alpha_{\max}}\!\!\int_0^{2\pi}\,
  [ |A_\alpha(\alpha,\beta)|^2 +| A_\beta(\alpha,\beta)|^2] \,
 \cos^2 \alpha \sin \alpha\,  d\alpha\, d\beta.
 \nonumber \\
 \label{eq.flowenergy}
 \end{eqnarray}
 This
 is independent of the plane $z=\mbox{constant}$,
 as should  be in a medium without losses.

The quantity $ F(\mathbf{A})=\mathbf{E}(\mathbf{0})\cdot \hat{\mathbf{v}}$ is the
complex  electric field component in the direction of $\hat{\mathbf{v}}$ at time $t=0$.
Without restricting the generality  we may assume that
$F(\mathbf{A})$
  is real.  If it were not real, a time
shift could be applied to make it  real. Hence we may assume  that
\begin{equation}
\mbox{Im}\left[ \mathbf{E}(\mathbf{0})\cdot \hat{\mathbf{v}}\right]=
  \int_0^{\alpha_{\max}}\!\int_0^{2\pi}\,
        \mbox{Im }\left[ A_\alpha(\alpha,\beta)v_\alpha  + A_\beta(\alpha,\beta) v_\beta \right] \cos \alpha\, \sin \alpha
   \,d \alpha \, d\beta =0.
   \label{eq.real}
   \end{equation}

The optimization problem is to find the plane wave amplitudes
$\mathbf{A}=A_\alpha\hat{\bfalpha} +  A_\beta \hat{\bfbeta}$ for
which the electric field component at the origin $\mathbf{0}$
that is parallel to the direction  of $\hat{\mathbf{v}}$
is larger than for  any other field with the same mean power flow
through a plane $z=\mbox{constant}$ and the same numerical aperture.
To formulate this problem  mathematically,
 we introduce  the space ${\cal H}$ of plane wave amplitudes
$\mathbf{A} =A_\alpha\hat{\bfalpha} + A_\beta \hat{\bfbeta}$
 which have finite mean flow of power through the planes
$z=\mbox{constant}$:
\begin{eqnarray}
{\cal H}& = & \{\mathbf{A}= A_\alpha\hat{\bfalpha} +  A_\beta  \hat{\bfbeta};\;\;\;
   \int_0^{\alpha_{\max}}\!\int_0^{2\pi}\,
   \left[ |A_\alpha(\alpha,\beta)|^2 + |A_\beta(\alpha, \beta)|^2 \right]  \,
 \cos^2 \alpha \sin \alpha\,   d\alpha \, d\beta < \infty \},\nonumber \\
\label{eq.defH}
\end{eqnarray}
and we define ${\cal H}_0$ as the subspace of ${\cal H}$ consisting of all $\mathbf{A}$  which
satisfy (\ref{eq.real}). Then ${\cal H}$ is
 a Hilbert space with scalar product
\begin{eqnarray}
< \mathbf{A}, \, \mathbf{B}>_{\cal H} = \int_0^{\alpha_{\max}}\,\int_0^{2\pi}\!
   \left[ A_\alpha(\alpha,\beta)  \,B_\alpha(\alpha,\beta)^* + A_\beta(\alpha,\beta)\, B_\beta(\alpha,\beta)^*
   \right] \,
 \cos^2 \alpha \, \sin \alpha\,  d\alpha\, d\beta,
\label{eq.scalar}
\end{eqnarray}
and ${\cal H}_0$ is a closed subspace of ${\cal H}$. Note that the
constraint (\ref{eq.real}) means that $\mbox{Im}( \mathbf{A})$ is
perpendicular to the vector field $\hat{\mathbf{v}}/\cos \alpha$ in the space ${\cal H}$, i.e.
$\mbox{Im} (\mathbf{A})$ is perpendicular to $\hat{\mathbf{v}}/\cos \alpha$   in the sense of
scalar product (\ref{eq.scalar}).

Define the quadratic functional
\begin{eqnarray}
 P(\mathbf{A}) & \stackrel{\mbox{def.}}{=} &
\frac{n^3}{2 \lambda_0^2} \left( \frac{\epsilon_0}{\mu_0}\right)^{1/2}
       \, \int_0^{\alpha_{\max}}\!\int_0^{2\pi}\,
  \left[ |A_\alpha(\alpha,\beta)|^2 + |A_\beta(\alpha,\beta)|^2\right]  \,
 \cos^2 \alpha \sin \alpha\,  d\alpha\, d \beta,
 \label{eq.defP}
\end{eqnarray}
which is the mean  power flowing  through a plane
$z=\mbox{constant}$ for fields with plane wave amplitudes
$\mathbf{A}=A_\alpha \hat{\bfalpha}+A_\beta \hat{\bfbeta}$.
Then the optimization problem is to find,
 for given  $P_0 >0$, the
solution of
\begin{eqnarray}
(*) \;\;\; \max_{\mathbf{A}\in {\cal H}_0} F(\mathbf{A}), \;\;
\mbox{ under the constraint } P(\mathbf{A})\leq P_0. \nonumber
\end{eqnarray}
For any solution of problem  (*)  the equality $P(\mathbf{A})=P_0$ holds,
because otherwise $\mathbf{A}$ could be multiplied by the number
$(P_0/P(\mathbf{A}))^{1/2}>1$ and this would increase the value of $F$
without violating the constraint on the energy. Hence it does not matter
whether we impose the equality constraint $P(\mathbf{A})=P_0$ or
the inequality constraint $P(\mathbf{A}) \leq P_0$ on the mean flow
of energy.

It is not completely obvious that  problem (*) has a unique solution since it
is posed in a linear  space ${\cal H}_0$ of  infinite dimension.
However, there is a functional analytic theorem  which states that
a continuous real linear functional attains its supremum on  a
sphere in a Hilbert space and that the solution is unique
\cite{teman}. Since the functional $F$ is linear, real and
continuous with respect to the norm on ${\cal H}$, and since the
feasible set of problem (*) is a sphere in ${\cal H}_0$, this
theorem applies to our problem. Hence the optimization problem has a
unique solution.  In the next section we shall compute the solution.

\section{Optimum plane wave amplitudes}
Since $F$ is a linear functional, the Fr\'{e}chet derivative of $F$ at $\mathbf{A}$ in the direction of $\mathbf{B}$
is simply $F(\mathbf{B})$, i.e.
\begin{eqnarray}
\delta F(\mathbf{A})(\mathbf{B}) = F(\mathbf{B})
= \frac{n^2}{\lambda_0^2} \int_0^{\alpha_{\max}}\!\int_0^{2\pi}\,
   \left[B_\alpha(\alpha,\beta)v_\alpha
   + B_\beta(\alpha,\beta)v_\beta\right]\,
    \cos \alpha \sin \alpha \, d\alpha \, d\beta.
    \label{eq.dF}
    \end{eqnarray}
    The Fr\'{e}chet derivative of the quadratic functional $P(\mathbf{A})$ is:
\begin{eqnarray}
\delta P(A_\alpha)(B_\alpha)  =
\frac{n^3}{\lambda_0^2} \left( \frac{\epsilon_0}{\mu_0}\right)^{1/2}
\mbox{Re}\,
 \int_0^{\alpha_{\max}}\!\int_0^{2\pi}\,\left[
 A_\alpha(\alpha,\beta)\, B_\alpha(\alpha,\beta)^*\right. \nonumber \\
 \left.+ A_\beta(\alpha)B_\beta(\alpha,\beta)^*\right]
   \, \cos^2 \alpha \,\sin \alpha\,  d \alpha\, d\beta.
  \label{eq.dG}
  \end{eqnarray}
According to the
Lagrange multiplier rule for inequality constraints
 (also known as Kuhn-Tucker's theorem) \cite{luenberger},
 there exists
a Lagrange multiplier $\Lambda \geq 0$ such that, if $\mathbf{A}$ is the optimum field, we have:
\begin{equation}
 \delta F(\mathbf{A})(\mathbf{B}) - \Lambda\,
 \delta P(\mathbf{A})(\mathbf{B})=0, \;\; \mbox{ for all } \mathbf{B} \mbox{ in } {\cal H}_0,
 \label{eq.lagrange}
 \end{equation}
 and
 \begin{equation}
   \Lambda \left[P(\mathbf{A})-P_0\right]=0.
   \label{eq.constraint}
   \end{equation}
 In the previous section we have shown that $P(\mathbf{A})=P_0$, therefore the last equation does not give
 new information.
 By substituting (\ref{eq.dF}) and (\ref{eq.dG}) into the Lagrange multiplier rule, we get
\begin{eqnarray}
\frac{n^2}{\lambda_0^2}
  \int_0^{\alpha_{\max}}\, \int_0^{2\pi}\,
\left[B_\alpha(\alpha,\beta)v_\alpha + B_\beta(\alpha,\beta)v_\beta\right] \, \cos\alpha \sin \alpha \, d\alpha\, d\beta
\nonumber \\
-\Lambda
\frac{n^3}{\lambda_0^2}\left(\frac{\epsilon_0}{\mu_0}\right)^{1/2}
\mbox{Re}
 \int_0^{\alpha_{\max}}\!\int_0^{2\pi}\,
\left[  A_\alpha(\alpha,\beta)\, B_\alpha(\alpha,\beta)^* + A_\beta(\alpha)B_\beta(\alpha,\beta)^*\right]
 \, \cos^2 \alpha \, \sin \alpha
\, d \alpha\, d\beta =0, \nonumber\\
  \mbox{ for all } \mathbf{B} \mbox{ in } {\cal H}_0.  \hspace{1cm}.
\label{eq.lagrange2}
\end{eqnarray}
Because $\mathbf{B}$ satisfies  (\ref{eq.real}), it follows that in the
first integral we may replace
  $B_\alpha$ and $B_\beta$ by $B_\alpha^*$ and $B_\beta^*$, respectively.
Hence,
\begin{eqnarray}
\mbox{Re}
 \int_0^{\alpha_{\max}}\!\int_0^{2\pi}\,
  \left[ \left( \frac{v_\alpha}{\cos \alpha}  - \Lambda n \left(\frac{\epsilon_0}{\mu_0}\right)^{1/2}
  A_\alpha \right)
   B_\alpha^* \, \right.
   \nonumber \\
   \left. + \left(\frac{v_\beta}{\cos \alpha} - \Lambda n \left(\frac{\epsilon_0}{\mu_0}\right)^{1/2} A_\beta
  \right) B_\beta^*
  \right]
   \,\cos^2 \alpha \sin\alpha \,d\alpha\, d\beta =0,  \mbox{ for all } \mathbf{B} \mbox{ in } {\cal H}_0.
     \label{eq.lagrange3}
\end{eqnarray}
This is equivalent to
\begin{eqnarray}
  \int_0^{\alpha_{\max}}\!\int_0^{2\pi}\, \left[
    \left( \frac{v_\alpha}{\cos \alpha} - \Lambda n\left(\frac{\epsilon_0}{\mu_0}\right)^{1/2} \mbox{Re}(A_\alpha)  \right)
    \mbox{Re}(B_\alpha) - \Lambda n \left(\frac{\epsilon_0}{\mu_0}\right)^{1/2} \mbox{Im}(A_\alpha)
      \, \mbox{Im}(B_\alpha) \right.  \nonumber \\
  \left. +
     \left( \frac{v_\beta}{\cos \alpha} - \Lambda n\left(\frac{\epsilon_0}{\mu_0}\right)^{1/2} \mbox{Re}(A_\beta)  \right)
    \mbox{Re}(B_\beta) - \Lambda n \left(\frac{\epsilon_0}{\mu_0}\right)^{1/2} \mbox{Im}(A_\beta) \,
    \mbox{Im}(B_\beta)\right]
    \nonumber \\
    \times \cos^2\alpha \sin \alpha \, d\alpha\, d\beta=0,
    \nonumber \\
    \label{eq.LM}
    \end{eqnarray}
    for all $\mathbf{B} \mbox{ in } {\cal H}_0$, i.e. for all $\mathbf{B}$ for which
\begin{equation}
 \int_0^{\alpha_{\max}}\!\int_0^{2\pi}\, \left[ \mbox{Im}( B_\alpha) v_\alpha +
 \mbox{Im}(B_\beta) v_\beta \right]\, \cos^2 \alpha \sin \alpha \, d\alpha\, d\beta=0.
 \label{eq.IMB}
 \end{equation}
 Choose first $B_\alpha$ and $B_\beta$ real. Then (\ref{eq.IMB}) is obviously satisfied and
 (\ref{eq.LM}) implies:
 \begin{eqnarray}
   \mbox{Re}(A_\alpha) & = &\frac{1}{\Lambda n} \left(\frac{\mu_0}{\epsilon_0}\right)^{1/2} \frac{v_\alpha}{\cos \alpha},
   \label{eq.ReAalpha} \\
  \mbox{Re}(A_\beta) & = &\frac{1}{\Lambda n} \left(\frac{\mu_0}{\epsilon_0}\right)^{1/2} \frac{v_\beta}{\cos \alpha}.
   \label{eq.ReAbeta}
   \end{eqnarray}
   By substituting this in (\ref{eq.LM}) it follows that
   \begin{eqnarray}
    \int_0^{\alpha_{\max}}\!\int_0^{2\pi}\, \left[ \mbox{Im}(A_\alpha)  \, \mbox{Im}(B_\alpha)
    + \mbox{Im}(A_\alpha)  \, \mbox{Im}(B_\alpha) \right]\, \cos^2 \alpha \sin \alpha =0,
       \label{eq.LM2}
    \end{eqnarray}
   for all $B_\alpha$, $B_\beta$ that satisfy (\ref{eq.IMB}).
   This can be stated alternatively by saying that if $\mathbf{B}=B_\alpha \hat{\bfalpha}
   + B_\beta \hat{\bfbeta} $ is perpendicular to
   $(v_\alpha/\cos\alpha) \hat{\bfalpha} + (v_\beta /\cos \alpha)\hat{\bfbeta}$,
   then $\mathbf{B}$ is perpendicular to
   $\mbox{Im}(A_\alpha) \hat{\bfalpha} + \mbox{Im}(A_\beta)\hat{\bfbeta}$ (perpendicular means here
   of course with respect to scalar product (\ref{eq.scalar})).
   We conclude that  $\mbox{Im}(A_\alpha) \hat{\bfalpha}+\mbox{Im}(A_\beta)\hat{\bfbeta}$ is
   proportional to $(v_\alpha/\cos\alpha) \hat{\bfalpha} + (v_\beta /\cos \alpha)\hat{\bfbeta}$:
      \begin{eqnarray}
   \mbox{Im}(A_\alpha) & = & C \frac{v_\alpha}{\cos \alpha}, \label{eq.ImAalpha}\\
   \mbox{Im}(A_\beta) &=& C \frac{v_\beta}{\cos\alpha},\label{eq.ImAbeta}
   \end{eqnarray}
   for some constant $C$.
   We shall now show that $C=0$. By substitution of (\ref{eq.ImAalpha}) and (\ref{eq.ImAbeta})
   into (\ref{eq.real}) we get
   \begin{equation}
     C  \int_0^{\alpha_{\max}}\!\int_0^{2\pi}\, \left( v_\alpha^2 +v_\beta^2\right) \, \cos \alpha \sin \alpha \,
     d \alpha\, d\beta=0.
     \label{eq.IMB3}
     \end{equation}
     If $C\neq 0$, then we must have
     \begin{equation}
     v_\alpha=v_\beta=0, \mbox{ for all } \alpha, \beta \mbox{ with } 0 \leq \alpha \leq \alpha_{\max}, \;
     \; 0 \leq \beta \leq 2\pi.
     \label{eq.hoera}
     \end{equation}
                Use the expressions  (\ref{eq.palpha}), (\ref{eq.pbeta}) for
   $v_\alpha$ and $v_\beta$ in terms of the cartesian components $v_x, v_y$ and $v_z$.
   It is then easily seen that (\ref{eq.hoera}) implies: $v_x=v_y=v_z=0$. This  contradicts the assumption
   that $\mathbf{v}$ is a unit vector. Hence
   $C=0$.

   We thus conclude that the plane wave amplitudes of the optimum field are given by
   \begin{eqnarray}
    A_\alpha&  = &  \frac{1}{\Lambda n} \left(\frac{\mu_0}{\epsilon_0}\right)^{1/2}
                    \, \frac{v_\alpha}{\cos \alpha}, \label{eq.Aalpha} \\
                    A_\beta & = & \frac{1}{\Lambda n} \left(\frac{\mu_0}{\epsilon_0}\right)^{1/2}
                    \, \frac{v_\beta}{\cos \alpha}.  \label{eq.Abeta}
   \end{eqnarray}
        The Lagrange multiplier $\Lambda$ can be determined by substituting
        (\ref{eq.Aalpha}) and  (\ref{eq.Abeta}) into $P(\mathbf{A})=P_0$ and then using
        (\ref{eq.palpha}), (\ref{eq.pbeta}). We find
                \begin{eqnarray}
      P(\mathbf{A}) & = &
        \frac{n}{2 \Lambda^2 \lambda_0^2} \left( \frac{\mu_0}{\epsilon_0}\right)^{1/2}
       \, \int_0^{\alpha_{\max}}\!\int_0^{2\pi}\,
  \left( v_\alpha^2 + v_\beta^2\right)  \,
  \sin \alpha\,  d\alpha\, d \beta \label{eq.PmaxI} \\
 & = &
 \frac{n}{2 \Lambda^2 \lambda_0^2} \left( \frac{\mu_0}{\epsilon_0}\right)^{1/2}
       \left\{ v_x^2\, \int_0^{\alpha_{\max}}\!\int_0^{2\pi}\, (\cos^2\alpha \cos^2\beta +\sin^2\beta) \,
       \sin \alpha \,  d\alpha \, d\beta
       \right. \nonumber \\
       & &  \left. +  v_y^2\, \int_0^{\alpha_{\max}}\!\int_0^{2\pi}\, (\cos^2\alpha \sin^2\beta +\cos^2\beta) \,
         \sin \alpha \, d\alpha \, d\beta
       \right. \nonumber \\
      & &  \left. +  v_z^2 \, \int_0^{\alpha_{\max}}\!\int_0^{2\pi}\, \sin^3 \alpha \,   d\alpha \, d\beta
      \right. \nonumber \\
      & & \left. - 2 v_x v_y\,\int_0^{\alpha_{\max}}\!\int_0^{2\pi}\,  \sin^3 \alpha \cos\beta \sin \beta\,   d\alpha \, d\beta \right. \nonumber \\
      & & \left.
      - 2 v_x v_z\,  \int_0^{\alpha_{\max}}\!\int_0^{2\pi}\,  \cos\alpha \sin^2 \alpha \cos \beta \, d\alpha\, d\beta
      \right.
      \nonumber \\
            & & \left. -2 v_y v_z\,  \int_0^{\alpha_{\max}}\!\int_0^{2\pi}\, \cos \alpha \sin^2 \alpha \sin \beta \, d\alpha \, d\beta \,      \right\} \nonumber \\
      & = & \frac{\pi n}{2 \Lambda^2 \lambda_0^2} \left( \frac{\mu_0}{\epsilon_0}\right)^{1/2}
      \left[   \left(\frac{4}{3}-\cos \alpha_{\max} -\frac{1}{3}\cos^3\alpha_{\max}\right)(v_x^2 + v_y^2)
      \right. \nonumber \\
      & & \left. +  \left( \frac{4}{3} -2 \cos \alpha_{\max} + \frac{2}{3} \cos^3 \alpha_{\max} \right)
        v_z^2
      \right] \nonumber \\
      & = & \frac{\pi n}{2 \Lambda^2 \lambda_0^2} \left( \frac{\mu_0}{\epsilon_0}\right)^{1/2}
      \left[   \frac{4}{3}-\cos \alpha_{\max} -\frac{1}{3}\cos^3\alpha_{\max}
      - \sin^2\alpha_{\max} \cos \alpha_{\max} \, v_z^2 \right].
 \label{eq.Pmax}
\end{eqnarray}
It  follows from  $P(\mathbf{A})=P_0$ and $\Lambda \geq 0$,
that
\begin{eqnarray}
\Lambda  & = & \sqrt{\frac{\pi}{2}}\frac{n^{1/2}}{P_0^{1/2}
\lambda_0}\left( \frac{\mu_0}{\epsilon_0}\right)^{1/4}
\,
\left[  \frac{4}{3}-\cos \alpha_{\max} -\frac{1}{3}\cos^3\alpha_{\max}
       - \sin^2\alpha_{\max} \cos\alpha_{\max} v_z^2
       \right]^{1/2}.
      \label{eq.Lambda}
\end{eqnarray}
Herewith the derivation of the plane waves amplitudes  of the optimum field is complete.

The maximum of the field component at the origin, i.e. of $F$, is:
\begin{eqnarray}
F_{\max} = F(\mathbf{A}) & = &
   \frac{n^2}{\lambda_0^2} \int_0^{\alpha_{\max}}\!\int_0^{2\pi}\!
   \left[  A_\alpha(\alpha,\beta)\, v_\alpha + A_\beta(\alpha,\beta)
    v_\beta\right]\, \cos \alpha\, \sin \alpha \, d\alpha\, d\beta \nonumber \\
    & = & \frac{1}{\Lambda} \frac{n}{\lambda_0^2} \left(\frac{\mu_0}{\epsilon_0}\right)^{1/2}
    \int_0^{\alpha_{\max}}\!\int_0^{2\pi}\! \left( v_\alpha^2 + v_\beta^2\right) \, \sin\alpha\, d\alpha\, d\beta
    \nonumber \\
    & = &  2 \Lambda  P_0
    \nonumber \\
    & = &
         \sqrt{2\pi} P_0^{1/2} \frac{n^{1/2}}{\lambda_0} \left(\frac{\mu_0}{\epsilon_0}\right)^{1/4}
        \left[  \frac{4}{3}-\cos \alpha_{\max} -\frac{1}{3}\cos^3\alpha_{\max}
      -  \sin^2\alpha_{\max} \cos\alpha_{\max} v_z^2
           \right]^{1/2},\nonumber \\
              \label{eq.maxFII}
            \label{eq.maxF}
     \end{eqnarray}
where we used (\ref{eq.PmaxI}) and (\ref{eq.Lambda}).

        \section{The optimum electromagnetic field}
        \label{section.optimum}
   The electric field amplitudes of the plane waves of the optimum field are given by
    \begin{eqnarray}
    \mathbf{A}(\alpha,\beta)& = &
    A_\alpha(\alpha,\beta) \hat{\bfalpha} + A_\beta(\alpha,\beta) \hat{\bfbeta}
    \nonumber \\
        & = & \frac{1}{\Lambda n}\left(\frac{\mu_0}{\epsilon_0}\right)^{1/2}\left( v_\alpha
    \hat{\bfalpha} + v_\beta \hat{\bfbeta} \right) \frac{1}{\cos \alpha} \nonumber \\
    & = & \frac{1}{\Lambda n}\left(\frac{\mu_0}{\epsilon_0}\right)^{1/2}
    \left[ \left(\cos \beta \, \hat{\bfalpha} - \frac{\sin\beta}{\cos \alpha}  \, \hat{\bfbeta}\right) \,v_x
          +\left( \sin \beta \, \hat{\bfalpha} + \frac{\cos \beta}{\cos \alpha}\,\hat{\bfbeta} \right) \,v_y - \tan\alpha \,\hat{\bfalpha}\, v_z\right]
          \nonumber \\
          & = &\frac{1}{\Lambda n}\left(\frac{\mu_0}{\epsilon_0}\right)^{1/2}
          \left\{  \left[ \left( \cos \alpha \cos^2 \beta + \frac{\sin^2\beta}{\cos\alpha}\right) \hat{\mathbf{x}}
          -\frac{\sin^2 \alpha}{\cos \alpha} \cos \beta \sin\beta \, \hat{\mathbf{y}} - \sin \alpha \cos \beta \, \hat{\mathbf{z}} \right] \, v_x
          \right.
          \nonumber \\
          & & \left.+ \left[ -\frac{\sin^2 \alpha}{\cos \alpha} \cos \beta \sin \beta  \, \hat{\mathbf{x}}
           +  \left( \cos\alpha \sin^2 \beta + \frac{\cos^2 \beta}{\cos \alpha} \right) \, \hat{\mathbf{y}}
            - \sin \alpha \sin \beta \,
            \hat{\mathbf{z}} \right] \,v_y \right. \nonumber \\
            & & \left. + \left[ - \sin \alpha \cos \beta \hat{\mathbf{x}} - \sin \alpha \sin \beta \hat{\mathbf{y}}
            + \frac{\sin^2 \alpha}{\cos \alpha }\, \hat{\mathbf{z}} \right] \, v_z \right\}.
                        \label{eq.plw1}
    \end{eqnarray}
    If we write the right-hand side of (\ref{eq.plw1}) as the product of a matrix and the vector $\mathbf{v}$
    on the cartesian basis $\hat{\mathbf{x}}$, $\hat{\mathbf{y}}$, $ \hat{\mathbf{z}}$, we get
    \begin{equation}
      \mathbf{A}(\alpha,\beta) =  \frac{1}{\Lambda n}\left(\frac{\mu_0}{\epsilon_0}\right)^{1/2}
      \left( \begin{array}{ccc}
        \cos \alpha \cos^2 \beta + \frac{\sin^2\beta}{\cos\alpha} \;\;\;&  -\frac{\sin^2 \alpha}{\cos \alpha} \cos \beta \sin \beta  &   \;\;\; -\sin \alpha \cos \beta   \\
      \;\;\;  -\frac{\sin^2 \alpha}{\cos \alpha} \cos \beta \sin\beta                                     \;\;\;     &   \cos\alpha \sin^2 \beta + \frac{\cos^2 \beta}{\cos \alpha} & \;\;\;  -\sin \alpha \sin \beta \\
         -\sin \alpha \cos \beta                              \;\;\;        &  -\sin \alpha \sin \beta  & \frac{\sin^2 \alpha}{\cos \alpha }
         \end{array}\right) \left( \begin{array}{c}
                    v_x \\
                    v_y \\
                    v_z
                    \end{array}\right).\nonumber \\
                    \label{eq.plw2}
                        \end{equation}

       We shall use cylindrical coordinates $\varrho$, $\varphi$, $z$ for
   the  point of observation $\mathbf{r}$. There holds
\begin{equation}
\varrho=r\sin \varphi,   \;\;\; z=r \cos \varphi,
\label{eq.cyl}
\end{equation}
and the unit vectors
 $\hat{\bfphi}$, $\hat{\bfrho}$ are defined by
 \begin{eqnarray}
\hat{\bfrho}& = & \cos \varphi\, \hat{\mathbf{x}} + \sin
  \varphi \, \hat{\mathbf{y}},
  \label{eq.defhatrho} \nonumber \\
\hat{\bfphi}& = & -\sin \varphi\, \hat{\mathbf{x}} + \cos
  \varphi \, \hat{\mathbf{y}}.
  \label{eq.defhatphi}
 \end{eqnarray}
 Then
 \begin{eqnarray}
 \mathbf{k} \cdot \mathbf{r} & = & k_0 n ( x \sin \alpha \cos \beta + y \sin \alpha \sin \beta + z \cos \alpha)
 \nonumber \\
 & = & k_0 n ( \varrho \cos \varphi \sin \alpha \cos \beta +
                 \varrho \sin \varphi \sin \alpha \sin \beta + z \cos \alpha) \nonumber \\
                 & = & k_0 n   [ \varrho \sin \alpha \cos (\varphi -\beta) + z \cos \alpha] \nonumber \\
                 & = & k_0 n \varrho \sin \alpha \, \cos (\varphi-\beta) + k_0 n z \cos \alpha.
                 \label{eq.kr}
                 \end{eqnarray}
  The optimum electric field in a point $\mathbf{r}$ with cylindrical coordinates
  $\varrho, \varphi, z$ is then:
\begin{eqnarray}
     \mathbf{E}(\varrho,\varphi,z)  =
       \frac{n^2}{\lambda_0^2}
        \int_0^{\alpha_{\max}}\!\int_0^{2\pi}\,
    \left[ A_\alpha(\alpha,\beta) \hat{\bfalpha} + A_\beta(\alpha,\beta) \hat{\bfbeta}\right]
    \, e^{i \mathbf{k}\cdot\mathbf{r}}
    \, \sin\alpha \, \cos \alpha \, d\alpha\, d\beta \nonumber \\
       =  \frac{n}{\Lambda\lambda_0^2} \left(\frac{\mu_0}{\epsilon_0}\right)^{1/2}
        \int_0^{\alpha_{\max}} \sin \alpha \,
    e^{i k_0 n z\, \cos\alpha }\, d\alpha \nonumber \\
     \int_0^{2 \pi}
    \left( \begin{array}{ccc}
      1-\sin^2\alpha   \cos^2 \beta \;\;\;&  -\sin^2 \alpha\cos \beta \sin \beta  &   \;\;\; -\cos \alpha \sin \alpha \cos \beta   \\
      \;\;\;  -\sin^2 \alpha \cos \beta \sin\beta                & \;\;\; 1-\sin^2 \alpha \sin^2\beta & \;\;\;  -\cos \alpha \sin \alpha \sin \beta \\
         -\cos \alpha \sin \alpha \cos \beta                              \;\;\;        &  -\cos \alpha \sin \alpha \sin \beta  &\sin^2 \alpha
      \end{array} \right)
      \left( \begin{array}{c}
                v_x \\
                v_y \\
                v_z
                \end{array}\right)\nonumber \\
       \times e^{i k_0 n \varrho\, \sin \alpha\, \cos(\beta-\varphi) }
       \, d\beta. \nonumber \\
       \label{eq.Efield}
       \end{eqnarray}
              Furthermore, the magnetic field amplitudes of the plane waves are
       (\ref{eq.bfH}):
    \begin{eqnarray}
   -A_\beta(\alpha,\beta) \hat{\bfalpha} +  A_\alpha(\alpha,\beta) \hat{\bfbeta} =
    \frac{1}{\Lambda n} \left(\frac{\mu_0}{\epsilon_0}\right)^{1/2}
    \left( -v_\beta \hat{\bfalpha} + v_\alpha \hat{\bfbeta} \right) \frac{1}{\cos \alpha} \nonumber \\
    =   \frac{1}{\Lambda n} \left(\frac{\mu_0}{\epsilon_0}\right)^{1/2}
    \left[
   \left( - \frac{\sin \beta}{\cos \alpha} \hat{\bfalpha} + \cos \beta \hat{\bfbeta} \right)\,
    v_x + \left( \frac{\cos\beta}{\cos \alpha} \hat{\bfalpha} + \sin \beta \hat{\bfbeta} \right)\,v_y
    -\tan\alpha \, \hat{\bfbeta} \, v_z \right] \nonumber \\
     = \frac{1}{\Lambda n} \left(\frac{\mu_0}{\epsilon_0}\right)^{1/2} \left\{
    \left[- 2 \cos\beta \sin\beta \, \hat{\mathbf{x}} + (\cos^2\beta -\sin^2\beta) \, \hat{\mathbf{y}}
    +  \tan\alpha \, \sin  \beta \, \hat{\mathbf{z}}\right] v_x \right.\nonumber \\
     \left.+ \left[ (\cos^2\beta-\sin^2\beta) \hat{\mathbf{x}} + 2 \cos\beta\sin\beta \, \hat{\mathbf{y}}
    -\tan \alpha \cos \beta \, \hat{\mathbf{z}} \right] v_y \right.\nonumber \\
    \left. + \left(\sin\beta \, \hat{\mathbf{x}} - \cos\beta \hat{\mathbf{y}}\right)\,\tan\alpha \, v_z \right\}
           \label{eq.2}
    \end{eqnarray}
    Hence, on the cartesian basis
    \begin{eqnarray}
    -A_\beta(\alpha,\beta) \hat{\bfalpha} +  A_\alpha(\alpha,\beta) \hat{\bfbeta} = \nonumber \\
     \frac{1}{\Lambda n} \left(\frac{\mu_0}{\epsilon_0}\right)^{1/2}
     \left( \begin{array}{ccc}
        - 2 \cos\beta \sin\beta  & \;\;\;\cos^2\beta -\sin^2\beta & \;\;\;\tan\alpha \, \sin  \beta \\
        \cos^2\beta-\sin^2\beta&\;\;\; 2\cos \beta \sin \beta & \;\;\; - \tan\alpha \cos \beta \\
        \tan \alpha \sin \beta & \;\;\; - \tan \alpha \cos \beta &   \;\;\; 0
        \end{array}\right)\left( \begin{array}{c}
            v_x \\
            v_y \\
            v_z
            \end{array}\right).
            \label{eq.Hmatrix}
            \end{eqnarray}
            The optimum magnetic field is thus
            \begin{eqnarray}
     \mathbf{H}(\varrho, \varphi,z)  =
       \frac{n^3}{\Lambda \lambda_0^2}\left(\frac{\epsilon_0}{\mu_0}\right)^{1/2}
                \int_0^{\alpha_{\max}}\!\int_0^{2\pi}\,
    \left[ -A_\beta(\alpha,\beta) \hat{\bfalpha} + A_\alpha(\alpha,\beta) \hat{\bfbeta}\right]
    \, e^{i \mathbf{k}\cdot\mathbf{r}}
    \, \sin\alpha \, \cos \alpha \, d\alpha\, d\beta \nonumber \\
       =  \frac{n^2}{\Lambda\lambda_0^2}
        \int_0^{\alpha_{\max}} \sin \alpha \,
    e^{i k_0 n z\, \cos\alpha }\, d\alpha \nonumber \\
     \int_0^{2 \pi}
    \left( \begin{array}{ccc}
     -  \cos \alpha \sin(2\beta)  & \;\;\;\cos\alpha \cos(2\beta) & \;\;\;\sin\alpha \, \sin  \beta \\
        \cos\alpha \cos(2\beta)&\;\;\; \cos\alpha \sin (2\beta) & \;\;\; - \sin\alpha \cos \beta \\
        \sin\alpha \sin \beta & \;\;\; - \sin \alpha \cos \beta &   \;\;\; 0
      \end{array} \right)
      \left( \begin{array}{c}
                v_x \\
                v_y \\
                v_z
                \end{array}\right)
       \, e^{i k_0 n \varrho\, \sin \alpha\, \cos(\beta-\varphi) }
       \, d\beta.
       \nonumber \\
       \label{eq.Efield}
       \end{eqnarray}

The integrals over $\beta$ can be computed with the following  formulas \cite{prudnikov2}
\begin{eqnarray}
\int_0^{2\pi} e^{i \zeta \cos(\beta -\varphi)} \cos(m \beta) \, d\beta & = &
   2 \pi \, i^m J_m(\zeta) \cos (m \varphi), \label{eq.cos}  \\
\int_0^{2\pi} e^{i \zeta \cos(\beta -\varphi)}\, \sin(m\beta) \, d\beta & = &
   2 \pi \, i^m J_m(\zeta) \sin (m\varphi),
   \label{eq.sin}
  \end{eqnarray}
  for $m=0,1,2,\ldots$. Hence,
  \begin{eqnarray}
  \int_0^{2\pi} e^{i \zeta \cos(\beta -\varphi)} \cos^2 \beta \, d\beta & = &
  \pi \left[ J_0(\zeta) - J_2(\zeta) \cos(2\varphi)\right],
  \label{eq.cossq}\\
  \int_0^{2\pi} e^{i \zeta \cos(\beta -\varphi)} \sin^2 \beta \, d\beta & = &
  \pi \left[ J_0(\zeta) + J_2(\zeta) \cos(2\varphi)\right],
  \label{eq.sinsq}\\
  \int_0^{2\pi} e^{i \zeta \cos(\beta -\varphi)} \cos \beta \sin\beta\, d\beta & = &
  -\pi J_2(\zeta) \, \sin(2\varphi).
  \label{eq.cossin}
  \end{eqnarray}

 By using  the notation
\begin{eqnarray}
  g_l^{\nu,\mu}(\varrho, z) & = &  \int_0^{\alpha_{\max}}
  e^{i k_0 n z\, \cos\alpha }
  \cos^\nu \alpha \,\sin^\mu \alpha\, J_l(
 k_0 n \varrho \sin \alpha   ) \, d \alpha,
  \label{eq.glnumu}
  \end{eqnarray}
the electric and magnetic fields can be expressed on the cartesian basis as:
\begin{eqnarray}
\mathbf{E}(\varrho,\varphi,z)     =  \pi \frac{n}{\Lambda\lambda_0^2} \left(\frac{\mu_0}{\epsilon_0}\right)^{1/2}
   \nonumber \\
    \times  \left\{
            \left( \begin{array}{ccc}
    g_0^{0,1}(\varrho,z) + g_0^{2,1}(\varrho,z)  & \;\; 0 & \;\; 0 \\
    0 & \;\;  g_0^{0,1}(\varrho,z) + g_0^{2,1}(\varrho,z)& \;\; 0  \\
    0 & \;\; 0  &  \;\; 2 g_0^{0,3}(\varrho,z)
    \end{array}\right)
      \right.
                \nonumber \\
        \left.    +
        \left( \begin{array}{ccc}
                g_2^{0,3}(\varrho,z) \cos(2\varphi) & \;\; g_2^{0,3}(\varrho,z) \sin(2\varphi)  & \;\;
                -2 i g_1^{1,2}(\varrho,z)  \cos \varphi \\
                g_2^{0,3}(\varrho,z) \sin(2\varphi) & \;\;\; -g_2^{0,3}(\varrho,z) \cos(2\varphi) & \;\;\;
                 -2 i g_1^{1,2}(\varrho,z)  \sin \varphi  \\
                                            -2 i g_1^{1,2}(\varrho,z)  \cos \varphi &\;\;     -2 i g_1^{1,2}(\varrho,z)  \sin \varphi &\;\;             0  \end{array} \right)
           \right\}
                   \left( \begin{array}{c}
                v_x \\
           v_y \\
                v_z
                \end{array} \right),
                             \label{eq.Efield3}
       \end{eqnarray}
        \begin{eqnarray}
     \mathbf{H}(\varrho,\varphi,z)  =
       \frac{2\pi n^2}{\Lambda \lambda_0^2}
       \left( \begin{array}{ccc}
        g_2^{1,1}(\varrho,z) \sin(2\varphi) & \;\; -g_2^{1,1}(\varrho,z)\cos(2\varphi) & \;\;
                                                                       i g_1^{0,2}(\varrho,z) \sin \varphi \\
       -g_2^{1,1}(\varrho,z)\cos(2\varphi) & \;\;-g_2^{1,1}(\varrho,z) \sin(2\varphi) & \;\;-i g_1^{0,2}(\varrho,z) \cos\varphi \\
       ig_1^{0,2}(\varrho,z) \sin \varphi & \;\; -i g_1^{0,2}(\varrho,z) \cos\varphi & \;\; 0
       \end{array}\right)
        \left( \begin{array}{c}
                v_x \\
                v_y \\
                v_z
                \end{array} \right).\nonumber \\
               \label{eq.Hfield3}
    \end{eqnarray}
Alternative concise expressions are obtained  when the electric field, the magnetic field and
the vector $\mathbf{v}$ are all written on the
local cylindrical unit basis $\{\hat{\bfrho}, \hat{\bfphi}, \hat{\mathbf{z}}\}$ (attached
to the point of observation $\mathbf{r} = \varrho \hat{\bfrho} + z \hat{\mathbf{z}}$):
 \begin{eqnarray}
\mathbf{E}(\varrho,\varphi,z)     =   \frac{\pi n}{\Lambda\lambda_0^2} \left(\frac{\mu_0}{\epsilon_0}\right)^{1/2}
   \nonumber \\ \left\{ \left[ \left( g_0^{0,1}(\varrho,z) + g_0^{2,1}(\varrho,z) +  g_2^{0,3}(\varrho,z)\right)\,
   (v_x \cos \varphi + v_y \sin\varphi )
     - 2 i g_1^{1,2}(\varrho,z) \, v_z \right] \, \hat{\bfrho} \right. \nonumber \\
     \left. + \left[ g_0^{0,1}(\varrho,z)+ g_0^{2,1}(\varrho,z)-g_2^{0,3}(\varrho,z) \right] \,
     ( - v_x \sin \varphi + v_y \cos \varphi) \, \hat{\bfphi}
  \right.   \nonumber \\
    + \left. \left[ -2i g_1^{1,2}(\varrho,z) \,(v_x \cos \varphi + v_y \sin \varphi)
     + 2 g_0^{0,3}(\varrho,z) \,v_z\right]\, \hat{\mathbf{z}}
     \right\}\label{eq.Efield4} \\
=  \frac{\pi n}{\Lambda\lambda_0^2} \left(\frac{\mu_0}{\epsilon_0}\right)^{1/2}
   \left\{ \left[ \left( g_0^{0,1}(\varrho,z) + g_0^{2,1}(\varrho,z) +  g_2^{0,3}(\varrho,z)\right)\, v_\varrho
     - 2 i g_1^{1,2}(\varrho,z) \, v_z \right] \, \hat{\bfrho} \right. \nonumber \\
     \left. + \left[ g_0^{0,1}(\varrho,z)+ g_0^{2,1}(\varrho,z)-g_2^{0,3}(\varrho,z) \right] \,v_\varphi\, \hat{\bfphi}
  \right.   \nonumber \\
     \left. + \left[ -2i g_1^{1,2}(\varrho,z) \,v_\varrho + 2 g_0^{0,3}(\varrho,z) \,v_z\right]\, \hat{\mathbf{z}}
     \right\}
    \label{eq.Efield5}
       \end{eqnarray}
       \begin{eqnarray}
     \mathbf{H}(\varrho,\varphi,z)  =
      -\frac{2\pi n^2}{\Lambda \lambda_0^2}\, \left\{
          g_2^{1,1}(\varrho,z) \, (- v_x \sin\varphi + v_y \cos \varphi) \, \hat{\bfrho}
          \right. \nonumber \\
         \left.
          + [ g_2^{1,1}(\varrho,z) \, (v_x \cos \varphi + v_y \sin \varphi)
          + i g_1^{0,2}(\varrho,z) v_z] \, \hat{\bfphi}
          + i g_1^{0,2}(\varrho,z) \, (-v_x\sin \varphi + v_y \cos \varphi) \, \hat{\mathbf{z}}\, \right\}
          \label{eq.Hfield4}  \\
       = -\frac{2\pi n^2}{\Lambda \lambda_0^2}\, \left\{
          g_2^{1,1}(\varrho,z) \, v_\varphi\, \hat{\bfrho}
          + [ g_2^{1,1}(\varrho,z) \, v_\varrho + i g_1^{0,2}(\varrho,z) v_z] \, \hat{\bfphi}
          + i g_1^{0,2}(\varrho,z) \, v_\varphi \, \hat{\mathbf{z}}\, \right\}.
          \label{eq.Hfield5}
          \end{eqnarray}

The time averaged  Poynting vector of the optimum field is
\begin{eqnarray}
\mbox{Re} \, \mathbf{S}(\varrho,\varphi,z) & = &
\frac{1}{2}\mbox{Re}\left[ \mathbf{E}(\varrho,\varphi,z) \times \mathbf{H}(\varrho,\varphi,z)^* \right]
\nonumber \\
& = & \frac{1}{2} \mbox{Re}\left\{\left[E_\varphi(\varrho,\varphi,z) H_z(\varrho,\varphi,z)^* -
E_z(\varrho,\varphi,z) H_\varphi(\varrho,\varphi,z)^* \right]\,\hat{\bfrho}
\right. \nonumber \\ & & \left. +
\left[E_z(\varrho,\varphi,z) H_\varrho(\varrho,\varphi,z)^* -
E_\varrho(\varrho,\varphi,z) H_z(\varrho,\varphi,z)^* \right] \,\hat{\bfphi}
\right. \nonumber \\ & & \left.
+ \left[E_\varrho(\varrho,\varphi,z) H_\varphi(\varrho,\varphi,z)^*- E_\varphi(\varrho,\varphi,z) H_\varrho(\varrho,\varphi,z)^* \right] \,\hat{\mathbf{z}}\right\},
\label{eq.Svec}
\end{eqnarray}
with
\begin{eqnarray}
\mbox{Re} \, S_\varrho(\varrho,\varphi,z) &=&\frac{1}{2}  \mbox{Re}
\left[ E_\varphi(\varrho,\varphi,z) H_z(\varrho,\varphi,z)^* - E_z(\varrho,\varphi,z) H_\varphi(\varrho,\varphi,z)^*\right]
\nonumber \\
&=&
\pi^2 \frac{n^3}{\Lambda^2 \lambda_0^4} \left(\frac{\mu_0}{\epsilon_0}\right)^{1/2}
 \left\{ 2  \mbox{Im}\left[ g_1^{1,2}(g_2^{1,1})^*\right] \, v_\varrho^2 \right.
 \nonumber \\
 & &\left.
 -\mbox{Im}\left[(g_0^{0,1} + g_0^{2,1}-g_2^{0,3}) (g_1^{0,2})^* \right] \, v_\varphi^2
 +2 \mbox{Im}\left[ g_0^{0,3} (g_1^{0,2})^*\right]\, v_z^2  \right. \nonumber \\
 & & \left.
- 2\mbox{Re}\left[g_1^{1,2}(g_1^{0,2})^* - g_0^{0,3}(g_{2}^{1,1})^*\right] \, v_\varrho v_z\right\},
\label{eq.Srho}
\end{eqnarray}
\begin{eqnarray}
\mbox{Re}\, S_\varphi(\varrho,\varphi,z) & = &
\frac{1}{2}\mbox{Re}\left[ E_z(\varrho,\varphi,z) H_\varrho(\varrho,\varphi,z)^* -
 E_\varrho(\varrho,\varphi,z) H_z(\varrho,\varphi,z)^* \right]
 \nonumber \\
 &=&
\pi^2 \frac{n^3}{\Lambda^2 \lambda_0^4} \left(\frac{\mu_0}{\epsilon_0}\right)^{1/2}
\, \left\{
\mbox{Im}\left[ -2 g_1^{1,2}(g_2^{1,1})^* + (g_0^{0,1}+g_0^{2,1}+g_2^{0,3})(g_1^{0,2})^*\right] \, v_\varrho v_\varphi
\right. \nonumber \\ & & \left.
   - 2\mbox{Re}\left[ g_1^{1,2}(g_1^{0,2})^*+ g_0^{0,3}(g_2^{1,1})^*\right] \, v_\varphi v_z
  \right\},
  \nonumber \\
\label{eq.Sphi}
\end{eqnarray}
\begin{eqnarray}
\mbox{Re}\, S_z(\varrho,\varphi,z) & = & \frac{1}{2} \mbox{Re}
\left[E_\varrho(\varrho,\varphi,z) H_\varphi(\varrho,\varphi,z)^* - E_\varphi(\varrho,\varphi,z)
 H_\varrho(\varrho,\varphi,z)^* \right] \nonumber \\
 &   = &
\pi^2 \frac{n^3}{\Lambda^2 \lambda_0^4} \left(\frac{\mu_0}{\epsilon_0}\right)^{1/2}
\left\{ -\mbox{Re}\left[(g_0^{0,1}+g_0^{2,1}+g_1^{0,3}) ( g_2^{1,1})^*\right] \, v_\varrho^2 \right. \nonumber \\
& & \left.
+ \mbox{Re}\left[( g_0^{0,1} + g_0^{2,1} - g_2^{0,3}) (g_2^{1,1})^*\right] \, v_\varphi^2
 +2\mbox{Re}\left[g_1^{1,2}(g_1^{0,2})^*\right]\, v_z^2
\right. \nonumber \\
& & \left.  -
\mbox{Im}\left[ (g_0^{0,1} + g_0^{2,1} + g_2^{0,3})(g_1^{0,2})^* +  g_1^{1,2}(g_2^{1,1})^*)\right] \, v_\varrho v_z
\right\}.
\label{eq.Sz}
\end{eqnarray}

\section{Optimum field distributions}
We  will consider now in more detail  subsequently fields obtained by optimizing   the longitudinal,  the
transverse  and  an intermediate component.

\subsection{Optimum longitudinal component}
\begin{figure}
  \begin{center}
  \includegraphics[width=8.15cm]{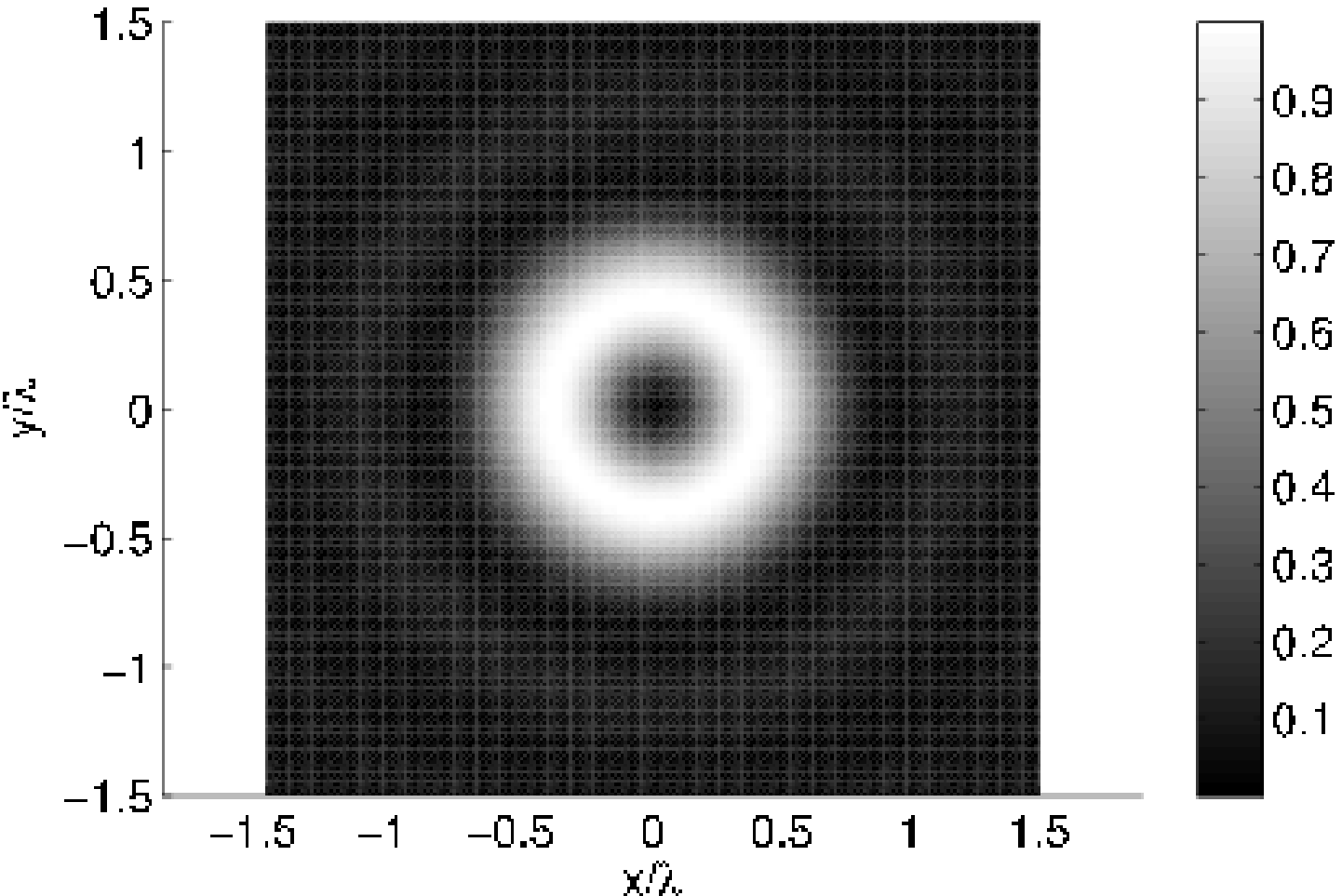}
 \includegraphics[width=8.15cm]{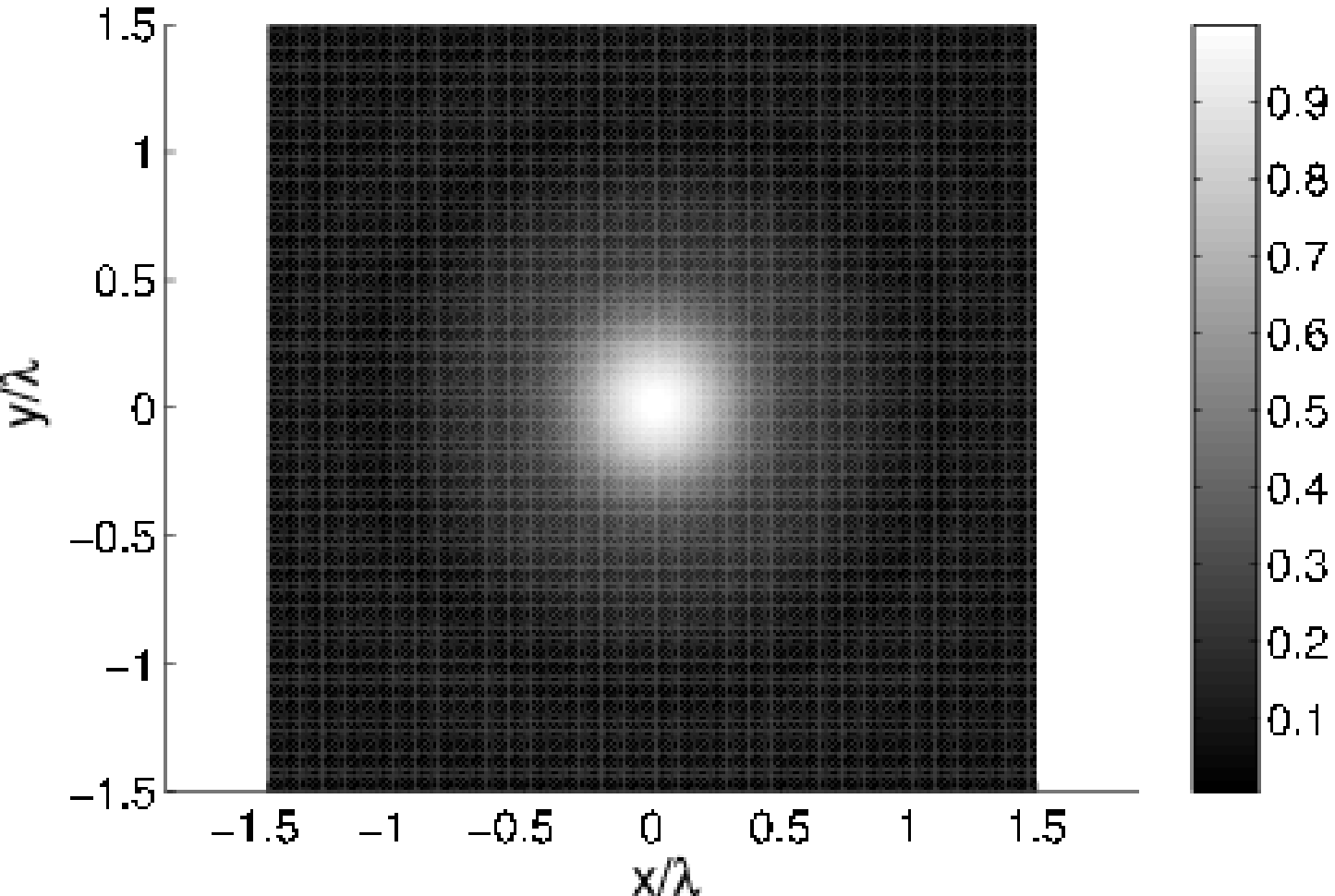}
   \caption{\label{Fig.ErhoE2}
The squared modulus $|E_\varrho|^2$ of the radial component and
 electric energy density $|\mathbf{E}|^2$ of
  the electric  field with maximum longitudinal component for  $\mbox{NA}/n=0.9$.
  The maximum values of $|E_\varrho|^2$
  is approximately $25 \%$ of the maximum of the squared modululus $|E_z|^2$ of
  the longitudinal component.}
   \end{center}
\end{figure}

In this case:
\begin{equation}
\hat{\mathbf{v}}=\hat{\mathbf{z}}.
\label{eq.plong}
\end{equation}
Then (\ref{eq.Efield5}) and (\ref{eq.Hfield5}) become:
\begin{eqnarray}
\mathbf{E}(\varrho,\varphi,z)
=  \frac{2\pi n}{\Lambda\lambda_0^2} \left(\frac{\mu_0}{\epsilon_0}\right)^{1/2}
   \left[   -  i g_1^{1,2}(\varrho,z) \,  \, \hat{\bfrho}
         +  g_0^{0,3}(\varrho,z) \, \hat{\mathbf{z}}
     \right],
    \label{eq.Efieldlong}
       \end{eqnarray}
       and
       \begin{eqnarray}
     \mathbf{H}(\varrho,\varphi,z)
            = -\frac{2\pi i n^2}{\Lambda \lambda_0^2}\,
                      g_1^{0,2}(\varrho,z) \, \hat{\bfphi}.
               \label{eq.Hfieldlong}
          \end{eqnarray}
   Since $g_1^{1,2}(0,0)=0$, the electric field in the origin is parallel to the
   $z$-axis, hence it is purely longitudinal in the origin.
In the $z=0$-plane the functions  $g_\ell^{\nu,\mu}$ are
       real and therefore (\ref{eq.Efieldlong}) implies that
        the polarization ellipse of
       the electric field in that plane has minor and major axis
       parallel  to the $\hat{\bfrho}$- and $\hat{\mathbf{z}}$-axis
       (which is the major and which  the minor axis
    depends on the relative values of $g_0^{0,3}(\varrho,0)$ and
   $g_1^{1,2}(\varrho,0)$).
   In the $z=0$-plane, the phase of $E_z$  is  $\pm \pi$  whereas
the other electric field components have phase  $\pm \pi/2$.
       The magnetic field is everywhere parallel to  $\hat{\bfphi}$, i.e. it is
       azimuthal.
  \begin{figure}
 \begin{center}
 \includegraphics[width=8.15cm]{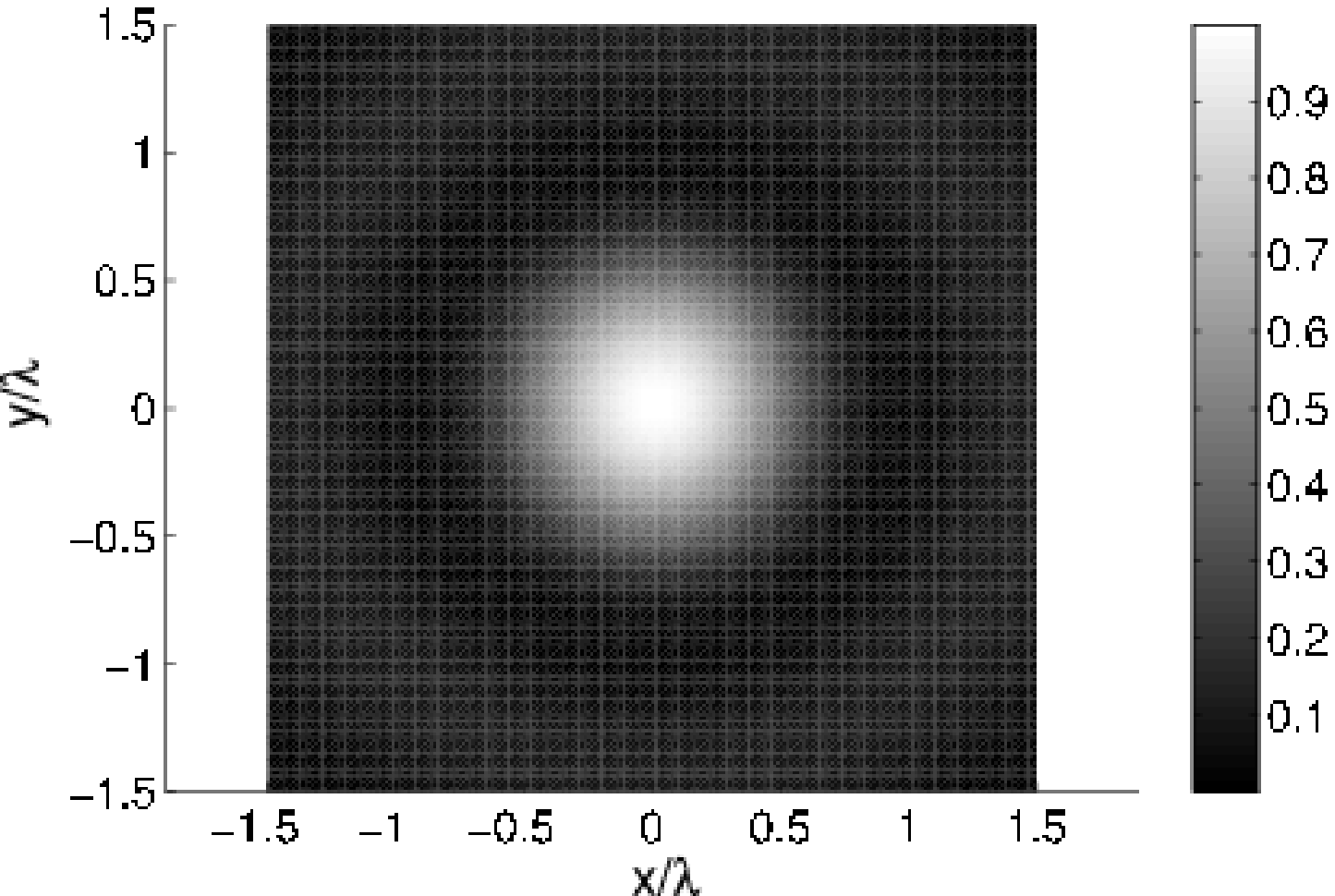}
  \includegraphics[width=8.15cm]{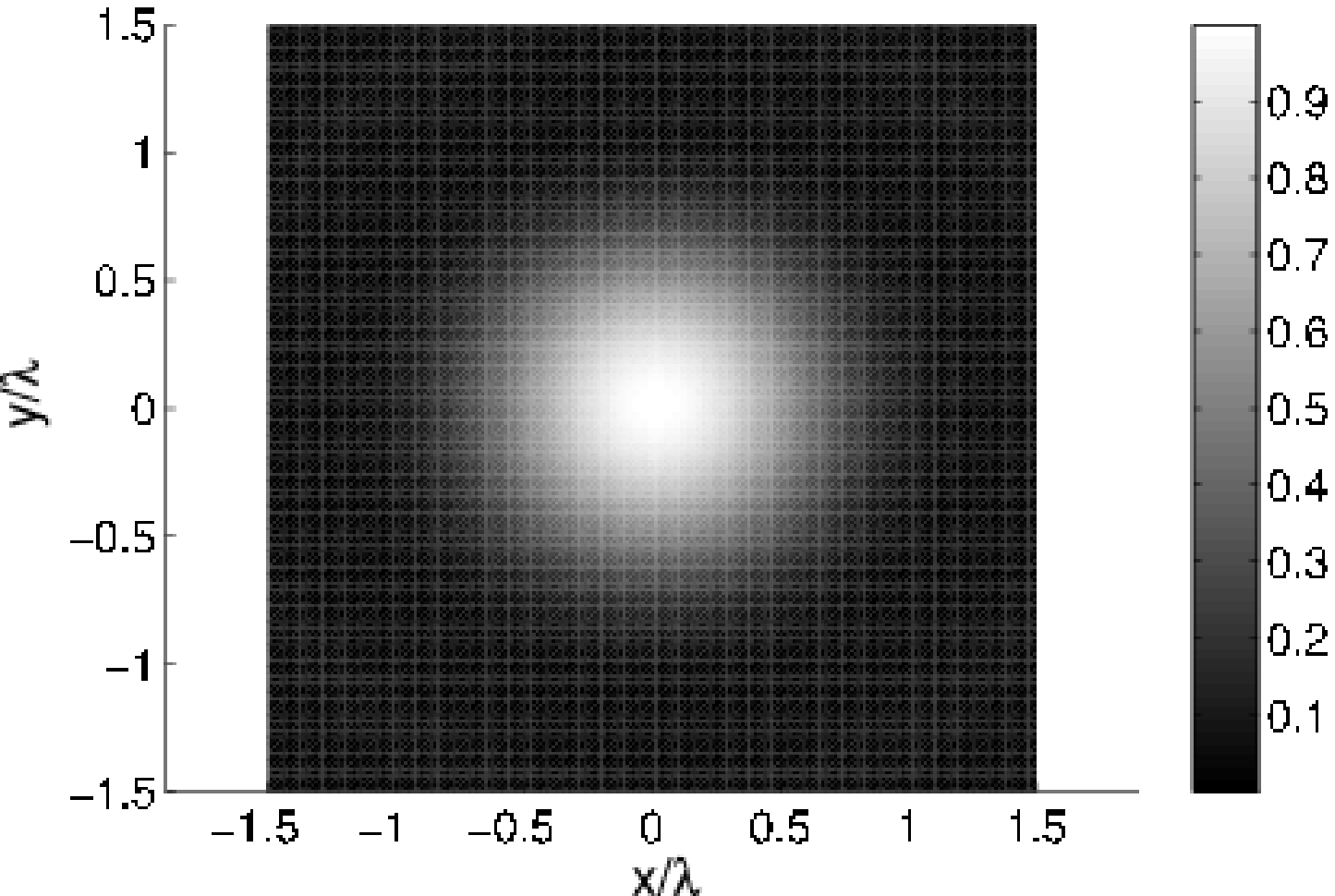}
   \caption{ \label{Fig.EzEfocalNA05}
 Left: the normalized distribution of $|E_z|^2$ in the $z=0$-plane
for the  field with maximum longitudinal
component when $\mbox{NA}/n=0.5$ and $P_0=1$ ($\lambda=\lambda_0/n$ is the wavelength in the material with
refractive index $n$).
    For comparison, the normalized electric energy distribution
   (\ref{eq.Edensity}) of the $x$-polarized focused plane wave is shown at the right
   for the same numerical aperture.
   The total flow of power in the $z$-direction is the same for both fields.  }
   \end{center}
\end{figure}
In Fig. \ref{Fig.EzEfocalNA05} the normalized distribution
$|E_z(x,y,0)|^2$ of the optimum field with maximum longitudinal
component for $\mbox{NA}/n=0.5$ in the $z=0$-plane,  is compared to
the normalized total electric energy density:
\begin{equation}
  |E_x(x,y,0)|^2 + |E_y(x,y,0)|^2 + |E_z(x,y,0)|^2,
  \label{eq.Edensity}
  \end{equation}
  in the focal plane  of the focused $x$-polarized plane wave.
  The coordinates $x, y$ are expressed in units of the wavelength $\lambda=\lambda_0/n$
  in the material with refractive index $n$.
  The
  Formulas for the  electric field of a focused plane wave
  are given in the appendix (see \ref{eq.Explw}), (\ref{eq.Eyplw}), (\ref{eq.Ezplw})).
  The  power flow in the $z$-direction of the optimum longitudinal field
  and the focused plane wave are the same. The distribution of the optimum $|E_z|^2$ is rotationally symmetric while that of
  the electric energy density Eq. (\ref{eq.Edensity}) of the focused
  linearly polarized plane wave
  is elliptical,  with short axis parallel to the $y$-direction
  (i.e. perpendicular to the direction of polarization of the
  focused plane wave). 
  In Fig. \ref{Fig.EzEfocalNA09} the distributions are compared for
   $\mbox{NA}/n=0.9$. In this case the short axis of the elliptic
  distribution is considerably shorter than the long axis.
  In Fig. \ref{Fig.ErhoE2} the squared modulus $|E_\varrho|^2$ and
  $|\mathbf{E}|^2$  are shown. Due to the broad doughnut shaped distribution
  of the radial component, $|\mathbf{E}|^2$ is broader than the Airy spot.

\begin{figure}
  \begin{center}
  \includegraphics[width=8.15cm]{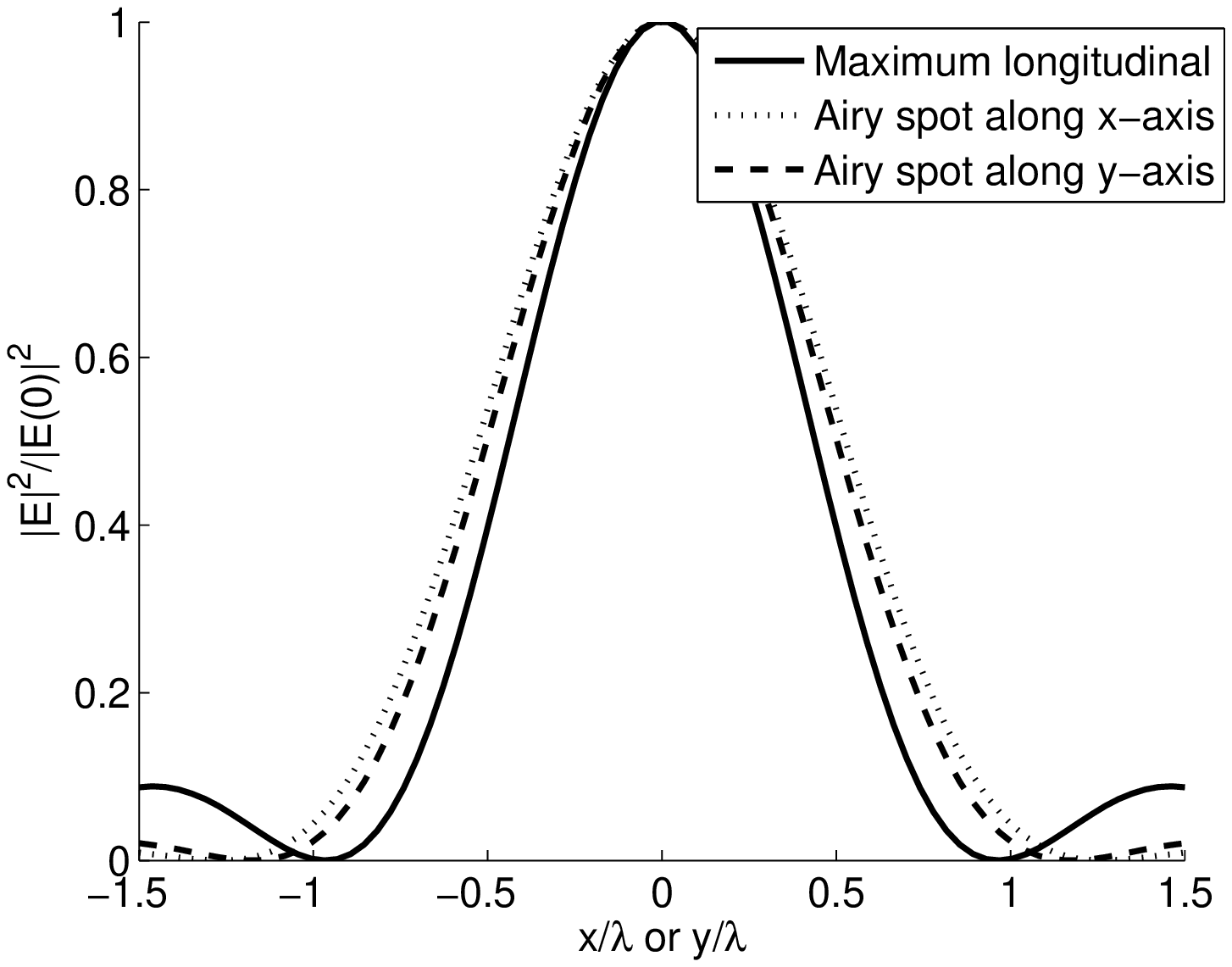}
  \includegraphics[width=8.15cm]{figure4b}
   \caption{\label{Fig.crosssection}
   Cross-section of the rotational symmetric $|E_z|^2$ (solid curves) of the field with maximum
   longitudinal component and
   cross-sections of $|\mathbf{E}(x,0,0)|^2$ (dashed) and
   $|\mathbf{E}(0,y,0)|^2$ (dotted) of the Airy spot for the $x$-polarized
   focused plane wave, for $\mbox{NA}/n=0.5$ (left) and $\mbox{NA}/n=0.9$ (right).
   $\lambda=\lambda_0/n$ and the
  maxima of all quantities are rescaled to unity.}
     \end{center}
   \end{figure}

  Cross-sections along the short and long axes are shown in Fig.
      \ref{Fig.crosssection} for both $\mbox{NA}/n=0.5$ and $\mbox{NA}/n=0.9$.
      To compare the shapes the maxima of all cross-sections are rescaled to 1.
      \begin{figure}
  \begin{center}
  \includegraphics[width=8.15cm]{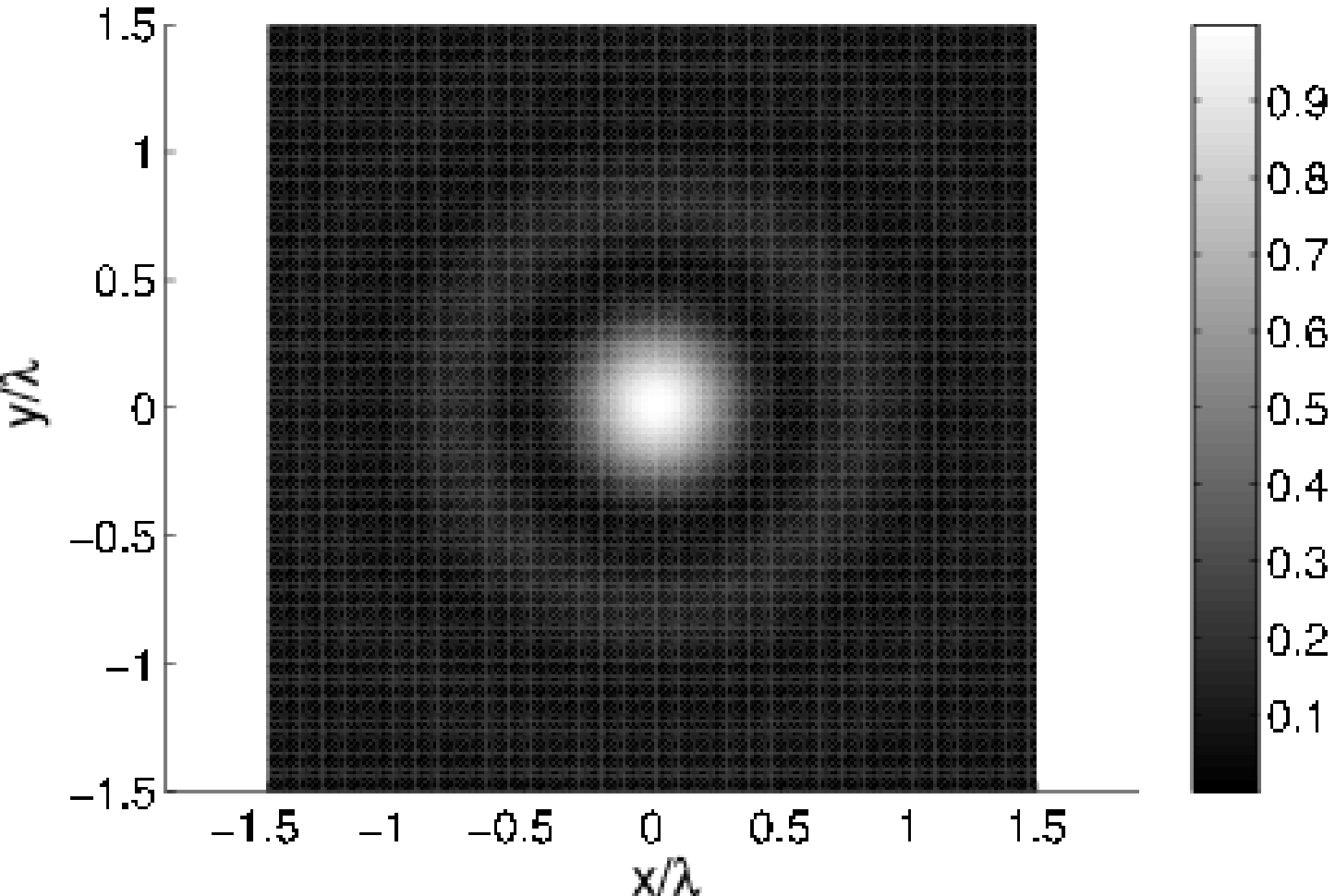}
 \includegraphics[width=8.15cm]{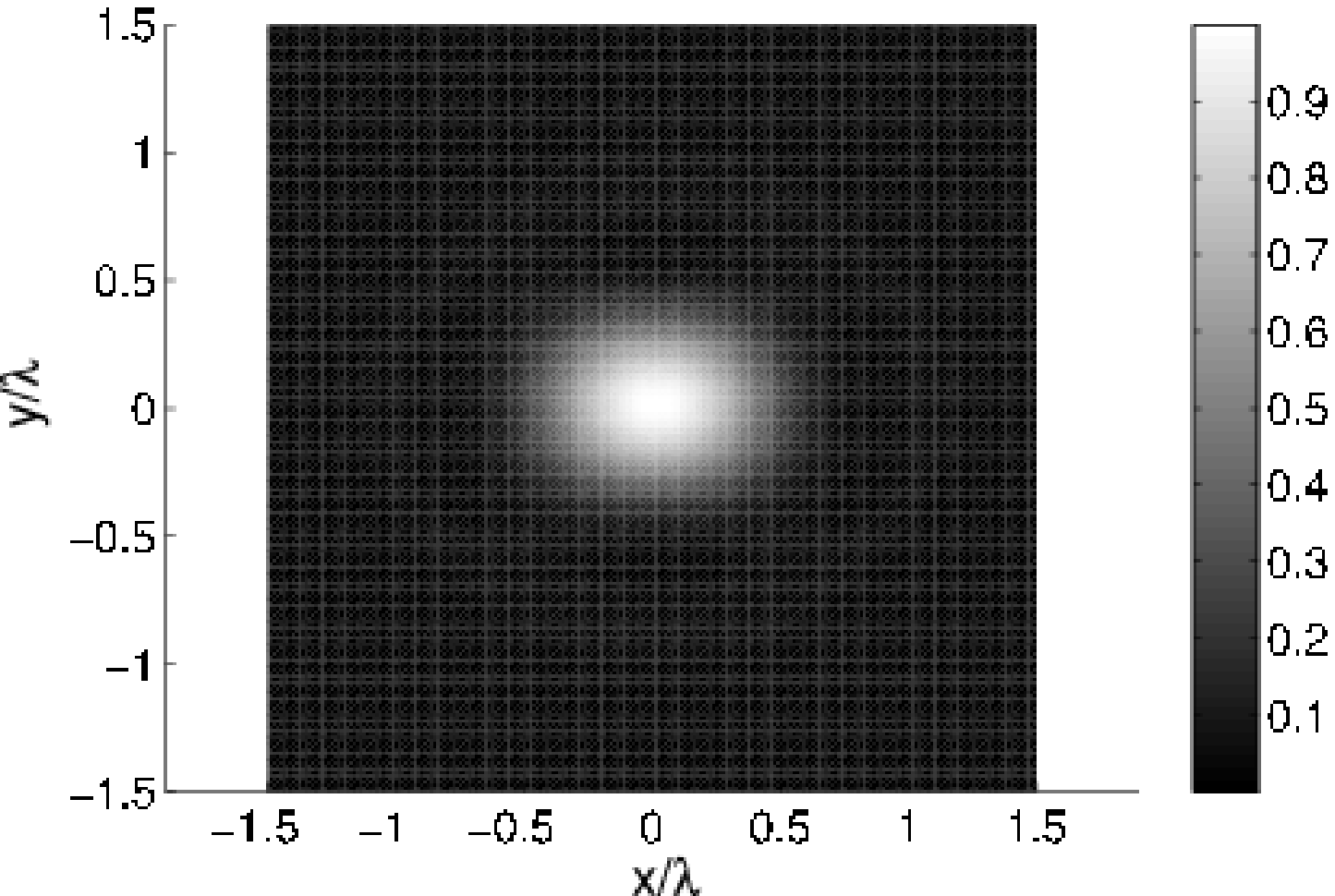}
   \caption{\label{Fig.EzEfocalNA09}
Same as Fig. \ref{Fig.EzEfocalNA05} except that now $\mbox{NA}/n=0.9$. }
   \end{center}
\end{figure}
 It is seen that the longitudinal component has smaller FWHM (Full Width at Half Maximum) but also higher secondary maxima.

The FWHM of the optimum longitudinal component is for $\mbox{NA}/n=1$ almost identical to
that of the longitudinal component in \cite{quabis_AP}, obtained by focusing a radially polarized beam using a ring mask function
(with radius $90 \%$ of the total pupil). However, the
side lobes are higher at the cost of the central maximum compared to our longitudinal component.
This is
of course not surprising because the longitudinal
component in \cite{quabis_AP} was not optimized  for a high maximum on the optical axis.

\begin{figure}
\begin{center}
  \includegraphics[width=8.15cm]{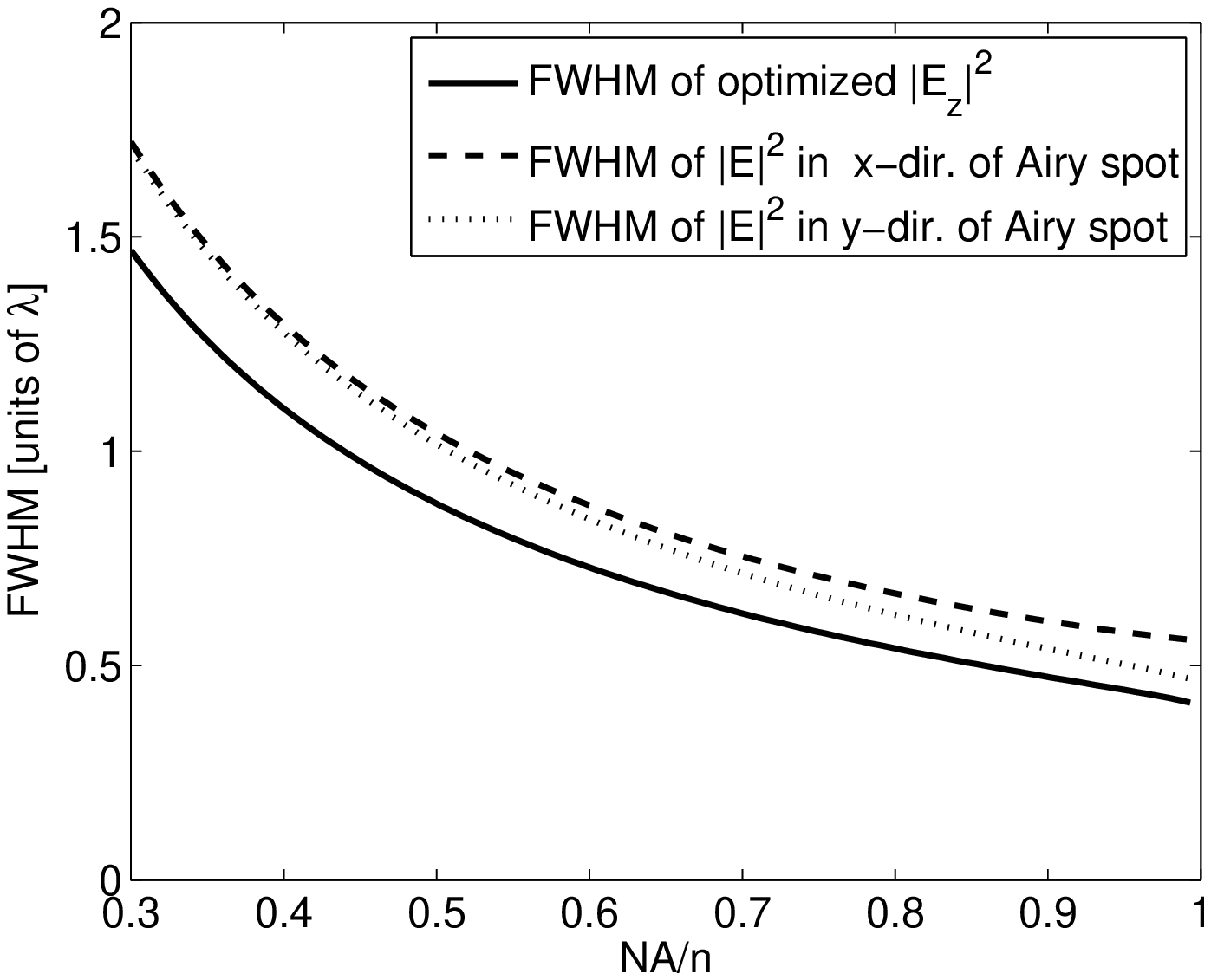}
   \includegraphics[width=8.15cm]{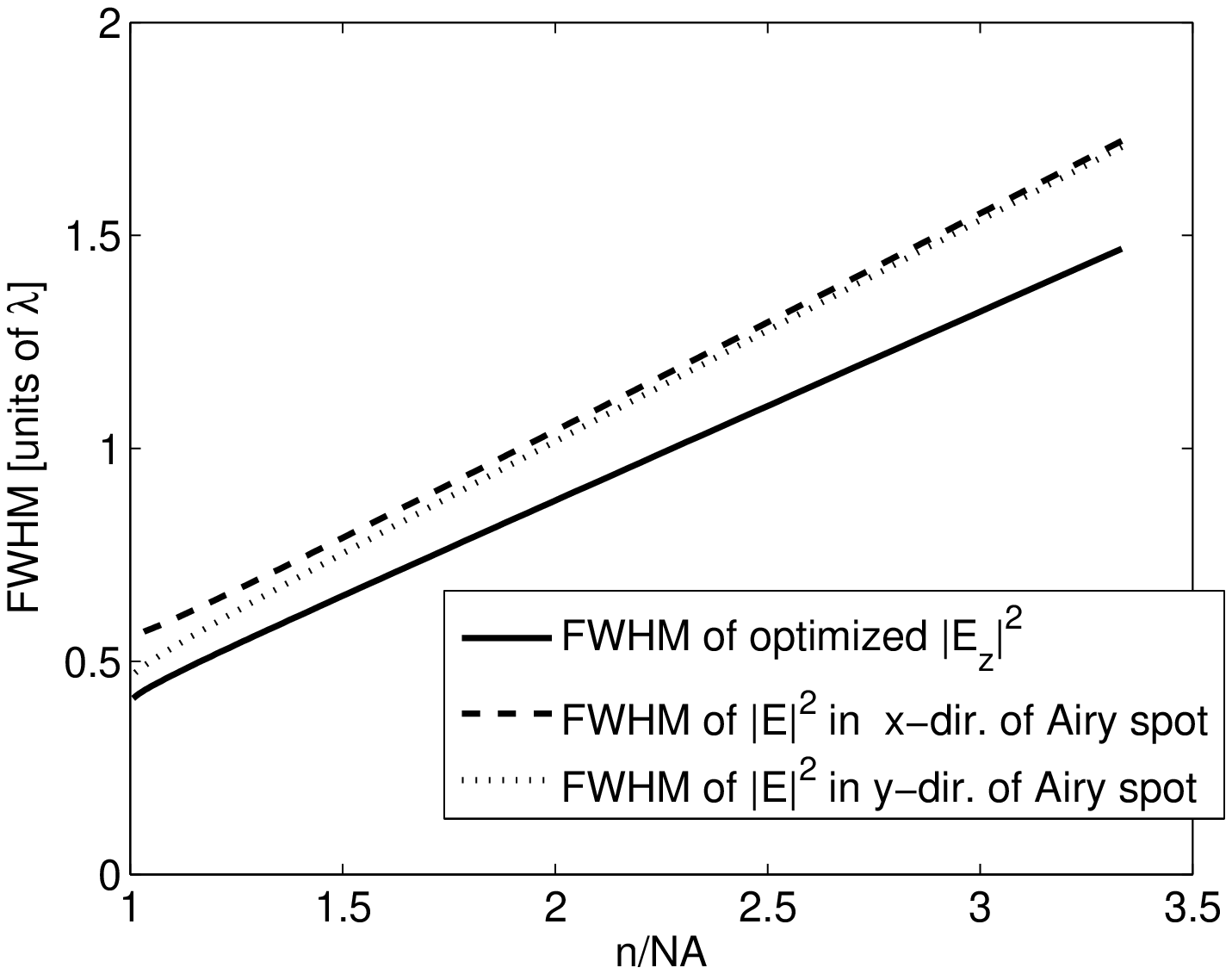}
   \caption{\label{Fig.FWHM} FWHM of $|E_z(x,y,0)|^2$ of the field with maximum
   longitudinal component and FWHM of $|\mathbf{E}(x,y,0)|^2$ along  the $x$- and $y$-directions
   of the focused $x$-polarized plane wave. At the left the FWHM is
   shown as function of $\mbox{NA}/n$, at the right as function of $n/\mbox{NA}$
   in units of  $\lambda =\lambda_0/n$.}
   \end{center}
   \end{figure}

\begin{figure}
   \begin{center}
  \includegraphics[width=10.cm]{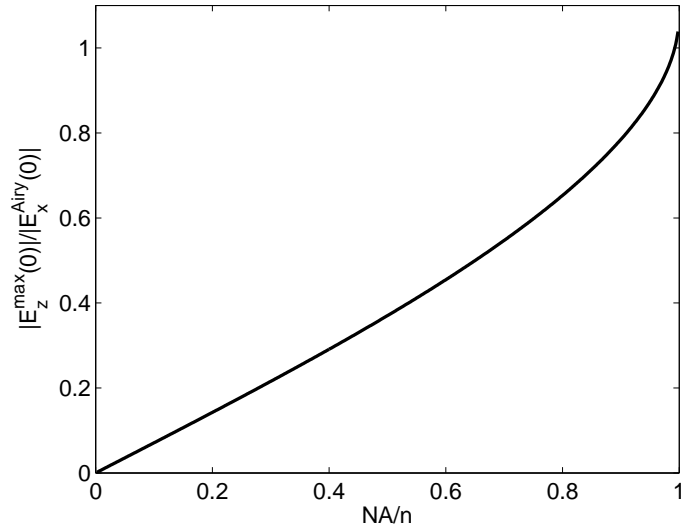}
      \caption{\label{Fig.ratio} Ratio of maximum longitudinal amplitude $|E_z|$
      and the amplitude $|E_x|$ of the focused $x$-polarized plane wave for the same
      power and as function of $\mbox{NA}/n$.}
           \end{center}
   \end{figure}

In Fig. \ref{Fig.FWHM} the FWHM of $|E_z(x,y,0)|^2$ of the optimum
field is compared to the FWHM in the $x$- and $y$-directions of the
electric energy density $|\mathbf{E}(x,y,z=0)|^2$ of the focal spot of the $x$-polarized
plane wave. At the left, the FWHM is shown as a function of the
numerical aperture $\mbox{NA}/n$, at the right as a function of $n/NA$, both in units
of wavelength. It is seen that the FWHM of the longitudinal component
is smaller than the FWHM in both the $x$- and $y$-directions of the focused
spot.
  The FWHM as a function of $n/\mbox{NA}$ is almost linear, but the slope
  is smaller for the longitudinal component. This means that
  the
  spot size of the optimum longitudinal $|E_z|^2$  is relatively smaller than the
  Airy spot when the numerical aperture is larger.
  Nevertheless, for all values of the numerical aperture, the longitudinal spot is narrower than the
  Airy spot.
    As shown in Fig. \ref{Fig.ratio},  the maximum amplitude $|E_z(\mathbf{0})|$
   of the optimum
   longitudinal component is for most values of the numerical aperture
   smaller than the maximum amplitude $|E_x(\mathbf{0})|$ of the
   focused $x$-polarized plane wave. But for $\mbox{NA}/n >0.65 $ this ratio is already more than
   $0.5$ and for $\mbox{NA}/n>0.994$  the maximum longitudinal component is even larger than the
   maximum amplitude $|E_x(\mathbf{0})|$ of the focused $x$-polarized plane wave.

Along the optical axis we have $\varrho=0$ and
 \begin{eqnarray}
g_0^{0,3}(0,z) & = & \int_0^{\alpha_{\max}} e^{ik_0 n z \cos \alpha }  \sin^3 \alpha
\, d \alpha \nonumber \\
& = & \frac{i}{k_0 nz} \left( 1 + \frac{2}{k_0^2n^2 z^2}\right)
  \left[e^{ik_0nz \cos \alpha_{\max}}-e^{ik_0n z}\right] \nonumber \\
  & & + 2\frac{ \cos \alpha_{\max} e^{ik_0 n z \cos \alpha_{\max}}-e^{ik_0n z} }
              { k_0^2 n^2 z^2} \nonumber \\
  & & - i  \frac{ \cos^2\alpha_{\max} e^{ik_0 n z \cos \alpha_{\max}}-e^{ik_0n z} }
                {k_0 n z}.
        \label{eq.g0r0}
        \end{eqnarray}
        Because $J_1(0)=0$:
\begin{equation}
g_1^{1,2}(0,z)=g_1^{1,2}(0,z) = 0.
\label{eq.g1r0}
\end{equation}
It thus follows from (\ref{eq.Efieldlong}) that along the optical axis the
electric field is parallel to the $z$-axis
 and that the modulus of the field is symmetric with respect to the
 focal plane.
Furthermore,
\begin{equation}
  \frac{|E_z(x=0,y=0,z)|^2}{|E_z(0,0,0)|^2} = \frac{|g_0^{0,3}(0,z)|^2}{|g_0^{0,3}(0,0)|^2},
  \;\;\;,
  \label{eq.ratio}
 \end{equation}
 where
 \begin{equation}
 g_0^{0,3}(0,0) = \frac{2}{3} - \cos \alpha_{\max}
 + \frac{1}{3} \cos^3 \alpha_{\max}.
 \label{eq.g003}
 \end{equation}
We define the focal depth of the optimum longitudinal component
 as the distance $\Delta z$ to the focal plane for which the ratio
 \begin{equation}
    \frac{|g_0^{0,3}(0,\Delta z)|^2}{|g_0^{0,3}(0,0)|^2} =0.8.
\label{eq.focaldepth}
\end{equation}

\begin{figure}
\begin{center}
  \includegraphics[width=10.cm]{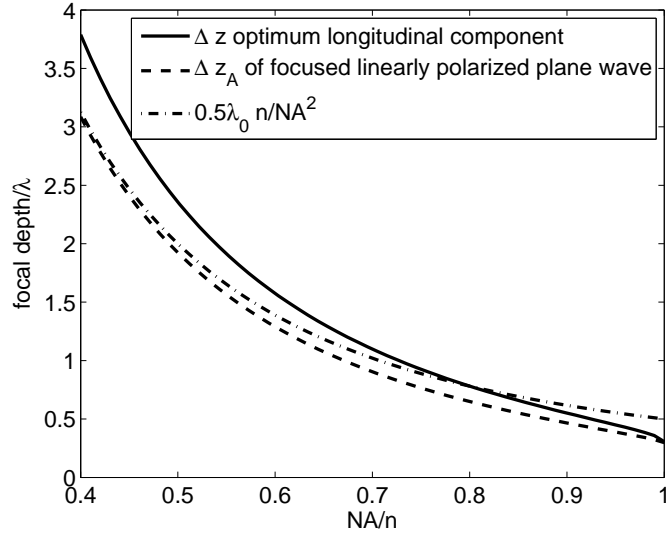}
      \caption{\label{Fig.focaldepth}Focal depth in unit of $\lambda=\lambda_0/n$
      as function of $\mbox{NA}/n$ for the optimum longitudinal
      component and for the energy density of the focused linear polarized plane wave.
      The approximative paraxial focal depth: $0.5\lambda_0 n /NA^2$ is also shown.}
     \end{center}
   \end{figure}

For the focused linearly polarized plane wave, the electric energy density on the
optical axis is given by (\ref{eq.EEdens}):
\begin{eqnarray}
|\mathbf{E}(x=0,y=0,z)|^2 & = & |E_x(x=0,y=0,z)|^2 + |E_y(x=0,y=0,z)|^2 + |E_z(x=0,y=0,z)|^2 \nonumber \\
& = &
  \frac{\pi^2 n^2 f^2}{\lambda_0^2} | g_0^{\frac{1}{2},1}(0, z) + g_0^{\frac{3}{2},1}(0,z)|^2.
 \label{eq.intenszaxis}
\end{eqnarray}
In the origin we have (\ref{eq.EEdens0}):
\begin{equation}
|\mathbf{E}(0,0,0)|^2  =
  \frac{ \pi^2 n^2 f^2}{\lambda_0^2}  \left[
  \frac{2}{3} (1-\cos^{3/2}\alpha_{\max}) + \frac{2}{5} ( 1-\cos^{5/2}\alpha_{\max})
  \right]^2.
  \label{eq.EEdensfocalpoint}
  \end{equation}
The energy density on the optical axis is again symmetric around the
focal plane and  the focal depth is the distance $\Delta z_A$ such
that
\begin{equation}
\frac{|\mathbf{E}(x=0,y=0,\Delta z_A)|^2}{
     |\mathbf{E}(0,0,0)|^2} =0.8.
     \label{eq.focaldepthA}
     \end{equation}
     Both  $\Delta z$ and $\Delta z_A$ are shown as functions of
     $\mbox{NA}/n$  in Fig. \ref{Fig.focaldepth}. The approximative
     focal depth $0.5 \lambda_0 n/\mbox{NA}^2$, which is valid  in the scalar
     paraxial theory, is also shown. It is seen that for $\mbox{NA} /n<0.4$
     the focal depths of the focused linearly polarized plane wave
    calculated  in the vectorial and the scalar paraxial theory are almost
     identical. The focal depth of the optimized longitudinal spot is for $\mbox{NA}/n < 0.8$ larger than
     that of the Airy spot, while for $\mbox{NA}/n>0.8$ it is smaller.

\begin{figure}
\begin{center}
  \includegraphics[width=8.15cm]{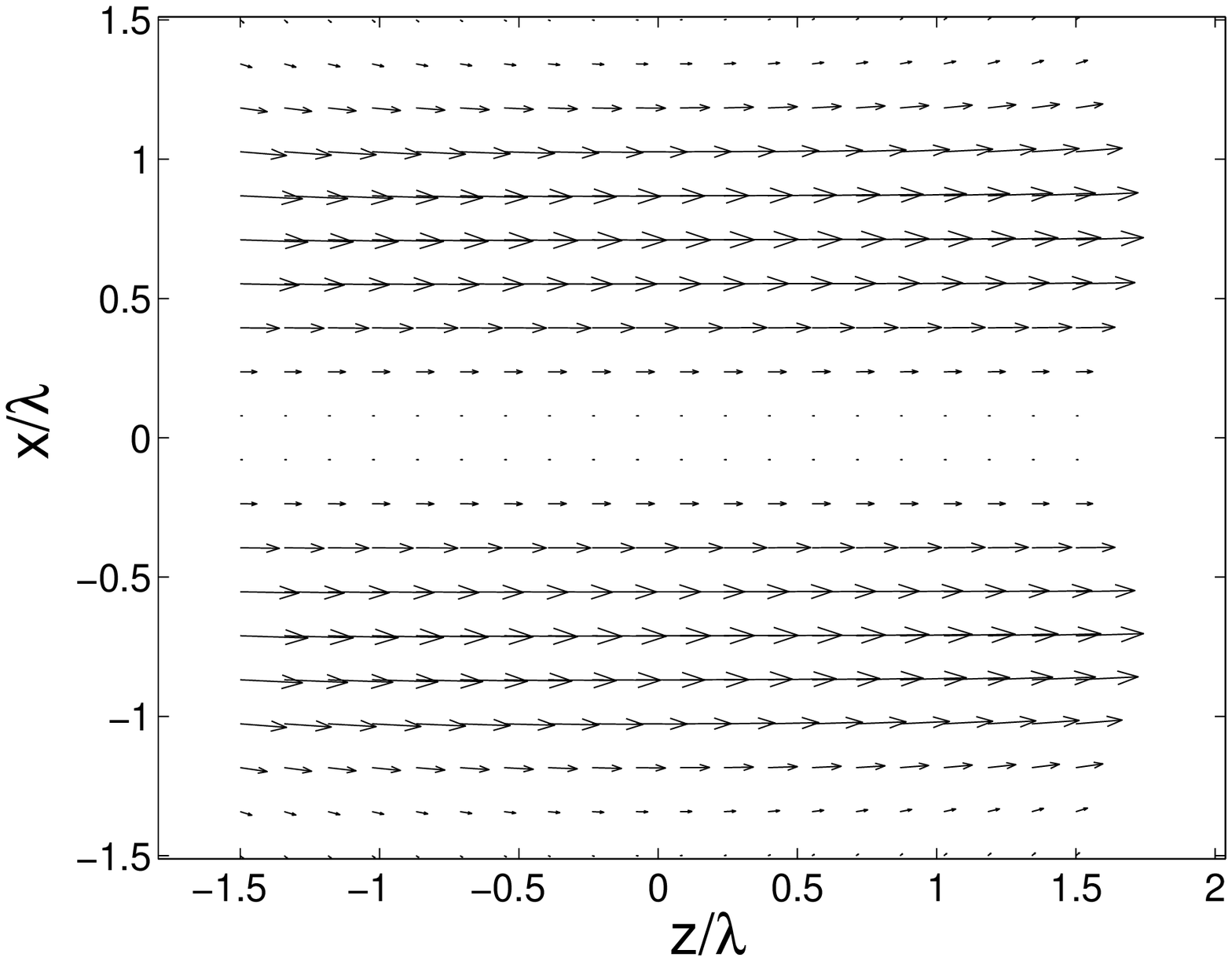}
   \includegraphics[width=8.15cm]{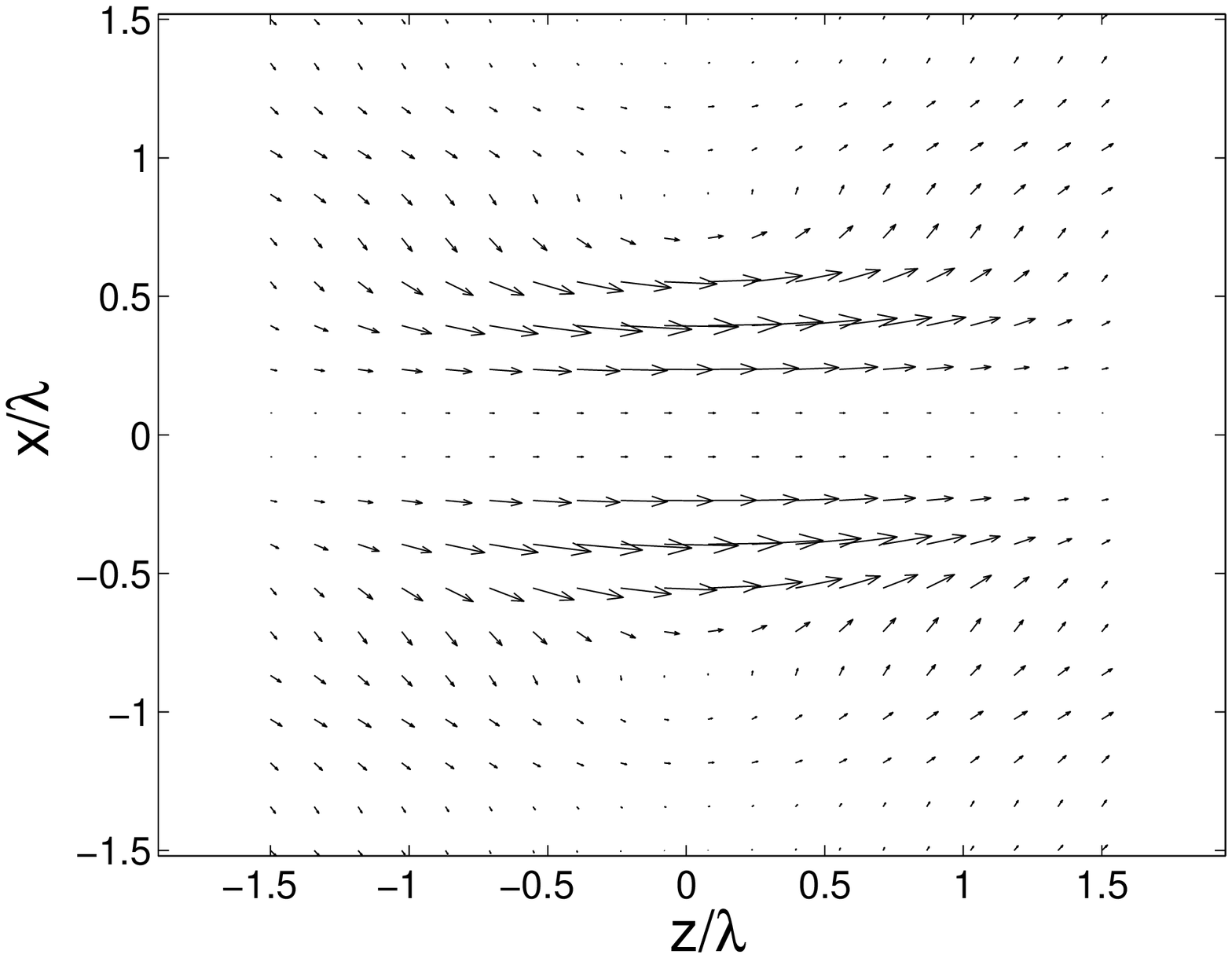}
   \caption{\label{Fig.Poynting} Poynting vector $\mathbf{S}=S_\varrho \, \hat{\bfrho} + S_z \, \hat{\mathbf{z}}$
    for the fields with maximum longitudinal component, in the $(z,x)$-plane (i.e. the $\varphi=\pi/2$-plane)  for
   $\mbox{NA}/n=0.5$ (left) and for $\mbox{NA}/n=0.9$ (right). The Poynting vector is
   independent of the angular coordinate $\varphi$.}
     \end{center}
   \end{figure}

 Eqs. ( \ref{eq.Srho}), (\ref{eq.Sphi}) and (\ref{eq.Sz})  imply
         \begin{eqnarray}
         \mathbf{S}(\varrho,\varphi,z) =
         \frac{2\pi^2n^3}{\Lambda^2 \lambda_0^4} \left(\frac{\mu_0}{\epsilon_0}\right)^{1/2}
         \left\{ \mbox{Im}\left[ g_0^{0,3}(\varrho,z) (g_1^{0,2}(\varrho,z))^*\right]\,
         \hat{\bfrho} + \mbox{Re}\left[ g_1^{1,2}(\varrho,z) (g_1^{0,2}(\varrho,z))^*\right]
         \hat{\mathbf{z}}\, \right\}.
         \label{eq.Svecleng}
         \end{eqnarray}
         The $\varphi$-component of the Poynting vector thus vanishes and the Poynting vector
         is  independent of the angle $\varphi$.
         In the  ($z=0$-)plane the energy flows in the $z$-direction.
In Fig. \ref{Fig.Poynting} the time averaged Poynting vector of
the field with maximum longitudinal component is shown in the
$(z,\varrho)$-plane for the cases $\mbox{NA}/n=0.5$ and $\mbox{NA}/n=0.9$. The
$\varphi$-component of the Poynting vector vanishes and the other
components $S_\varrho$ and $S_z$ are independent of rotation angle
$\varphi$. The Poynting vector vanishes on the $z$-axis.

\subsection{Optimum transverse component}
\label{subsection_transverse}
\begin{figure}
\begin{center}
    \includegraphics[width=8.15cm]{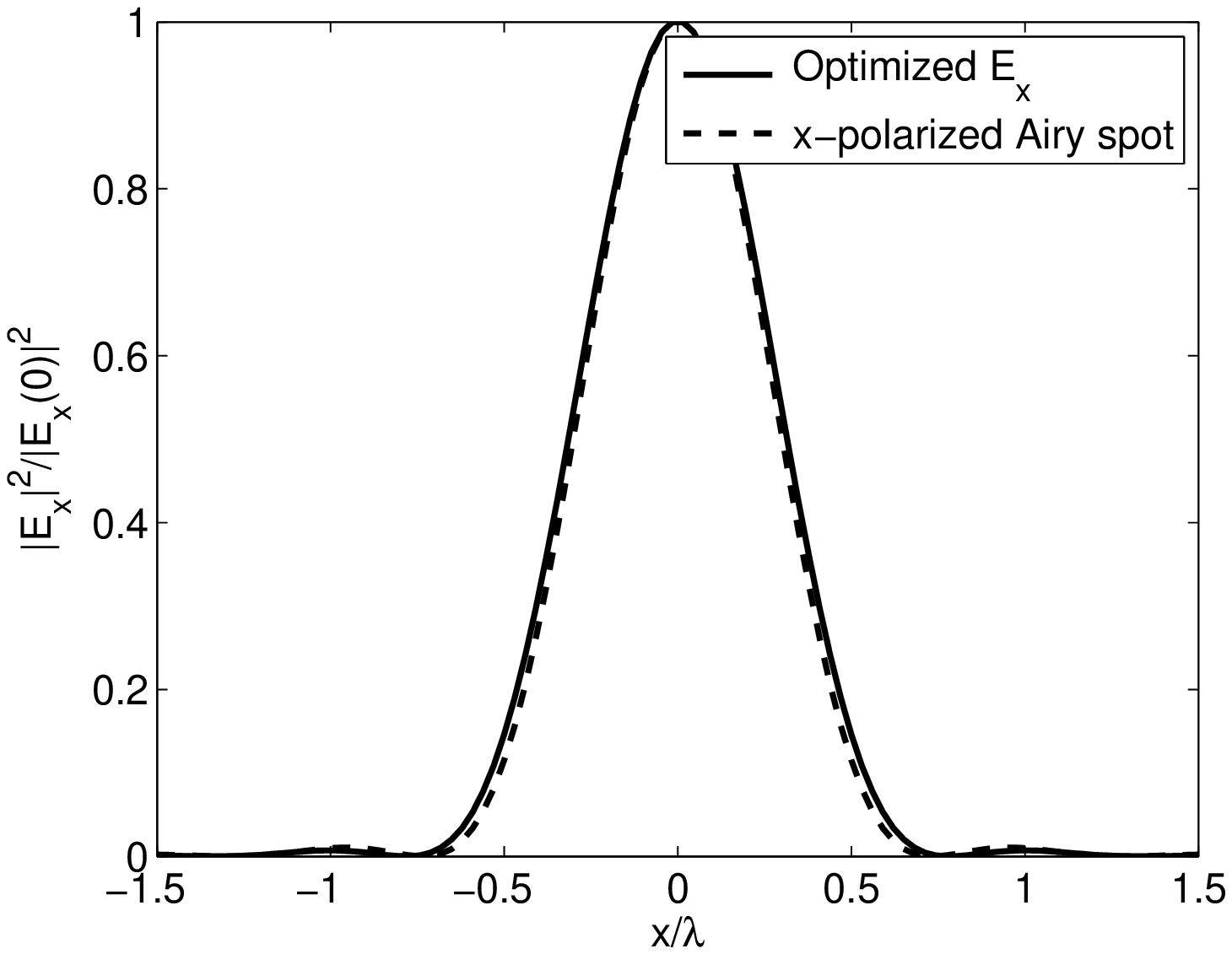}
 \includegraphics[width=8.15cm]{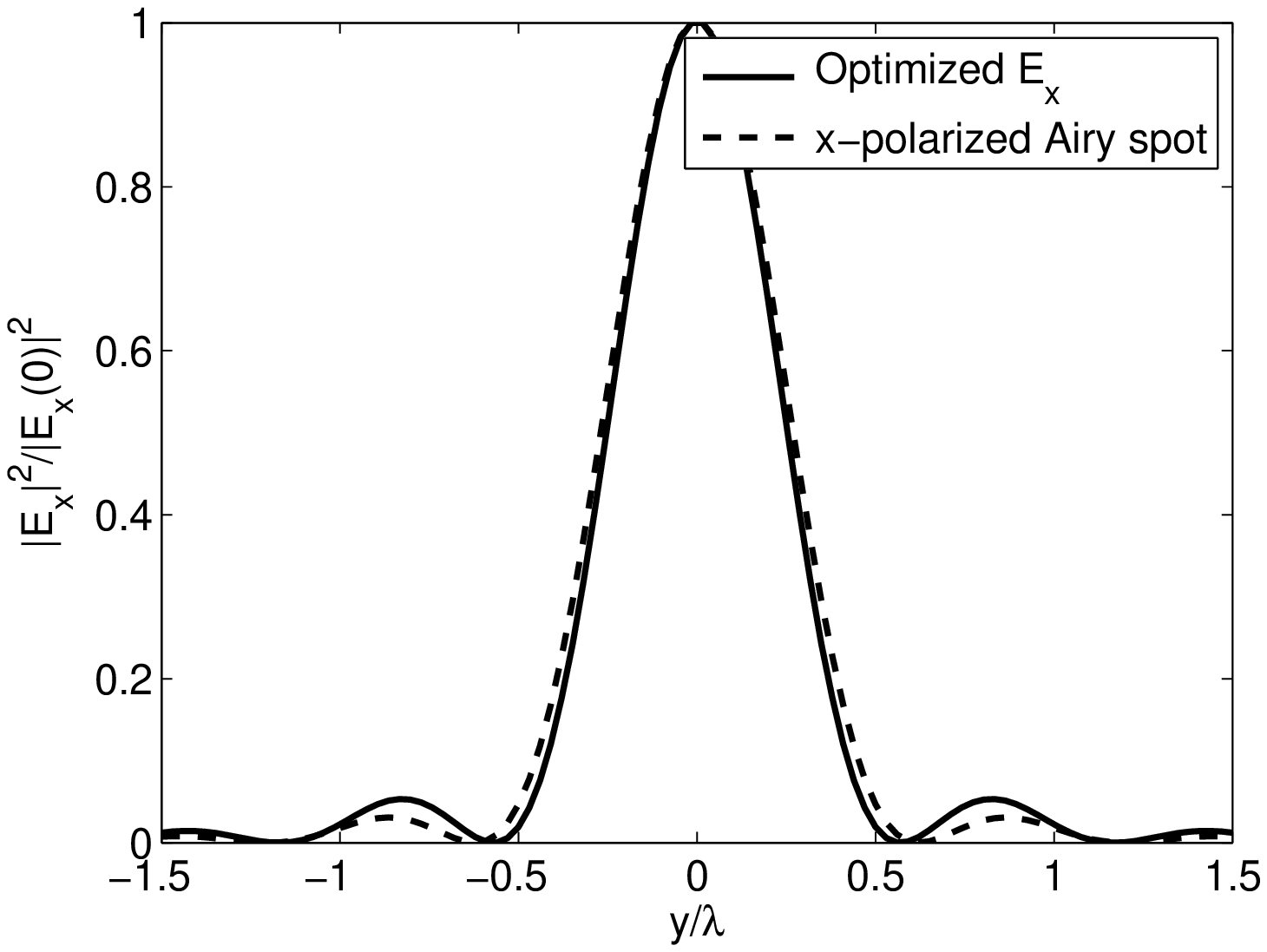}
 \end{center}
\caption{\label{fig.crossExmaxEA}Cross-sections of
$|E_x|^2$ along the $x-$ (left) and $y-$axis (right) for the
field with maximum $x$-component and for the focused $x$-polarized plane wave with the same power.
The numerical aperture is
 NA/n=0.9. The maxima of all cross-sections are rescaled to 1.}
\end{figure}

\begin{figure}
\begin{center}
   \includegraphics[width=10cm]{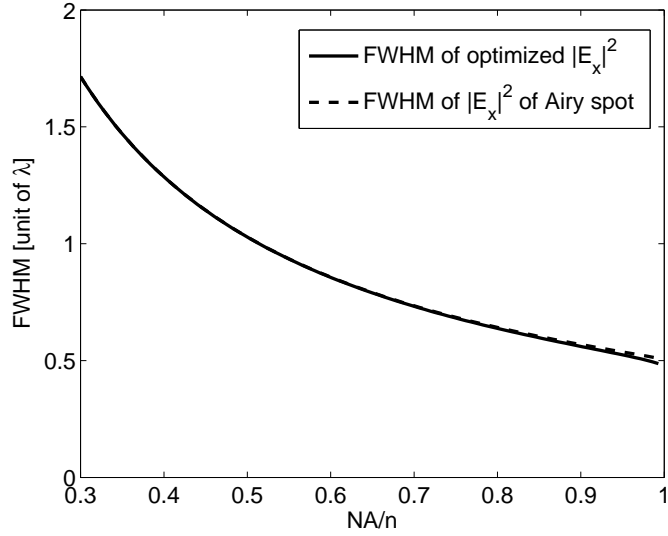}
         \caption{ The FWHM of $|E_x|^2$ along the $x$-axis of the optimized transverse field
and the FWHM of the averaged $|E_x|^2$ of the $x$-polarized
focused plane wave.}
\label{fig.FWHMEx}
\end{center}
\end{figure}

We choose the transverse component parallel to the $x$-axis:
\begin{equation}
\hat{\mathbf{v}}=\hat{\mathbf{x}},
\label{eq.ptrans}
\end{equation}
i.e. the $x$-component of the electric field in the origin is optimized.
On the local basis (attached to the point of observation) of cylindrical coordinates
$\varrho, \varphi, z$
we have: $v_\varrho=\cos\varphi$, $v_\varphi = - \sin \varphi$, $v_z=0$. Hence
 (\ref{eq.Efield3}), (\ref{eq.Efield5}), (\ref{eq.Hfield3}) and
(\ref{eq.Hfield5}) imply that the optimum  electromagnetic field becomes:
\begin{eqnarray}
\mathbf{E}(\varrho,\varphi,z) & = &  \frac{\pi n}{\Lambda \lambda_0^2}
\left(\frac{\mu_0}{\epsilon_0}\right)^{1/2}\,
\left\{ \left[ g_0^{0,1}(\varrho,z) + g_0^{2,1}(\varrho,z) +g_2^{0,3}(\varrho, z) \cos(2\varphi)\right]\, \hat{\mathbf{x}} \right. \nonumber \\ & &  \left. +
   g_2^{0,3}(\varrho, z) \sin(2\varphi)\, \hat{\mathbf{y}}
- 2 i g_1^{0,2}(\varrho,z) \cos \varphi\, \hat{\mathbf{z}} \right\}
\label{eq.Etrans1} \\
& = & \frac{\pi n}{\Lambda\lambda_0^2} \left(\frac{\mu_0}{\epsilon_0}\right)^{1/2}\,
\left\{ \left[ g_0^{0,1}(\varrho, z) + g_0^{2,1}(\varrho, z)
+ g_2^{0,3}(\varrho, z) \right]\, \cos \varphi  \, \hat{\bfrho}  \right. \nonumber \\
& & \left. - \left[ g_0^{0,1}(\varrho, z) + g_0^{2,1}(\varrho, z)
- g_2^{0,3}(\varrho, z) \right]\, \sin \varphi  \, \hat{\bfphi}
 -  2 i g_1^{1,2}(\varrho,z) \, \cos \varphi
 \, \hat{\mathbf{z}} \right\},
\label{eq.Etrans2}
\end{eqnarray}
  \begin{eqnarray}
     \mathbf{H}(\varrho,\varphi,z)  &= &
     -\frac{2\pi n^2}{\Lambda \lambda_0^2}\,\left[
     g_2^{1,1}(\varrho,z) \sin 2\varphi \, \hat{\mathbf{x}}
     - g_2^{1,1}(\varrho,z) \cos 2\varphi \, \hat{\mathbf{y}}
     - i g_1^{0,2}(\varrho,z) \sin \varphi\, \hat{\mathbf{z}}\right]
          \label{eq.Htrans1}\\
     & = &
      \frac{2\pi n^2}{\Lambda \lambda_0^2}\, \left[
          g_2^{1,1}(\varrho,z) \, \sin\varphi \, \hat{\bfrho}
                    - g_2^{1,1}(\varrho,z) \, \cos \varphi
           \, \hat{\bfphi}
          + i g_1^{0,2}(\varrho,z) \, \sin \varphi  \, \hat{\mathbf{z}}\, \right].
          \label{eq.Htrans2}
                 \end{eqnarray}
  \begin{figure}
 \begin{center}
 \includegraphics[width=8.15cm]{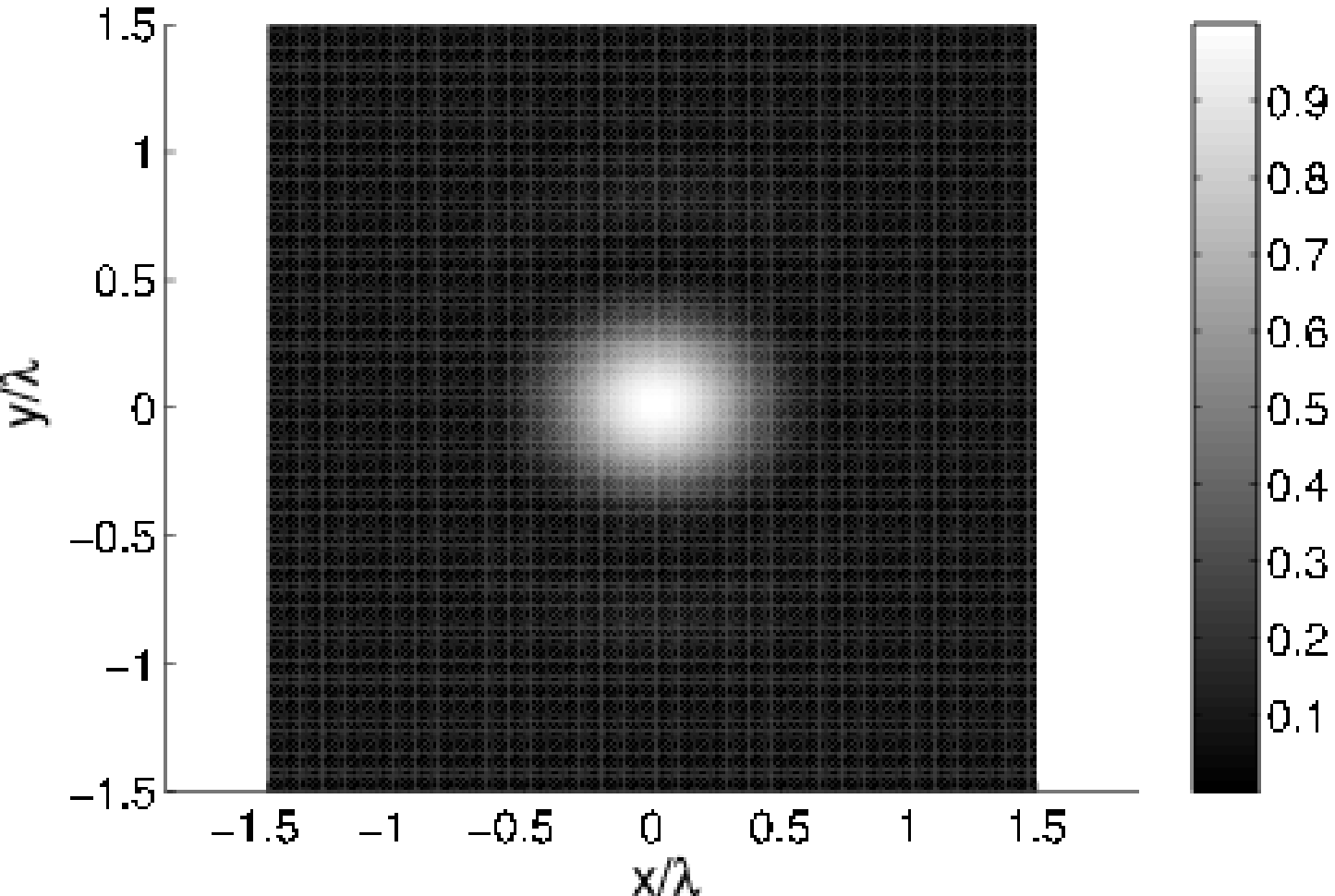}
  \includegraphics[width=8.15cm]{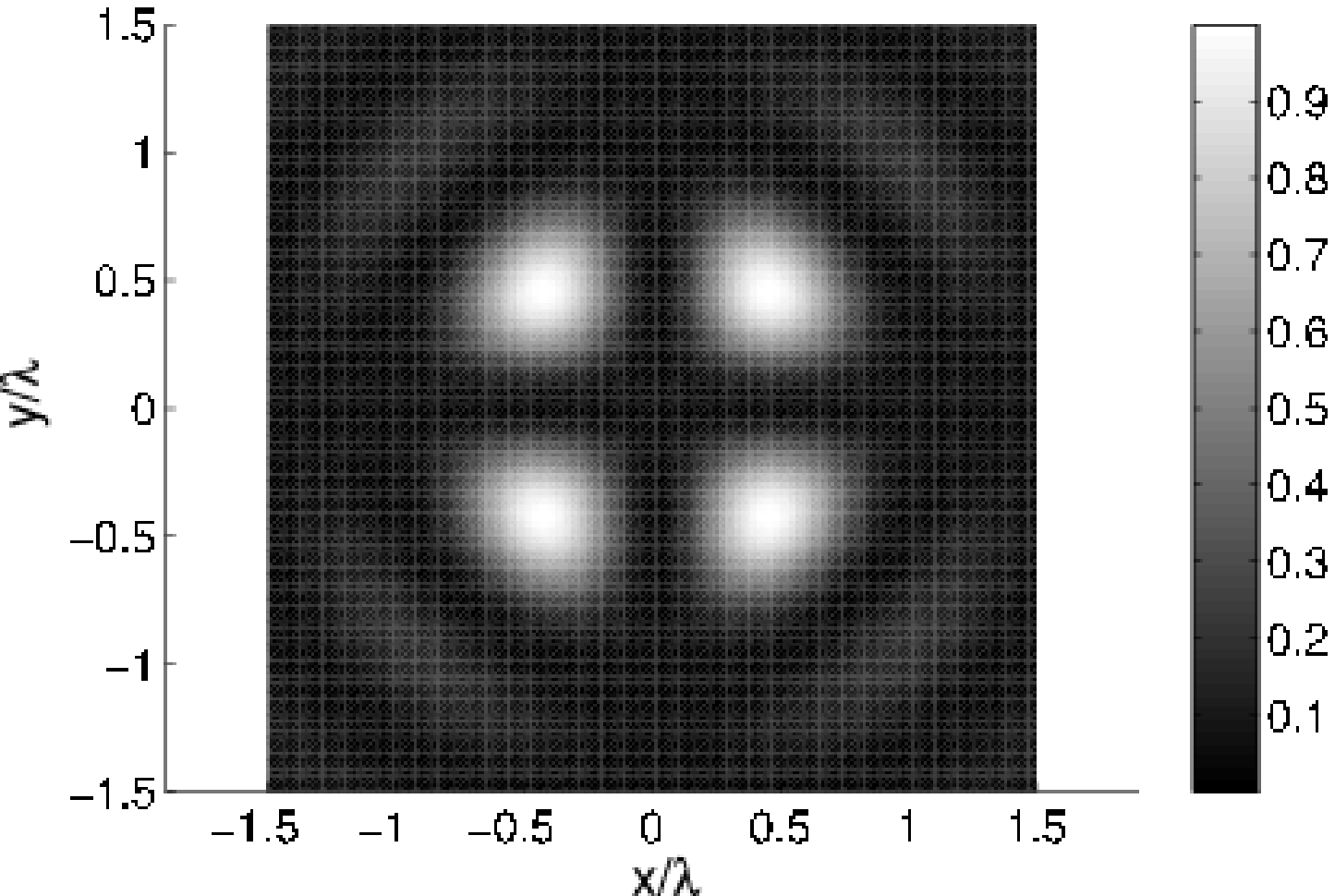}
   \caption{ \label{Fig.EvEy90}
 The normalized distribution of  $|E_x|^2$ (left)  and $|E_y|^2$ (right) in the $z=0$-plane
 for the field with optimized $x$-component when $\mbox{NA}/n=0.9$.
   }
   \end{center}
\end{figure}
    \begin{figure}
  \begin{center}
  \includegraphics[width=8.15cm]{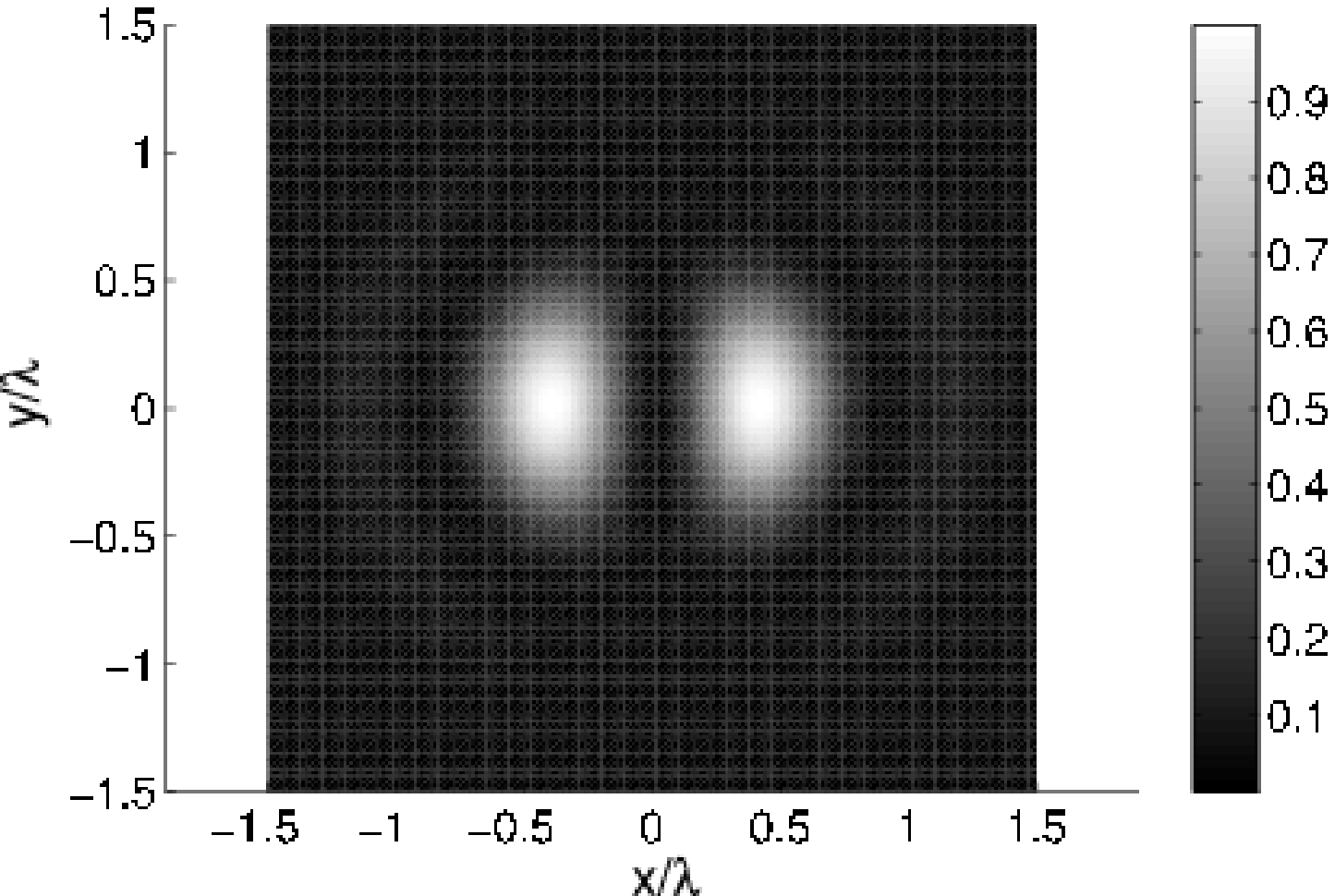}
 \includegraphics[width=8.15cm]{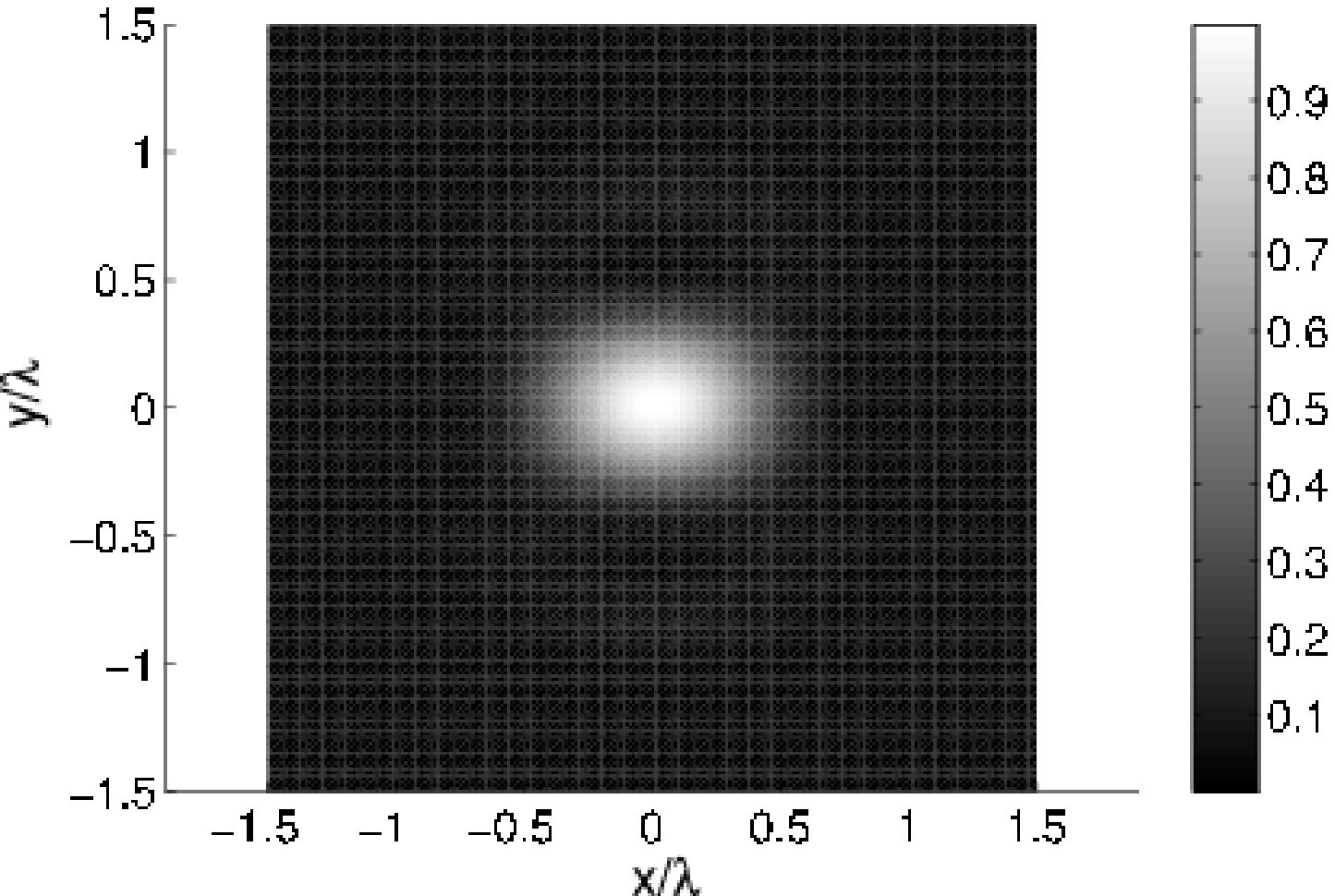}
   \caption{\label{Fig.EvyAE90}
Normalized distribution of $|E_z|^2$ and  of
 the electric energy density $|\mathbf{E}|^2$ for
  the  field with maximum $E_x$-component  $\mbox{NA}/n=0.9$.}
   \end{center}
\end{figure}
The squared amplitudes of the electric field components and the electric energy density in the $z=0$-plane
 are shown in
Figs. \ref{Fig.EvEy90} and \ref{Fig.EvyAE90} when $\mbox{NA}/n=0.9$.
%
                These field components  are very similar to
                the components of the focused $x$-polarized
                  plane wave for the same $\mbox{NA}/n$.
 In
Fig. \ref{fig.crossExmaxEA}
 cross-sections  along the $x$- and
$y$-axes  of $|E_x|^2$ for  the maximum transverse field are  the focused linearly polarized plane wave
pattern are compared for $\mbox{NA}/n=0.9$.
The maxima of all functions are normalized to 1.
Along the $y$-axis, the  $E_x$-spot of the optimized transverse field is slightly narrower than
for  the Airy spot, but
  has  also higher secondary maxima.
 It can be shown that, just as for the focused $x$-polarized plane wave,
   $E_x$ of  the optimized field is widest along the $x$-axis.
    In Fig. \ref{fig.FWHMEx} the FWHM as function of $\mbox{NA}/n$ of  $|E_x|^2$
 of the optimized field is compared to the FWHM of $|E_x|^2$
 of the $x$-polarized plane wave. The field distributions have  elliptical shape. The FWHM is therefore defined
 with respect to the radial distribution obtained by averaging $|E_x(\varrho,\varphi,z=0)|^2$
  over the angle $0 < \varphi < 2 \pi$, as explained
 in (\ref{eq.FHMAiry}) in the Appendix.
Note that, in contrast to Fig. \ref{Fig.FWHM}, where the optimum $E_z$ component was compared with the
{\em intensity} of the Airy spot, we here compare the squared moduli of the $x$-components of the electric fields.
It is seen that the FWHM of the optimized and the focused plane wave are almost identical.

\begin{figure}
\begin{center}
    \includegraphics[width=9cm]{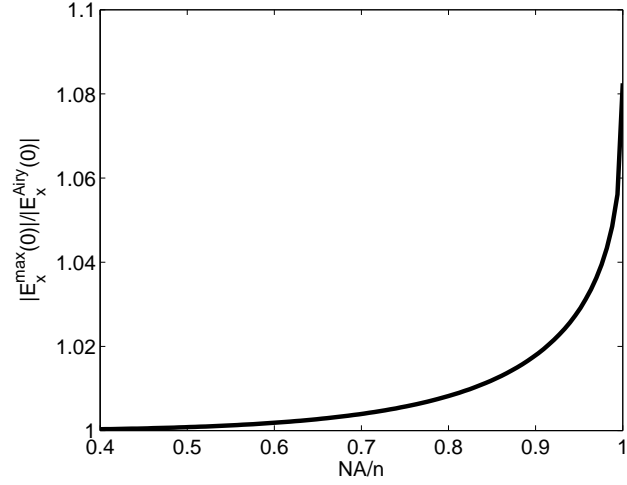}
  \end{center}
\caption{\label{Fig.RatioExTransAiry}  Ratio of the amplitudes of $E_x(0,0,0)$
at the focal point of the field with maximum $|E_x(0,0,0)|$ and of the focused $x$-polarized plane wave with the
same power. The ratio is for $\mbox{NA}/n=1$  maximum with value $15/(8\sqrt{3})$.}
\end{figure}

The optimized electric field  at the focal point
is pointing in the $x$-direction and is given by
\begin{equation}
E_x(0,0,0)= \sqrt{2\pi P_0} \frac{\sqrt{n}}{\lambda_0}
\left( \frac{\mu_0}{\epsilon_0}\right)^{1/4} \,
\left( \frac{4}{3} - \cos \alpha_{\max} - \frac{1}{3} \cos^3\alpha_{\max} \right)^{1/2}.
\label{eq.maxFtransv}
\end{equation}
Formulas for the field distributions of the focused $x$-polarized  plane wave are derived in the Appendix.
For a power given by (\ref{eq.P0plw}), the $x$-component of the electric field in focus is given by
(\ref{eq.EEdens0}). Hence if the power is $P_0$ we have for the focused $x$-polarized plane wave:
\begin{equation}
E_x^{plw}(0,0,0) = \sqrt{2\pi P_0}\frac{\sqrt{n}}{\lambda_0 } \left(\frac{\mu_0}{\epsilon_0}\right)^{1/4}
\frac{1}{\sin \alpha_{\max}}
\left( \frac{16}{15} - \frac{2}{3} \cos^{3/2} \alpha_{\max} - \frac{2}{5} \cos^{5/2}\alpha_{\max}\right).
\label{eq.Ex0plw}
\end{equation}
The ratio of (\ref{eq.maxFtransv}) and (\ref{eq.Ex0plw}):
\begin{equation}
   \sin \alpha_{\max} \frac{ \left( \frac{4}{3} - \cos \alpha_{\max} - \frac{1}{3} \cos^3\alpha_{\max} \right)^{1/2} }
            { \frac{16}{15} - \frac{2}{3} \cos^{3/2} \alpha_{\max} - \frac{2}{5} \cos^{5/2}\alpha_{\max} },
            \label{eq.ratio}
            \end{equation}
            is
 shown in Fig. \ref{Fig.RatioExTransAiry} as function of $\mbox{NA}/n=\sin \alpha_{\max}$. For $\mbox{NA}\rightarrow 0$ the ratio
 becomes 1, while for $\mbox{NA}/n \rightarrow 1$ it becomes $15/( 8 \sqrt{3}) =1.0825$. Hence, the optimized
 field can have more than $8\%$ higher amplitude $E_x$ than the $x$-polarized Airy spot of the same power.

\subsection{Intermediate case}
Without restricting the generality, we may assume that the vector $\mathbf{v}$ is in the
$(x,z)$-plane, i.e. $v_y=0$.
It is then convenient to write the optimum focused electric and magnetic field on the orthonormal
basis with unit vectors $\hat{\mathbf{v}}$, $\hat{\mathbf{y}}$ and $\hat{\mathbf{v}}\times \hat{\mathbf{y}}$.
We have
\begin{equation}
\hat{\mathbf{v}} = v_x \hat{\mathbf{x}} + v_z \hat{\mathbf{z}}, \;\;\;\;
\hat{\mathbf{v}}\times \hat{\mathbf{y}} = -v_z \hat{\mathbf{x}} + v_x \hat{\mathbf{z}},
\label{eq.basis1}
\end{equation}
and therefore
\begin{equation}
\hat{\mathbf{x}} = v_x \hat{\mathbf{v}} - v_z (\hat{\mathbf{v}}\times \hat{\mathbf{y}}),
\;\;\;\; \hat{\mathbf{z}} = v_z \hat{\mathbf{v}} + v_x ( \hat{\mathbf{v}}\times \hat{\mathbf{y}}).
\label{eq.basis2}
\end{equation}
Hence, (\ref{eq.Efield3}) and (\ref{eq.Hfield3})  become:
\begin{eqnarray}
\mathbf{E}(\varrho,\varphi,z)    & = &  \pi \frac{n}{\Lambda\lambda_0^2} \left(\frac{\mu_0}{\epsilon_0}\right)^{1/2}
\left\{ [(g_0^{0,1}+g_0^{2,1} + g_2^{0,3} \cos(2\varphi))\, v_x - 2 i g_1^{1,2} \cos \varphi\, v_z]
\, \hat{\mathbf{x}} \right. \nonumber \\
& &  \left. + [ g_2^{0,3} \sin(2\varphi) \,v_x - 2ig_1^{1,2} \sin\varphi \,v_z]\, \hat{\mathbf{y}}
+[ - 2 i g_1^{1,2} \cos \varphi \, v_x + 2 g_0^{0,3} \, v_z ] \, \hat{\mathbf{z}} \right\}
\nonumber \\
   & = &  \pi \frac{n}{\Lambda\lambda_0^2} \left(\frac{\mu_0}{\epsilon_0}\right)^{1/2}
   \left\{ \left[
   (g_0^{0,1}+g_0^{2,1} + g_2^{0,3} \cos(2\varphi))\, v_x^2 - 4i g_1^{1,2} \cos\varphi\, v_x v_z +
   2g_0^{0,3} \,v_z^2 \right] \hat{\mathbf{v}} \right. \nonumber \\
   & & \left. + [ g_2^{0,3} \sin(2\varphi)\, v_x - 2ig_1^{1,2} \sin\varphi \,v_z]\, \hat{\mathbf{y}}
   \right. \nonumber \\
   & & \left. + \left[ -2i g_1^{1,2} \cos \varphi\, v_x^2
   -( g_0^{0,1}+g_0^{2,1}-2g_0^{0,3}+g_2^{0,3} \cos(2\varphi) ) \, v_x v_z + 2ig_1^{1,2} \cos \varphi\, v_z^2
   \right] \, \hat{\mathbf{v}}\times \hat{\mathbf{y}} \right\},
   \nonumber \\
   \label{eq.Efield10}
   \end{eqnarray}
  and
  \begin{eqnarray}
\mathbf{H}(\varrho,\varphi,z)    & = &
\frac{2\pi n^2}{\Lambda \lambda_0^2} \left\{
[ g_2^{1,1} \sin(2\varphi)\, v_x + i g_1^{0,2}\sin\varphi\, v_z ] \, \hat{\mathbf{x}} \right.
\nonumber \\
& & \left. - [ g_2^{1,1}\cos(2\varphi) \, v_x + i g_1^{0,2} \cos \varphi\, v_z] \, \hat{\mathbf{y}}
+ i g_1^{0,2} \, \sin\varphi\, v_x \, \hat{\mathbf{z}} \right\} \nonumber \\
& = & \frac{2\pi n^2}{\Lambda \lambda_0^2} \left\{
[ g_2^{1,1} \sin(2\varphi) \, v_x^2 +2ig_1^{0,2}\, \sin\varphi\, v_x v_z ] \hat{\mathbf{v}} \right.
\nonumber \\
& & \left. - [ g_2^{1,1}\cos(2\varphi) \, v_x + i g_1^{0,1} \cos \varphi\, v_z] \, \hat{\mathbf{y}}
\right. \nonumber \\
& & \left. +
[ i g_1^{0,2} \, \sin \varphi\, v_x^2 - g_2^{1,1}\sin(2\varphi)\, v_x v_z - i g_1^{0,2} \, \sin \varphi\, v_z^2 ]\,
\hat{\mathbf{v}}\times \hat{\mathbf{y}} \right\}.
\label{eq.Hfield10}
\end{eqnarray}
   Because $g_\ell^{\nu,\mu}(0,z)=0$ when $\ell\geq1$, it follows that on the optical axis: $\varrho=0$,
we have
\begin{eqnarray}
\mathbf{E}(0,\varphi,z)    & = &  \pi \frac{n}{\Lambda\lambda_0^2} \left(\frac{\mu_0}{\epsilon_0}\right)^{1/2}
\left\{ \left[
   (g_0^{0,1}(0,z)+g_0^{2,1}(0,z)) \, v_x^2 +
   2g_0^{0,3}(0,z) \,v_z^2 \right] \hat{\mathbf{v}} \right. \nonumber \\
& & \left. - \left[
   g_0^{0,1}(0,z)+g_0^{2,1}(0,z)-2g_0^{0,3}(0,z) \right] \, v_x v_z
    \, \hat{\mathbf{v}}\times \hat{\mathbf{y}} \right\},
     \label{eq.Efield11}
   \end{eqnarray}
while the magnetic field vanishes there.
It follows in particular that the electric field in the origin
    $\varrho=z=0$ is parallel to
   $\hat{\mathbf{v}}$ only when  $\mathbf{v}$ is perpendicular to the optical axis
($v_z=0$, transverse case) or parallel to it ($v_x=0$ longitudinal case). In all other cases the projection
   of the electric field at the origin  on the plane perpendicular to $\hat{\mathbf{v}}$ is nonzero.

\begin{figure}
 \begin{center}
 \includegraphics[width=8.15cm]{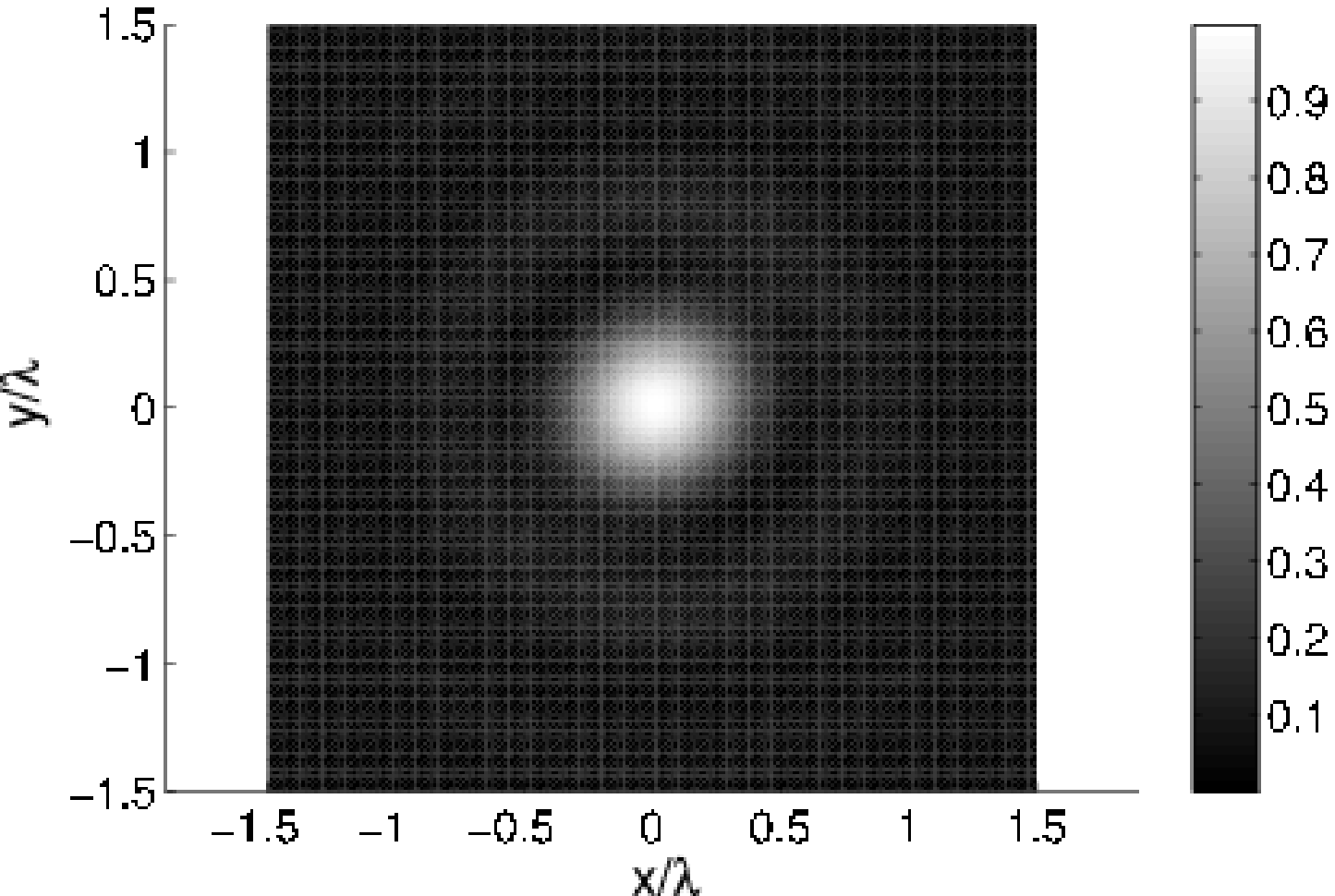}
  \includegraphics[width=8.15cm]{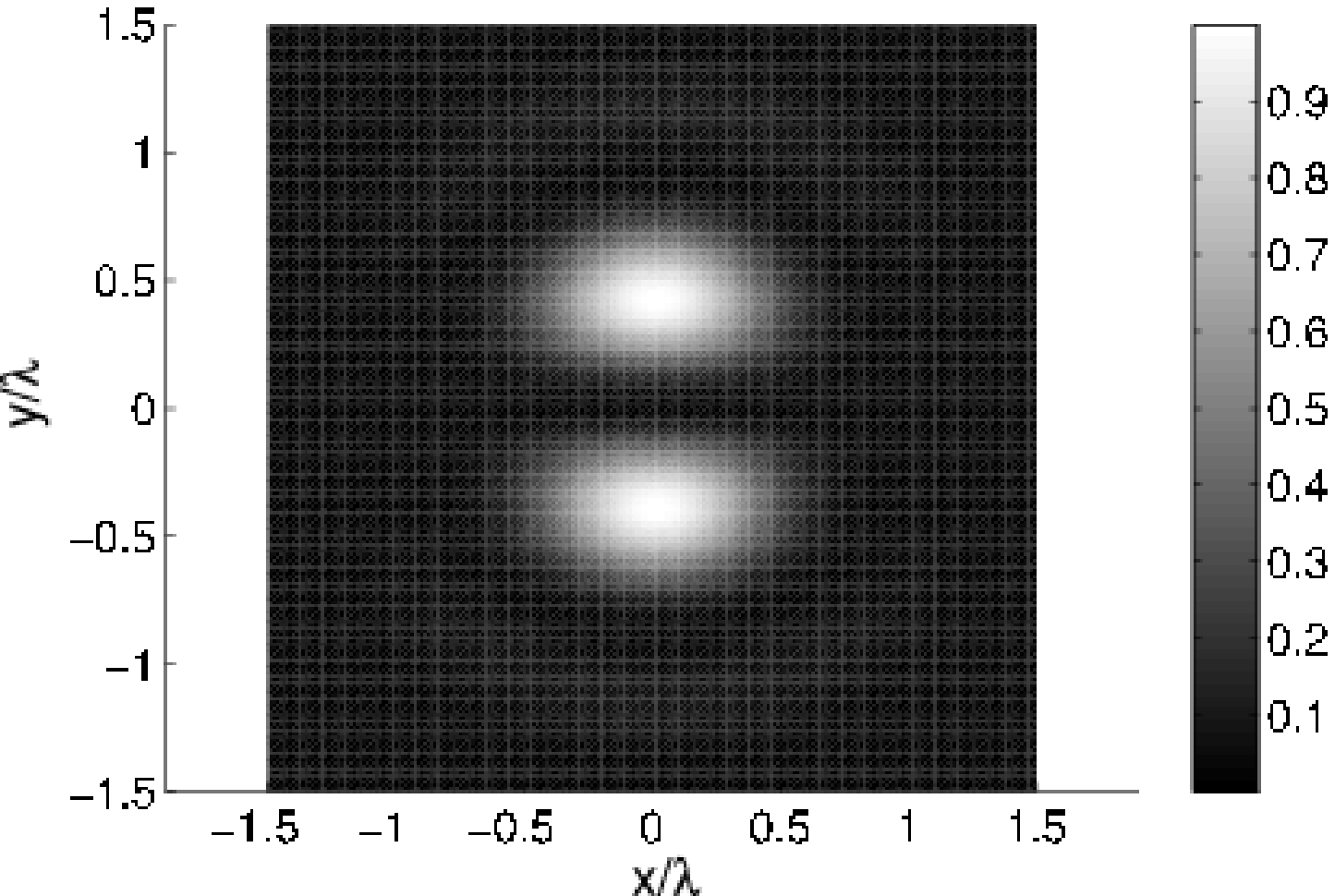}
   \caption{ \label{Fig.EvEy30}
 The normalized distribution of  $|E_v|^2$ (left)  and $|E_y|^2$ (right) in the $z=0$-plane
 for the field with optimum  field with $\vartheta_v=30^o$ when $\mbox{NA}/n=0.9$.
   }
   \end{center}
\end{figure}

\begin{figure}
  \begin{center}
  \includegraphics[width=8.15cm]{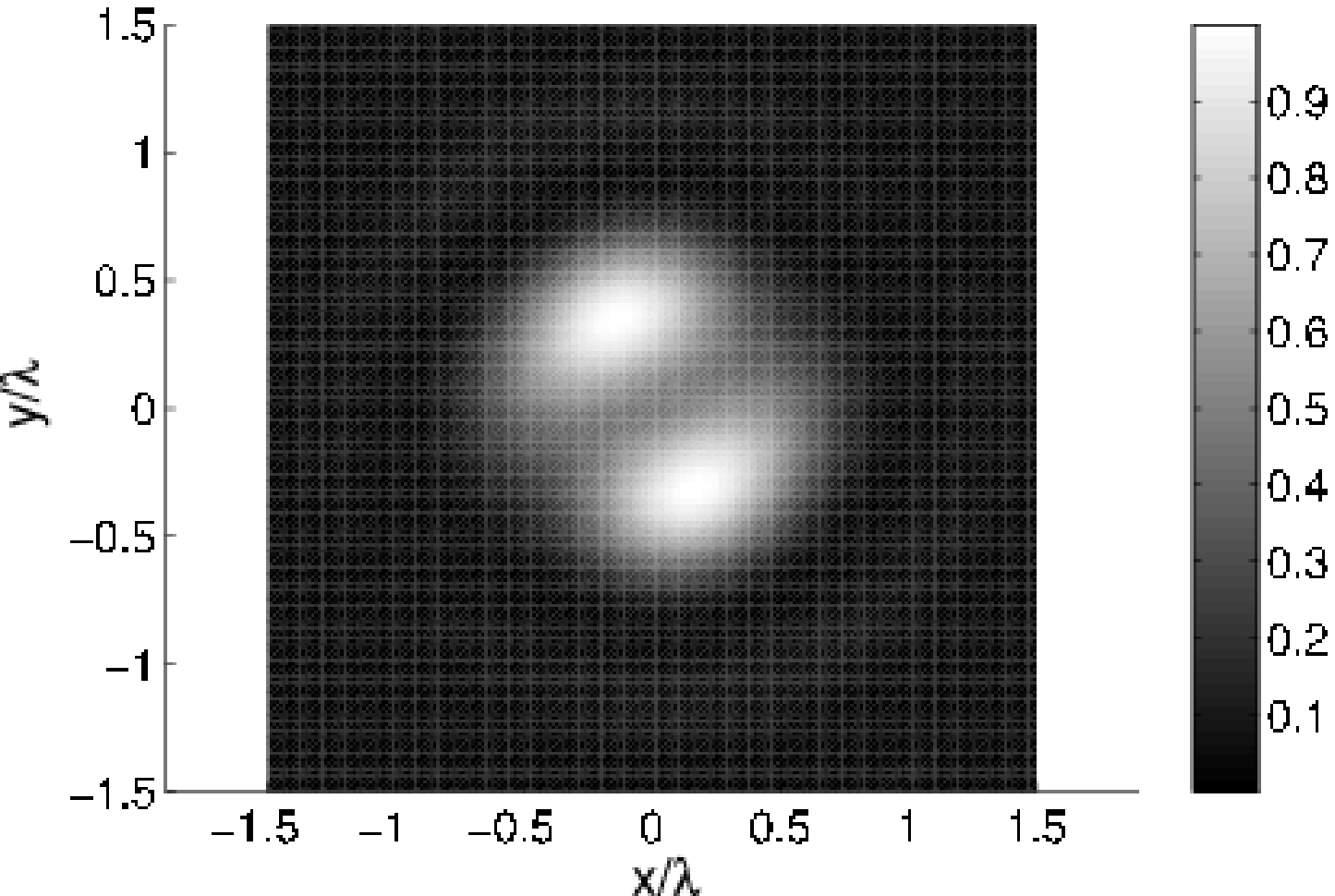}
 \includegraphics[width=8.15cm]{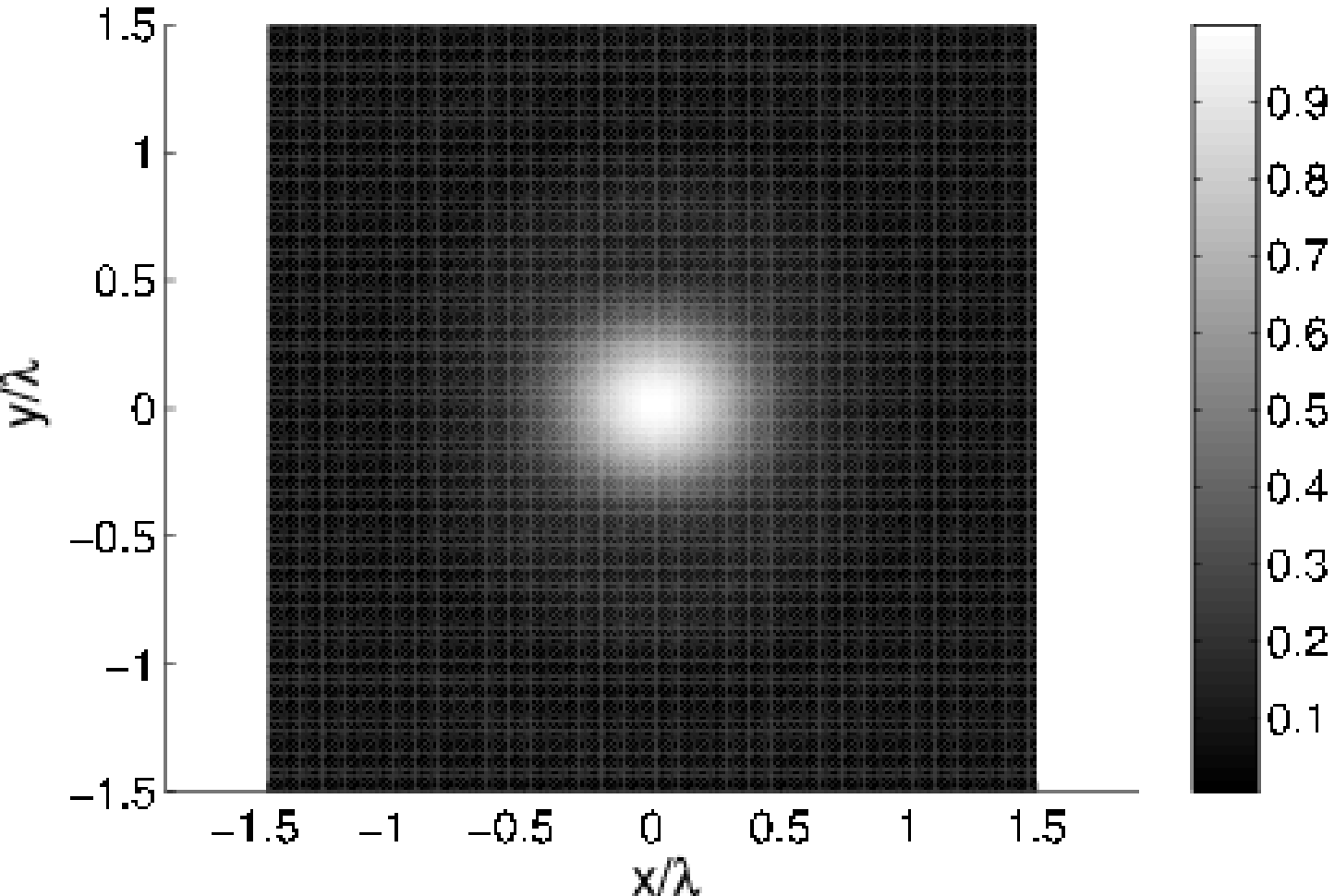}
   \caption{\label{Fig.EvyAE30}
Normalized distribution of $|E\cdot(\hat{\mathbf{v}}\times \hat{\mathbf{y}})|^2$ and  of
 the electric energy density $|\mathbf{E}|^2$ of
  the field with maximum $v$-component for $\vartheta_v=30^o$ and  $\mbox{NA}/n=0.9$.}
   \end{center}
\end{figure}

\begin{figure}
 \begin{center}
 \includegraphics[width=8.15cm]{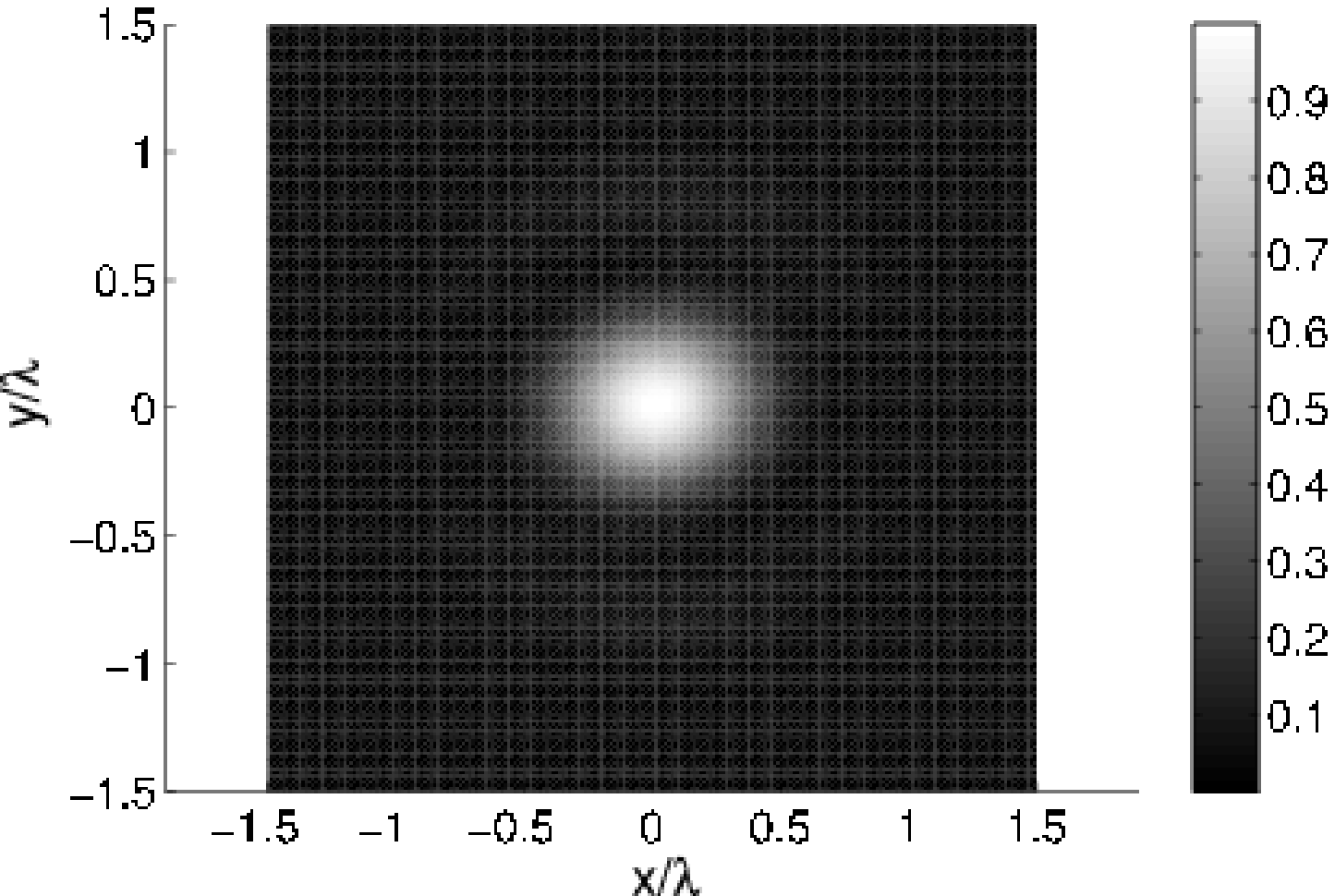}
  \includegraphics[width=8.15cm]{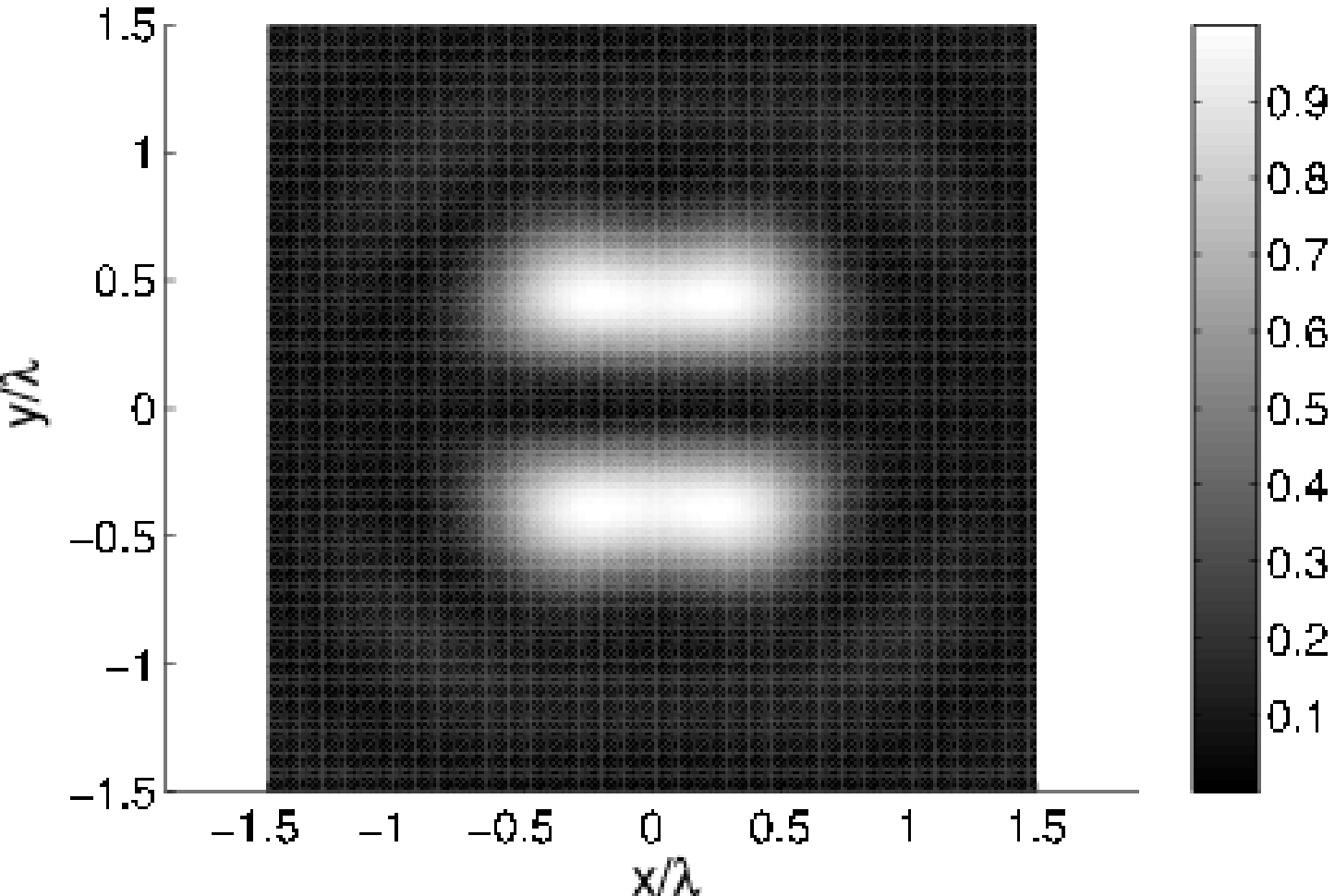}
   \caption{ \label{Fig.EvEy60}
 The normalized distribution of  $|E_v|^2$ (left)  and $|E_y|^2$ (right) in the $z=0$-plane
 for the field with optimum  field with $\vartheta_v=60^o$ when $\mbox{NA}/n=0.9$.
   }
   \end{center}
\end{figure}

\begin{figure}
  \begin{center}
  \includegraphics[width=8.15cm]{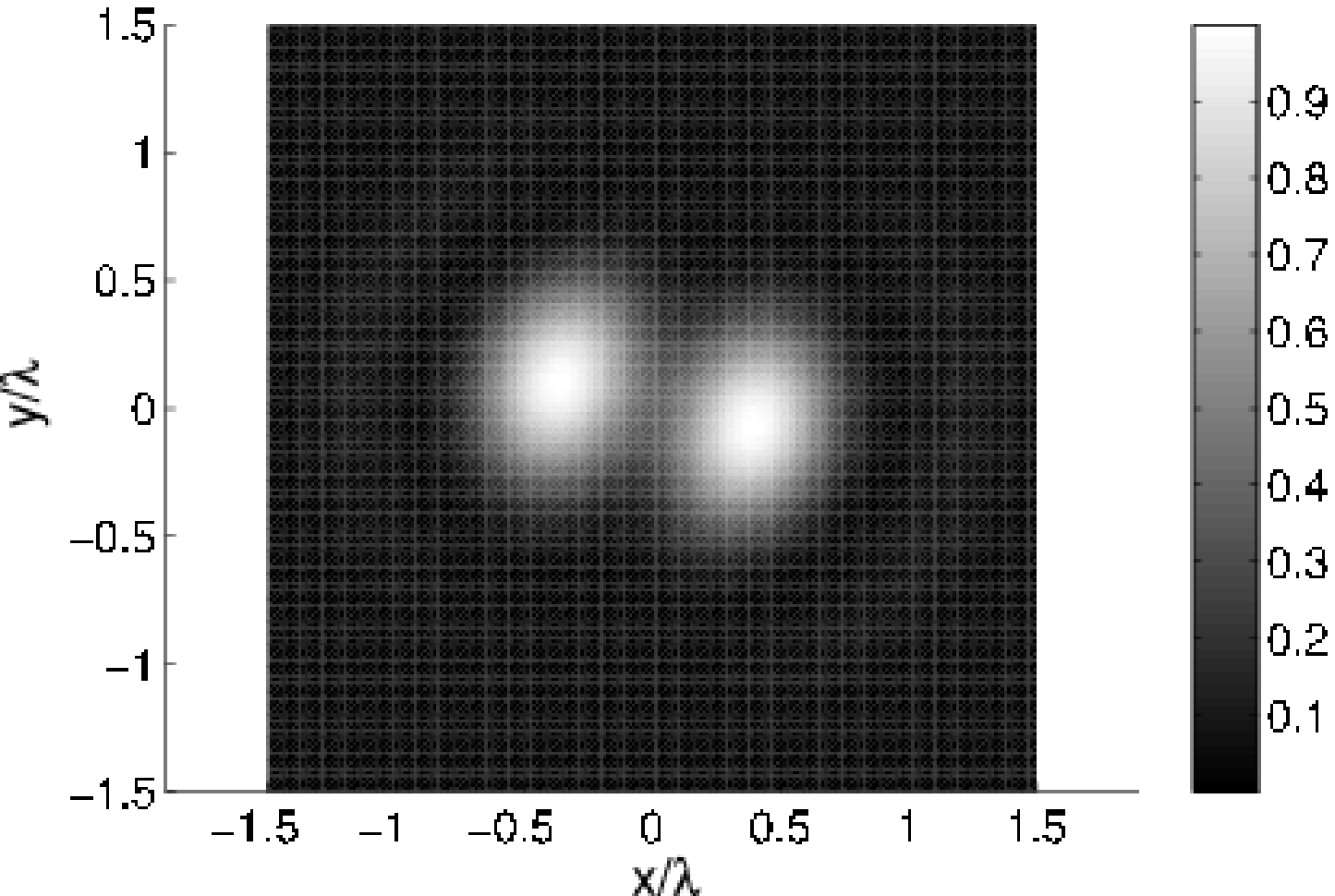}
 \includegraphics[width=8.15cm]{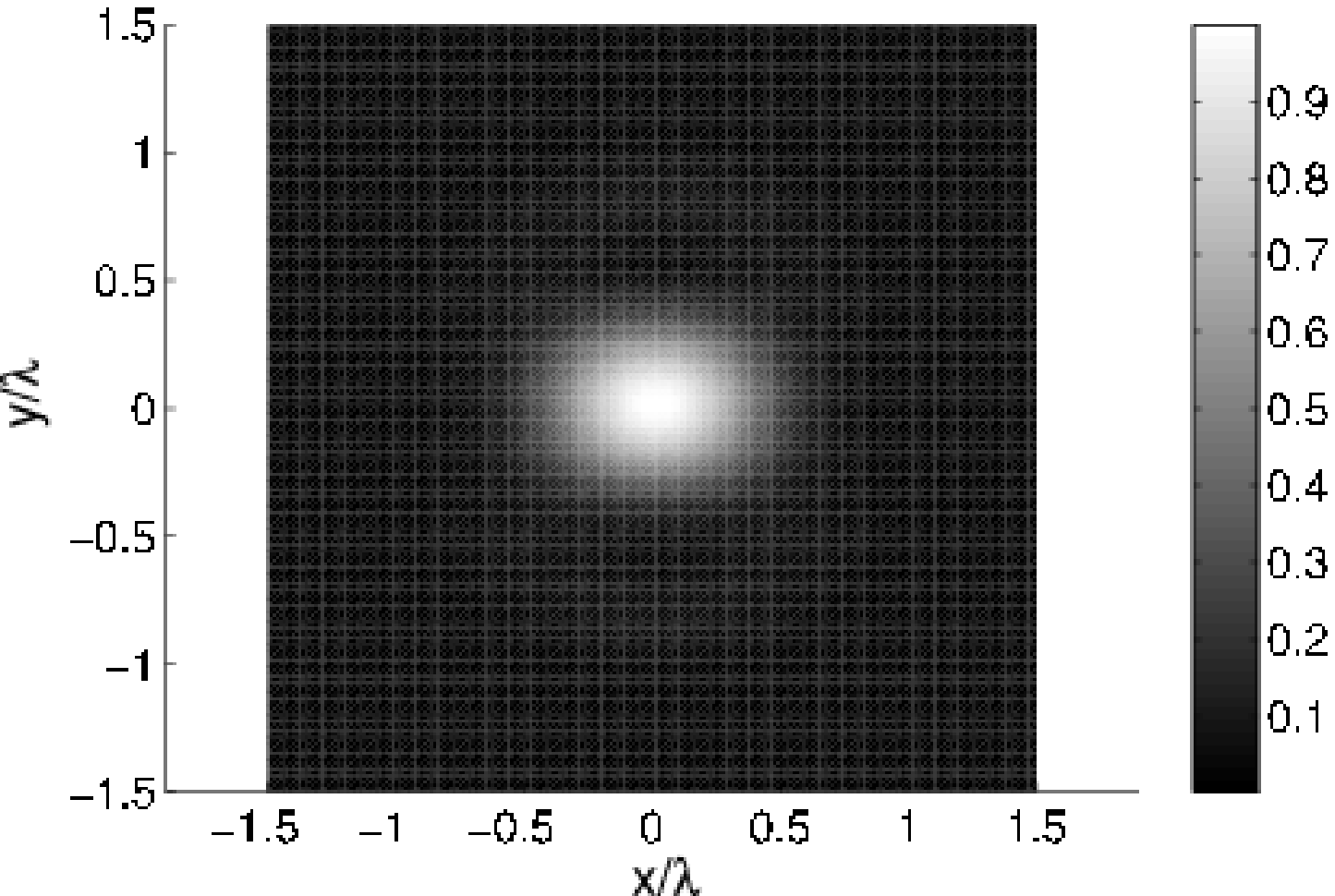}
   \caption{\label{Fig.EvyAE60}
Normalized distribution of $|E \cdot (\hat{\mathbf{v}}\times \hat{\mathbf{y}})|^2$ and  of
 the electric energy density $|\mathbf{E}|^2$ of
  the field with maximum $v$-component for $\vartheta_v=60^o$ and  $\mbox{NA}/n=0.9$.}
   \end{center}
\end{figure}

The $v$-component of the electric field is
\begin{eqnarray}
\mathbf{E}(\varrho,\varphi,z)\cdot \hat{\mathbf{v}} & = &
 \pi \frac{n}{\Lambda\lambda_0^2} \left(\frac{\mu_0}{\epsilon_0}\right)^{1/2}
 \left\{
 [g_0^{0,1}+g_0^{2,1} + g_2^{0,3} \cos(2\varphi)]\, v_x^2 - 4i g_1^{1,2} \cos\varphi\, v_x v_z +
   2g_0^{0,3} \,v_z^2 \right\}. \nonumber \\
   \label{eq.Ev}
\end{eqnarray}
In particular, since the functions $g^{\nu,\mu}_\ell$ are all real for $z=0$, we find for the
squared modulus of the $v$-component in the  $z=0$-plane:
\begin{eqnarray}
|\mathbf{E}(\varrho,\varphi,0)\cdot \hat{\mathbf{v}}|^2 & = &
 \pi^2 \frac{n^2}{\Lambda^2\lambda_0^4} \left(\frac{\mu_0}{\epsilon_0}\right)
 \left\{  [ (g_0^{0,1} + g_0^{2,1}-g_2^{0,3}) v_x^2  + 2 g_0^{0,3} v_z^2 ]^2 \right.\nonumber \\
  & & \left. + 4 \left[ (g_0^{0,1} + g_0^{2,1} - g_2^{0,3}) g_2^{0,3} v_x^4 +
  2(  g_0^{0,3}g_2^{0,3} + 2 (g_1^{1,2})^2 ) v_x^2 v_z^2 \right] \cos^2 \varphi \right. \nonumber\\
  & & \left. + 4 (g_2^{0,3})^2 \, v_x^4 \cos^4 \varphi \right\},
  \label{eq.Ev2}
 \end{eqnarray}
 where all $g_l^{\nu, \mu}$ are evaluated at $\varrho$ and $z=0$.
 If $\vartheta_v$ is the angle between the vector $\hat{\mathbf{v}}$ and the positive $z$-axis, we have
\begin{equation}
v_x = \sin \vartheta_v, \;\;\;
v_z =  \cos \vartheta_v.
\label{eq.vtheta}
\end{equation}
  In Figs. \ref{Fig.EvEy30} and \ref{Fig.EvyAE30},  $|\mathbf{E}\cdot \mathbf{v}|^2$,
$|E_y|^2$, $|\mathbf{E}\cdot(\mathbf{v}\times \hat{\mathbf{y}})|^2$ and the
electric energy density $|\mathbf{E}|^2$ are shown in the $z=0$-plane for the optimum field with
$\vartheta_v=30^o$ and $\mbox{NA}/n=0.9$.
Figs. \ref{Fig.EvEy60} and \ref{Fig.EvyAE60} correspond to $\vartheta_v=60^o$.

In Table 1 the maxima of the indicated field components in the focal plane are listed for
the optimum fields corresponding to $\mbox{NA}/n=0.9$ for several choices of $\vartheta_v$.
In all cases the total flow of power  in the pupil of the lens is 1 W.

   Eqn. (\ref{eq.Ev2}) is
 a quadratic function of $\cos^2 \varphi$ and since the coefficient of $\cos^4 \varphi$ is non-negative,
 it follows that,  for every $\varrho$, the maximum is attained when $\cos^2 \varphi =1$, i.e. when
 $\varphi=0$ and when $\varphi=\pi$. To obtain a useful measure of the spot size, we average
 (\ref{eq.Ev2}) over $0 < \varphi < 2 \pi$:
 \begin{eqnarray}
  \frac{1}{2\pi} \int_0^{2\pi} |\mathbf{E}(\varrho,\varphi,0)\cdot \hat{\mathbf{v}}|^2 \, d\varphi
  & = & \pi^2 \frac{n^2}{\Lambda^2\lambda_0^4} \left(\frac{\mu_0}{\epsilon_0}\right)
  \left\{  [ (g_0^{0,1} + g_0^{2,1}-g_2^{0,3}) v_x^2  + 2 g_0^{0,3} v_z^2 ]^2 \right.\nonumber \\
  & & \left. + 2\left[ (g_0^{0,1} + g_0^{2,1} - g_2^{0,3}) g_2^{0,3} v_x^4 +
  2 (  g_0^{0,3}g_2^{0,3} +  2 (g_1^{1,2})^2) v_x^2 v_z^2 \right] \right. \nonumber\\
  & & \left. + \frac{3}{2} \,(g_2^{0,3})^2 v_x^4  \right\},
  \label{eq.avEv2}
 \end{eqnarray}
 where we used
 \begin{equation}
 \frac{1}{2\pi} \int_0^{2\pi}\cos^2 \varphi \, d \varphi = \frac{1}{2}, \;\;\;
 \frac{1}{2\pi} \int_0^{2\pi} \cos^4 \varphi \, d \varphi = \frac{3}{8}.
 \label{eq.averages}
 \end{equation}
  The FWHM in  the $z=0$-plane is then defined as the value $2\varrho_0$ such that
 \begin{equation}
    \frac{1}{2\pi} \int_0^{2\pi} |\mathbf{E}(\varrho_0,\varphi,0)\cdot \hat{\mathbf{v}}|^2\,
     d \varphi = \frac{1}{2}\, |\mathbf{E}(0,0,0)\cdot \hat{\mathbf{v}}|^2.
    \label{eq.FWHMv}
    \end{equation}

In Figs. \label{Fig.FWHMEv_thetav} and \label{Fig.Evmax_thetav}
the FWHM of $E_v$ and the maximum $|E_v(\mathbf{0})|$ are shown as   as function of $\vartheta_v$
for several values of the numerical aperture and for $P_0=1$ W. In Fig. \label{Fig.Evmax_thetav}
 the amplitude of $|E_x(\mathbf{0})|$ of the focused $x$-polarized plane wave is shown for the
same power are indicated by dots.
In contrast to the FWHM, the maximum
$E_v(\mathbf{0})$ is a monotonically
increasing function of $\vartheta_v$.



\begin{table}
\begin{tabular}{||c|c|c|c||}
\hline \;\; $\vartheta_v$\;\;\;\; [degrees] &   $\;\; |E_v| \;\;\;\;[V/m] $  & \;\; $ |E_y| \;\;\;\;[V/m] $  &  \;\; $ |E\cdot( \hat{\mathbf{v}}\times \hat{\mathbf{y}})|$ \;\;\;\; [V/m] \\  \hline
$90^o$   &  45.28 &  6.03 &  13.67    \\ \hline
$60^o$ &     42.93 &   7.33 & 13.81    \\ \hline
$30^o$    &  37.76 &  14.17   &  14.04  \\ \hline
$0^o$    &   34.87 &  17.72  &  17.72           \\  \hline \hline
\end{tabular}
\caption{Maxmima of the optimum electric field components (not of their squares)
 in the focal plane when $\mbox{NA}/n=0.9$ and for several
choices of the angle $\vartheta_v$. The total power in the lens pupil is in all cases unity.
Note that the maximum of $|E_y|$ and
$|E\cdot( \hat{\mathbf{v}}\times \hat{\mathbf{y}})|$ are not  attained at the focal point.}
\end{table}

\begin{figure}
  \begin{center}
  \includegraphics[width=9cm]{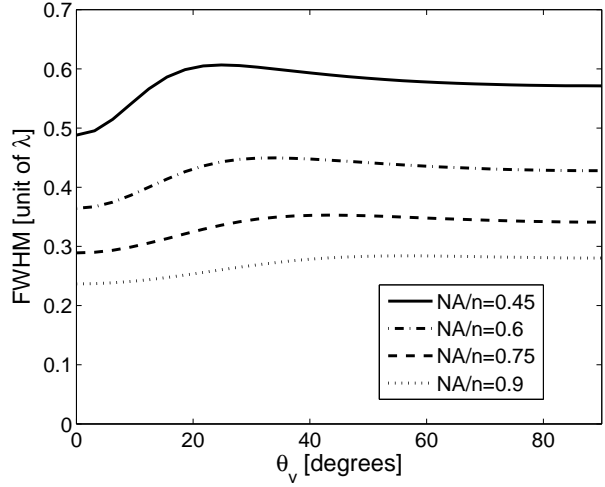}
    \caption{\label{Fig.FWHMEv_thetav}
FWHM expressed in units of $\lambda=\lambda_0/n$ of the optimum $\hat{\mathbf{v}}$-component as function of $\vartheta_v$ for several values of
$\mbox{NA}/n$.}
    \end{center}
\end{figure}

 \begin{figure}
  \begin{center}
  \includegraphics[width=9cm]{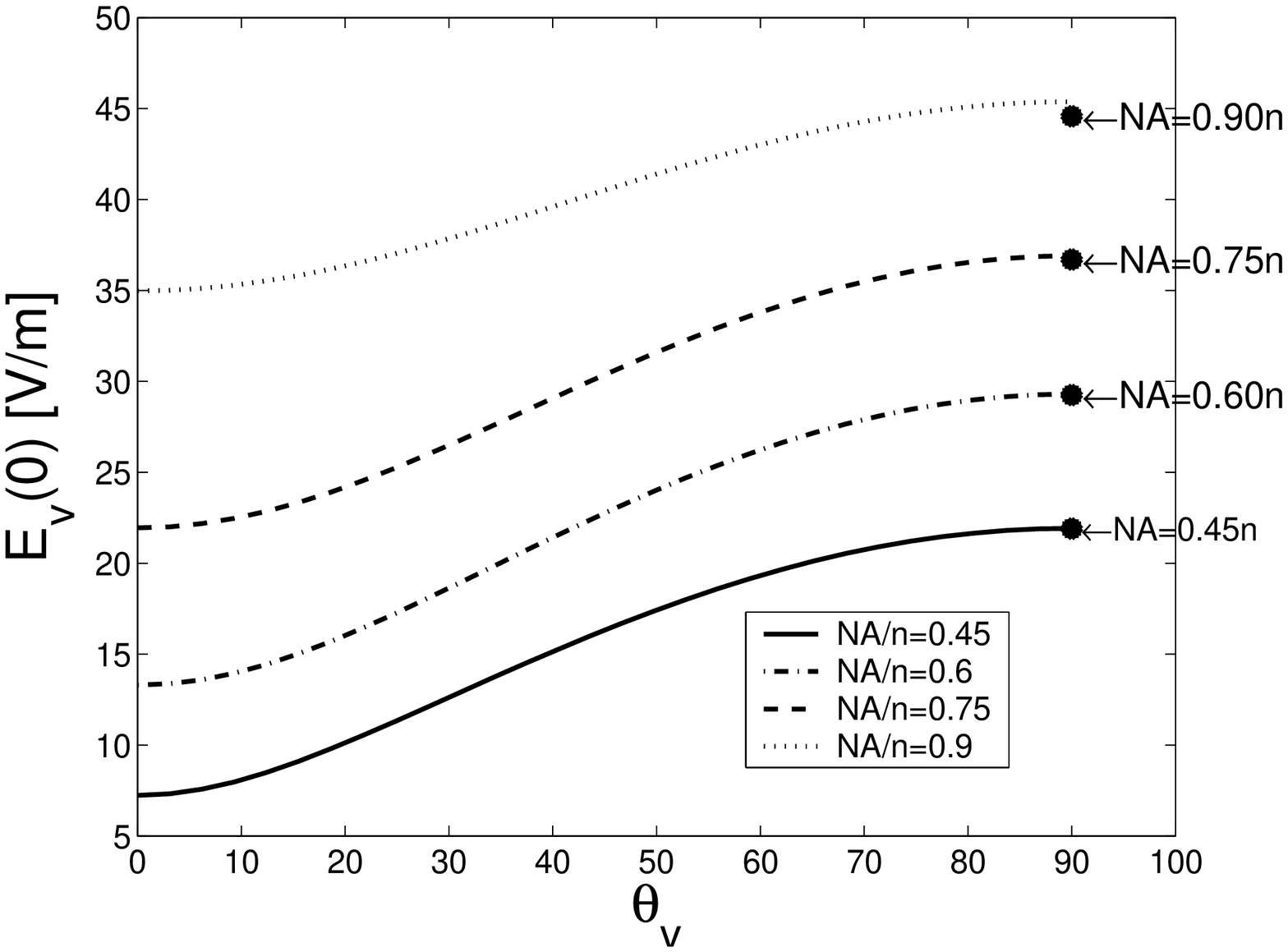}
    \caption{\label{Fig.Evmax_thetav}
$E_v(\mathbf{0})$ in V/m of the field with optimum $\hat{\mathbf{v}}-$component as function of
the angle between $\hat{\mathbf{v}}$ and  the $z$-axis (i.e. $\hat{\mathbf{v}}=\sin\vartheta_v \hat{\mathbf{x}}
+ \cos \vartheta_\vartheta \hat{\mathbf{z}}$), for several $\mbox{NA}$. The power incident on the lens
is $P_0=1$ W. The dots are the amplitudes  $|E_x(\mathbf{0})|$ of the focused $x$-polarized plane wave
for the same power and the same values of the numerical aperture.}
    \end{center}
\end{figure}

\section{The optimum field in the lens pupil}
\label{section.focusing}
 The optimum fields  in the  $z=0$-plane  can be obtained by focusing
 appropriate pupil distributions.
We derive these pupil distributions in this section, using the vector diffraction theory of
Ignatowksy and Richards and Wolf. This theory is based on 1. Debye's approximation which
expresses   the plane wave
amplitudes in image space to the field vectors in the entrance pupil, and 2. the Abbe's sine condition
to guarantee conservation of energy.

Consider a  beam that is incident on the pupil of a diffraction limited lens with
numerical aperture $\mbox{NA}$. The electric field in the pupil is given by
  \begin{equation}
  \mathbf{E}^p(x_p, y_p) =  E_{\varphi}^p(x_p,y_p)\hat{\bfphi}_p
  +  E_\varrho^p(x_p,y_p)\hat{\bfrho}_p,
  \label{eq.pupil2}
  \end{equation}
  where $x_p$ and $y_p$ are cartesian coordinates  in the lens pupil, which are
  parallel to the $x$- and $y$-coordinates  of the cartesian  system $(x,y,z)$
   in the focal region,   and $\varphi_p$
  and $\varrho_p$ are polar coordinates in the pupil:
 \begin{eqnarray}
   x_p & = & \varrho_p \cos \varphi_p,  \label{eq.xp} \\
   y_p & = & \varrho_p \sin \varphi_p,  \label{eq.yp}
   \end{eqnarray}
   with unit vectors
   \begin{eqnarray}
   \hat{\bfrho}_p & = &   \cos \varphi_p\hat{\bf x} + \sin \varphi_p \hat{\bf y}, \label{eq.bfrho}\\
   \hat{\bfphi}_p & = & - \sin \varphi_p\hat{\bf x} + \cos \varphi_p \hat{\bf y}.
   \end{eqnarray}

  According to the theory of Ignatowsky \cite{ignatowsky}, \cite{ignatowskyb}  and
  Richards and Wolf \cite{wolf}, \cite{richards} the
   field in the pupil of the lens  and the amplitude of the
   plane waves in the focal plane  are related by
   \begin{equation}
      \mathbf{A}(k_x, k_y) = - 2 \pi i \frac{f}{(k_0 n)^{1/2} k_z^{1/2} }  \,
      \tensR(\mathbf{k}) \mathbf{E}^p\left(-\frac{f k_x}{k_0 n}, - \frac{f k_y}{k_0 n} \right),
      \label{eq.richards1}
      \end{equation}
      where $f$ is the focal distance of the lens
      and $\tensR(\mathbf{k})$ is a matrix which rotates the electric field
      in the pupil plane in the direction perpendicular to the wave vector
$\mathbf{k}=k_0 n (\sin \alpha \cos \beta \,\hat{\mathbf{k}}+
\sin \alpha \sin \beta\,\hat{\bfalpha}+  \cos
\alpha\,\hat{\bfbeta})$:
 \begin{eqnarray}
   \tensR\left( \mathbf{k} \right) \mathbf{E}^p(x_p, y_p)
     & = &- E_{\varrho}^p(x_p, y_p) \, \hat{\bfalpha} -  E_{\varphi}^p(x_p, y_p) \,\hat{\bfbeta},
   \label{eq.tensR}
   \end{eqnarray}
   with
   \begin{eqnarray}
   x_p & = & \varrho_p \cos \varphi_p= - f \frac{k_x}{k_0 n}, \\
   y_p & = & \varrho_p \sin \varphi_p = -f \frac{k_y}{ k_0 n},
\end{eqnarray}
  and $\hat{\bfalpha}$ and $\hat{\bfbeta}$ are given by
               (\ref{eq.hatalpha}), (\ref{eq.hatbeta}).
Since $\alpha$ and $\beta$ are the polar and azimuthal  angles of the wave vector $\mathbf{k}$:
\begin{eqnarray}
  \varrho_p&  =  &f \sin \alpha,
  \label{eq.rhotheta} \\
  \varphi_p&  =  &  \beta + \pi, \label{eq.phi_pphi}
  \end{eqnarray}
  Hence, expressed in terms of $\alpha$ and $\beta$:
   \begin{eqnarray}
      \mathbf{A}(\alpha, \beta)& =&  - 2 \pi i \frac{f}{k_0 n \sqrt{\cos\alpha} }
      \tensR(\alpha, \beta) \mathbf{E}^p
      \left(-f \sin \alpha  \cos\beta, - f \sin\alpha \sin \beta \right) \nonumber \\
& = & - 2 \pi i \frac{f}{k_0 n \sqrt{\cos\alpha} }  \left[
    E^p_{\varrho}\left( -f \sin \alpha  \cos\beta, - f \sin\alpha \sin \beta \right)
\, \hat{\bfalpha} \right. \nonumber \\
& & \left. \hspace{2.5cm} +  E_{\varphi}^p\left( -f \sin \alpha
\cos\beta, - f \sin\alpha \sin \beta \right) \,
\hat{\bfbeta}\right].
      \label{eq.richards11}
      \end{eqnarray}

Conversely, given the plane waves amplitudes
$\mathbf{A}(k_x, k_y) = A_\alpha \hat{\bfalpha} + A_\beta \hat{\bfbeta}$
of the field in the focal plane $z=0$,
the corresponding field in the pupil of the lens is:
\begin{eqnarray}
\mathbf{E}^p(x_p,y_p) = \frac{ i}{2\pi} \frac{(k_0 n)^{1/2} k_z^{1/2}}{f}
 \tensR(\mathbf{k})^{-1} \mathbf{A}(k_x,k_y),
 \label{eq.richards2}
 \end{eqnarray}
 with
 \begin{equation}
 k_x=-k_0 n\frac{x_p}{f}, \;\;\;  k_y=-k_0 n\frac{y_p}{f},\;\;\;
 \label{eq.kvec}
 \end{equation}
 $(x_p^2 + y_p^2)^{1/2} \leq f \mbox{NA}/n$  and
 \begin{equation}
   \tensR\left( \mathbf{k} \right)^{-1} \mathbf{A}
    =  - A_\alpha \hat{\bfrho}_p- A_\beta \hat{\bfphi}_p.
   \label{eq.Rinverse}
   \end{equation}
     In terms of $\alpha$, $\beta$ we get:
\begin{eqnarray}
\mathbf{E}^p(\varrho_p,\varphi_p) = \frac{ i}{2\pi} \frac{k_0 n}{f} \cos^{1/2}\alpha  \,
 \tensR(\alpha,\beta)^{-1} \mathbf{A}(\alpha,\beta),
 \label{eq.richards22}
 \end{eqnarray}
 where $\varrho_p$, $\varphi_p$ and $\alpha$, $\beta$ are related by
 (\ref{eq.rhotheta}) and (\ref{eq.phi_pphi}).
By substituting the expressions for the optimum plane wave coordinates (\ref{eq.Aalpha}), (\ref{eq.Abeta}):
\begin{eqnarray}
    A_\alpha&  = &  \frac{1}{\Lambda n} \left(\frac{\mu_0}{\epsilon_0}\right)^{1/2}
                    \, \frac{v_\alpha}{\cos \alpha}, \label{eq.AalphaB}, \\
                    A_\beta & = & \frac{1}{\Lambda n} \left(\frac{\mu_0}{\epsilon_0}\right)^{1/2}
                    \, \frac{v_\beta}{\cos \alpha},  \label{eq.AbetaB}
   \end{eqnarray}
we find
\begin{eqnarray}
\mathbf{E}^p(\varrho_p,\varphi_p)& = &-\frac{ i}{2\pi} \frac{k_0 n}{f} \sqrt{\cos\alpha} \,
  \left( A_\alpha \hat{\bfrho_p} + A_\beta \hat{\bfphi_p}\right) \nonumber \\
  & = & -i \frac{1}{\Lambda \lambda_0 f} \left(\frac{\mu_0}{\epsilon_0}\right)^{1/2}
  \frac{1}{\sqrt{\cos \alpha}} \left( v_\alpha\hat{\bfrho_p} + v_\beta \hat{\bfphi_p}\right),
     \label{eq.Epupil1}
     \end{eqnarray}
where
\begin{eqnarray}
\Lambda  & = & \sqrt{\frac{\pi}{2}}\frac{n^{1/2}}{P_0^{1/2}
\lambda_0}\left( \frac{\mu_0}{\epsilon_0}\right)^{1/4}
\,
\left(\frac{4}{3}-\cos \alpha_{\max} -\frac{1}{3}\cos^3\alpha_{\max}
 -    \sin^2\alpha_{\max} \cos \alpha_{\max}
        v_z^2
      \right)^{1/2}. \nonumber \\
      \label{eq.LambdaB}
\end{eqnarray}

  Using (\ref{eq.palpha}), (\ref{eq.pbeta}) and
\begin{eqnarray}
\cos \alpha & =  &\frac{\sqrt{f^2 - \varrho_p^2}}{f}, \\
\sin \alpha & =  &\frac{\varrho_p}{f}, \\
\beta & = & \phi_p -\pi,
\end{eqnarray}
we obtain:
\begin{eqnarray}
\mathbf{E}^p(\varrho_p,\varphi_p)& = &
- \frac{i}{\Lambda \lambda_0 f} \left(\frac{\mu_0}{\epsilon_0}\right)^{1/2}
\left[ \left( v_x \cos\alpha \cos \beta + v_y \cos \alpha \sin \beta - v_z \sin \alpha \right)
\hat{\bfrho}_p   \right. \nonumber \\
& & \left. +\left(-v_x \sin \beta + v_y \cos \beta\right) \hat{\bfphi}_p  \right]\, \frac{1}{\sqrt{\cos \alpha}}
\nonumber \\
& = & \frac{i}{\Lambda \lambda_0 f} \left(\frac{\mu_0}{\epsilon_0}\right)^{1/2}
  \left\{ \left[ \frac{\sqrt{f^2 - \varrho_p^2}}{f} \left(v_x \cos\varphi_p + v_y \sin \varphi_p \right) +
  \frac{\varrho_p}{f^{1/2} (f^2 - \varrho_p^2)^{1/4} } v_z \right] \hat{\bfrho}_p \right. \nonumber \\
  & & \left. + \frac{f^{1/2}}{(f^2 -\varrho^2_p)^{1/4}} \left( -v_x \sin \varphi_p + v_y \cos \varphi_p \right)
  \hat{\bfphi}_p \right\} \nonumber \\
  & = &
  \frac{i}{\Lambda \lambda_0 f} \left(\frac{\mu_0}{\epsilon_0}\right)^{1/2}
  \left[  \left(\frac{\sqrt{f^2 - \varrho_p^2}}{f} \, v_{\varrho} +
  \frac{\varrho_p}{f^{1/2} (f^2 - \varrho_p^2)^{1/4} } v_z \right)\, \hat{\bfrho}_p
      + \frac{f^{1/2}}{(f^2 -\varrho_p^2)^{1/4}} \, v_{\varphi}\,  \hat{\bfphi}_p
    \right],\nonumber \\
    \label{eq.Epupil2}
    \end{eqnarray}
    where $v_\varrho$, $v_\varphi$ are the components of the vector $\mathbf{v}$ on the
   polar basis $\hat{\bfrho}_p$, $\hat{\bfphi}_p$ in the pupil.
It is seen that the optimum electric field in the pupil is linearly polarized
with direction of polarization and amplitude that depend on the radial coordinate only.

In the case of the maximum longitudinal component: $v_{\varrho}=v_{\varphi}=0$, $v_z=1$, this becomes
 \begin{eqnarray}
\mathbf{E}^p(\varrho_p,\varphi_p)& = &
\frac{i}{\Lambda \lambda_0 f} \left(\frac{\mu_0}{\epsilon_0}\right)^{1/2}
\frac{\varrho_p}{f^{1/2} (f^2 - \varrho_p^2)^{1/4} }  \,
\hat{\bfrho}_p,
\label{eq.Epupillong}
\end{eqnarray}
whereas for the maximum x-component: $v_{\varrho}=\cos \varphi_p, v_{\varphi}=-\sin \varphi_p,  v_z=0$:
\begin{eqnarray}
\mathbf{E}^p(\varrho_p,\varphi_p)& = &
\frac{i}{\Lambda \lambda_0 f} \left(\frac{\mu_0}{\epsilon_0}\right)^{1/2}
\left[ \frac{\sqrt{f^2 - \varrho_p^2}}{f} \, \cos \varphi_p\,  \hat{\bfrho}_p
 - \frac{f^{1/2}}{(f^2 -\varrho^2_p)^{1/4}} \, \sin\varphi_p\,  \hat{\bfphi}_p \right] \nonumber \\
 & = & \frac{i}{\Lambda \lambda_0 f} \left(\frac{\mu_0}{\epsilon_0}\right)^{1/2}
 \left[ \left( \frac{\sqrt{f^2 - \varrho_p^2}}{f} \, \cos^2 \varphi_p +
   \frac{f^{1/2}}{(f^2 -\varrho^2_p)^{1/4}} \,\sin^2 \varphi_p \right)\, \hat{\mathbf{x}} \right. \nonumber \\
   & & \left. + \frac{1}{2}\left( \frac{\sqrt{f^2 - \varrho_p^2}}{f} - \frac{f^{1/2}}{(f^2 -\varrho^2_p)^{1/4}}
   \right) \sin 2 \varphi_p \, \hat{\mathbf{y}}\right].
   \label{eq.Epupiltrans}
\end{eqnarray}
One may have expected that the maximum longitudinal component is obtained by concentrating the radially polarized
pupil field in a small annular ring at the rim of the pupil. Instead, the optimum field amplitude
(\ref{eq.Epupillong}) increases continuously
from zero at the center to a maximum at the rim of the pupil. This is due to the fact that the total power
is an integral over the pupil of the squared amplitude of the field while the longitudinal component in the focal
point is the integral over the pupil of the field itself. If the radially polarized pupil field is concentrated
in a ring of width $\delta \varrho_p=f \mbox{NA} \, \delta \alpha$, the longitudinal field component in the focal plane becomes
approximately
\begin{equation}
  E_z(\varrho,\phi,0) \propto J_0(k_0 n \varrho \mbox{NA})\, (\delta \alpha)^{1/2}.
  \label{eq.Ezring}
  \end{equation}
 The longitudinal component  thus becomes in the limit $\delta \alpha \rightarrow 0$
  proportional to  $J_0(k_0 n \varrho \mbox{NA})$ and hence  decreases slowly as
 $1/\sqrt{k_0 \varrho}$ for increasing distance $\varrho$ to the optical axis.
 To keep the total power finite, the
 amplitude vanishes  with the square root of the width of the annular ring.
 The FWHM of  the squared amplitude decreases with the width of the ring and has minimum
 value of $0.35 \lambda_0/(n \mbox{NA})$ for $\delta \alpha \rightarrow 0$.
  But for small widths the side lobes are relatively strong, the first maximum adjacent to the center
 having squared amplitude that is $16 \%$ of the central maximum.

\begin{figure}
\begin{center}
    \includegraphics[width=8cm]{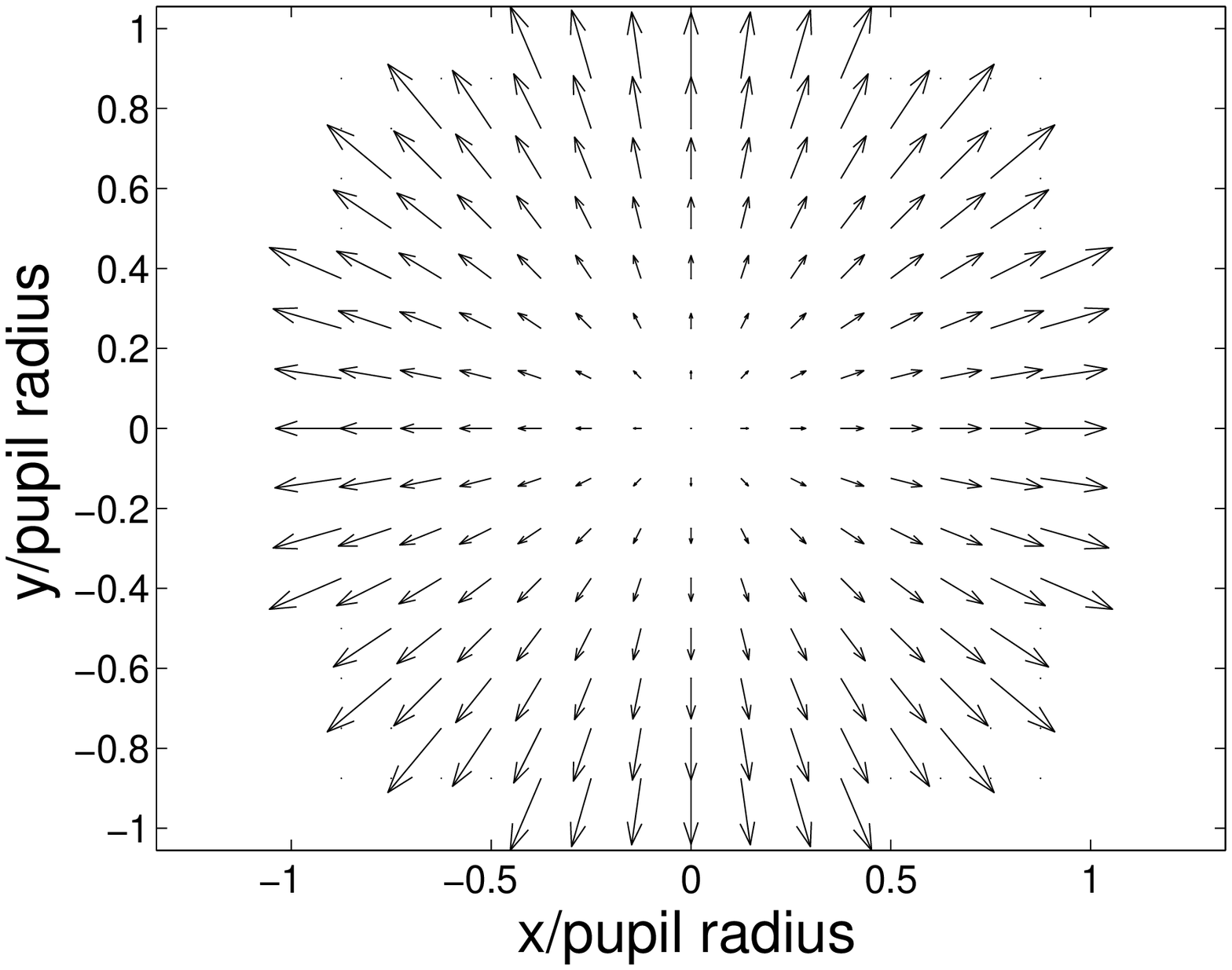} \includegraphics[width=8cm]{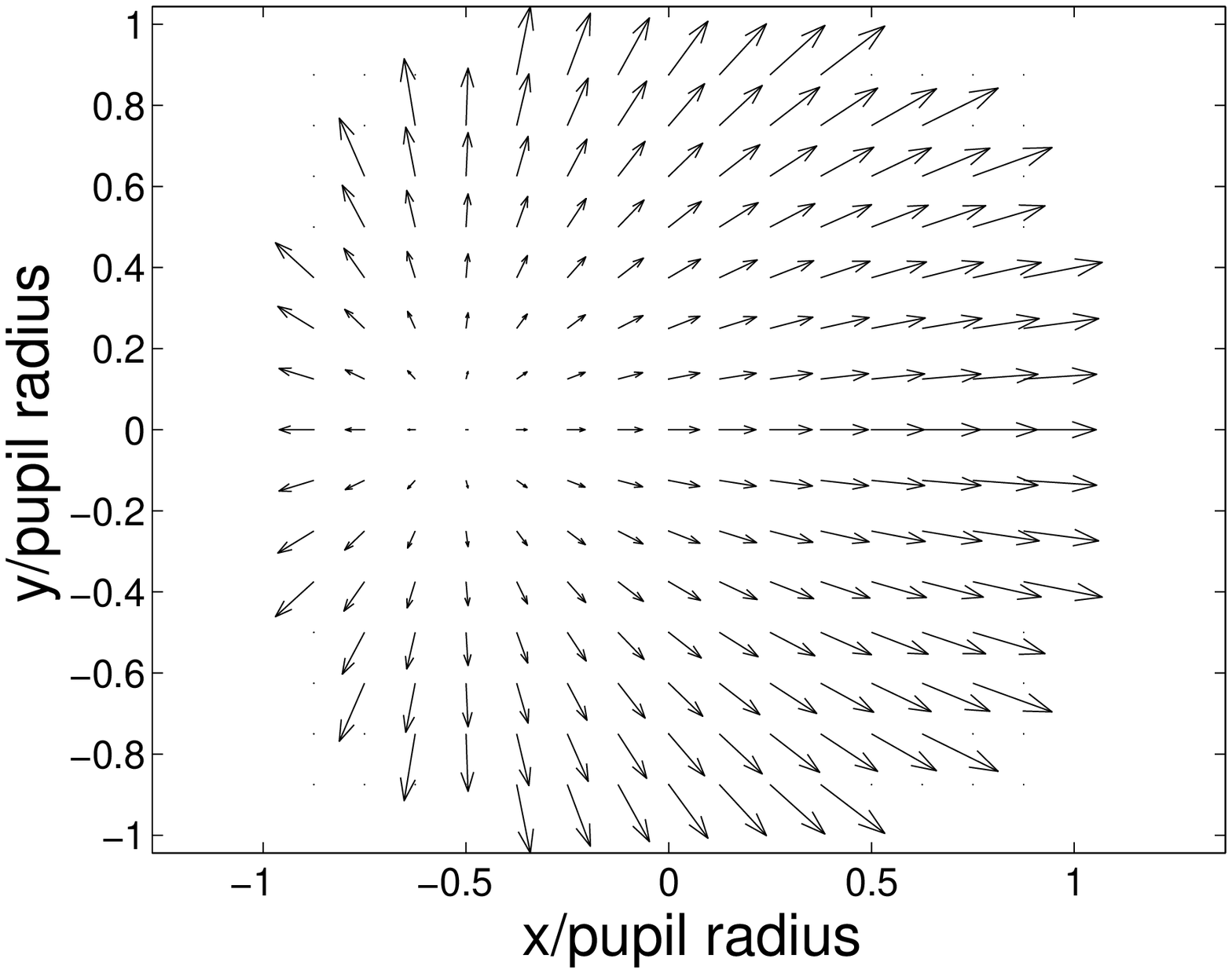} \\
    \includegraphics[width=8cm]{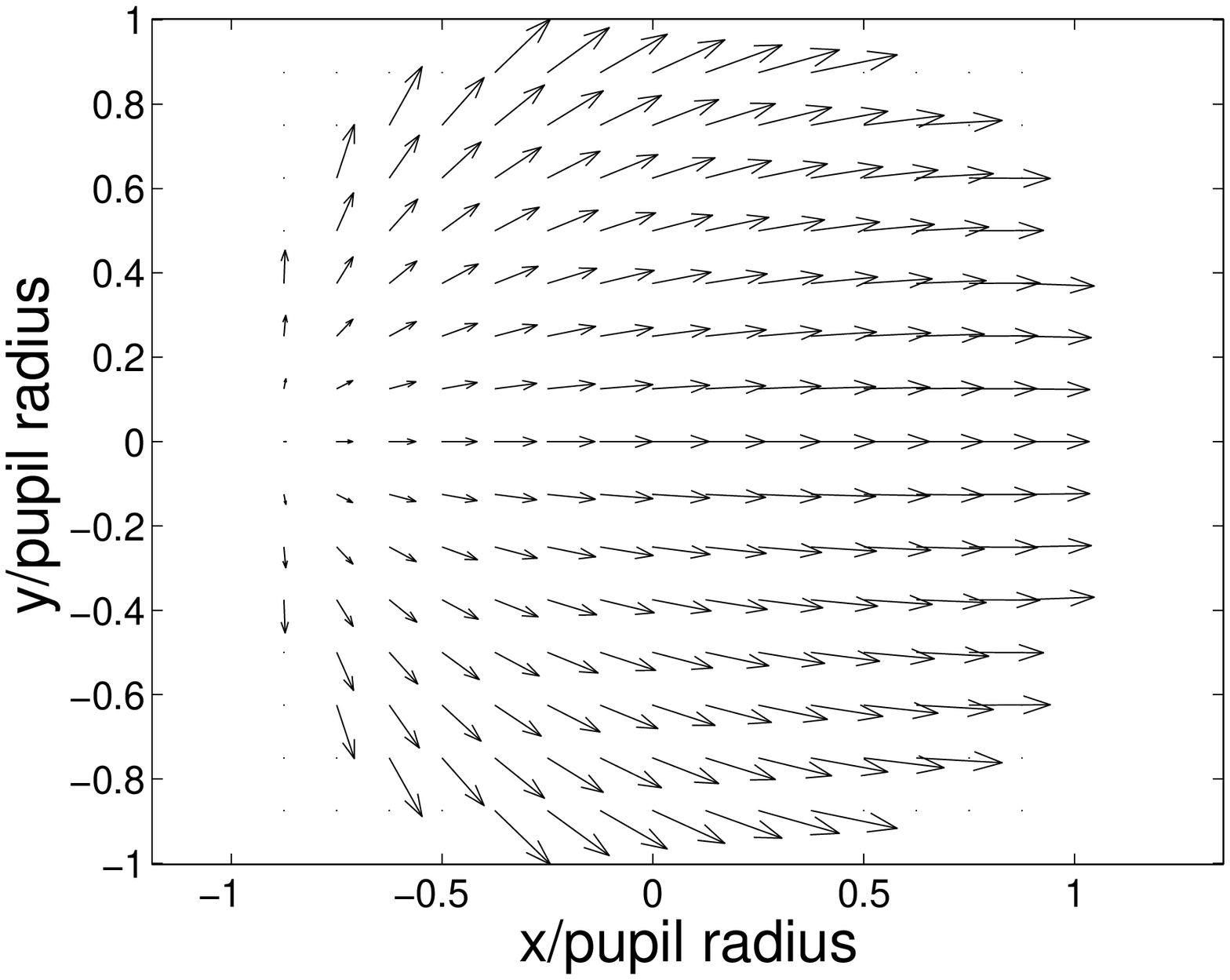} \includegraphics[width=8cm]{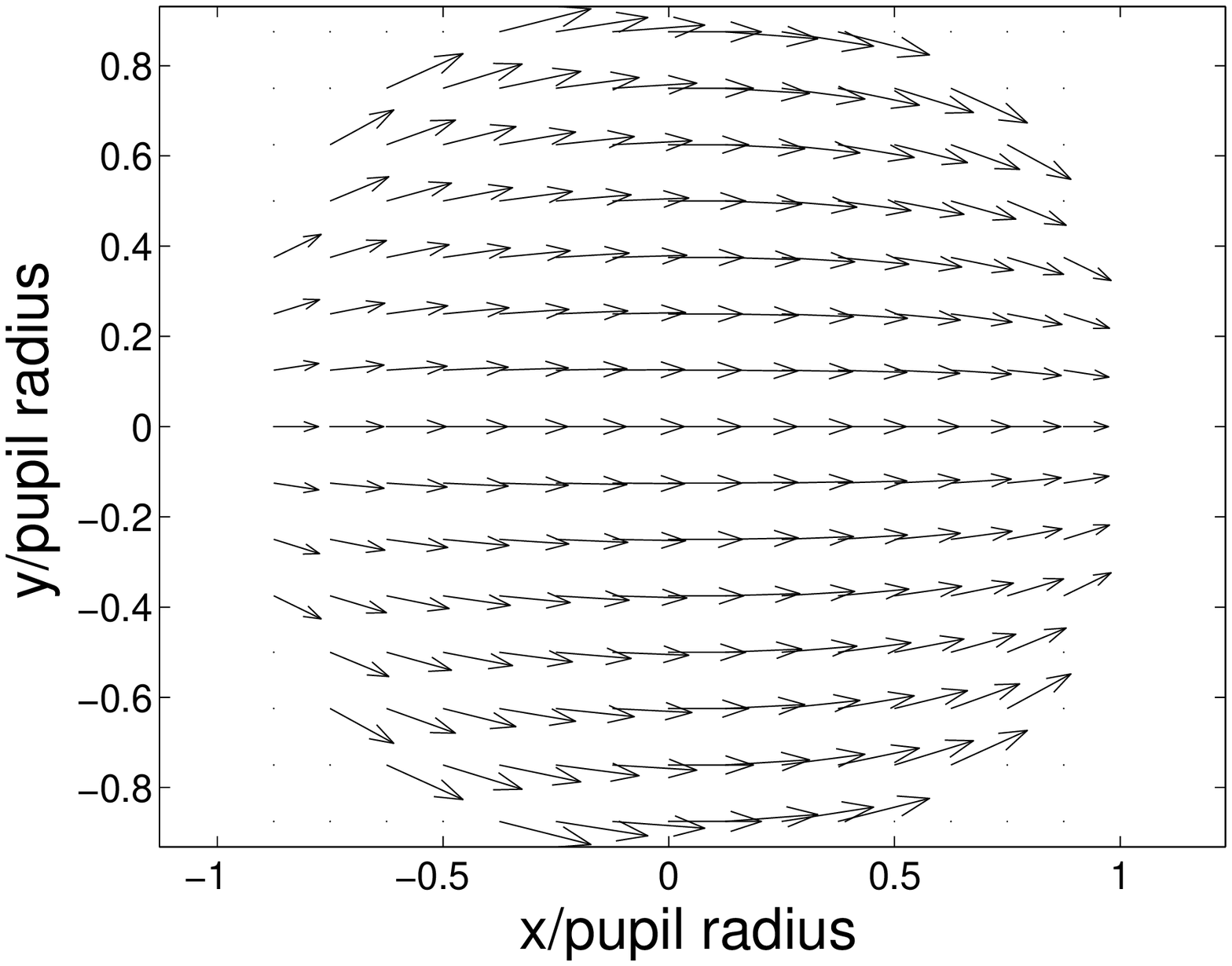}
   \caption{\label{Fig.quiverpupil} Snapshot of the electric field in the pupil that, when focused, yields  the
field with maximum $v$-component in focus when
$\mbox{NA}/n=0.9$. The upper left figure corresponds to  $\vartheta_v=0^o$, the upper right to $\vartheta_v=30^o$,
the lower left to $\vartheta_v=60^o$ and the lower right to
$\vartheta_v=90^o$.}
\end{center}
\end{figure}

\begin{figure}
\begin{center}
    \includegraphics[width=10cm]{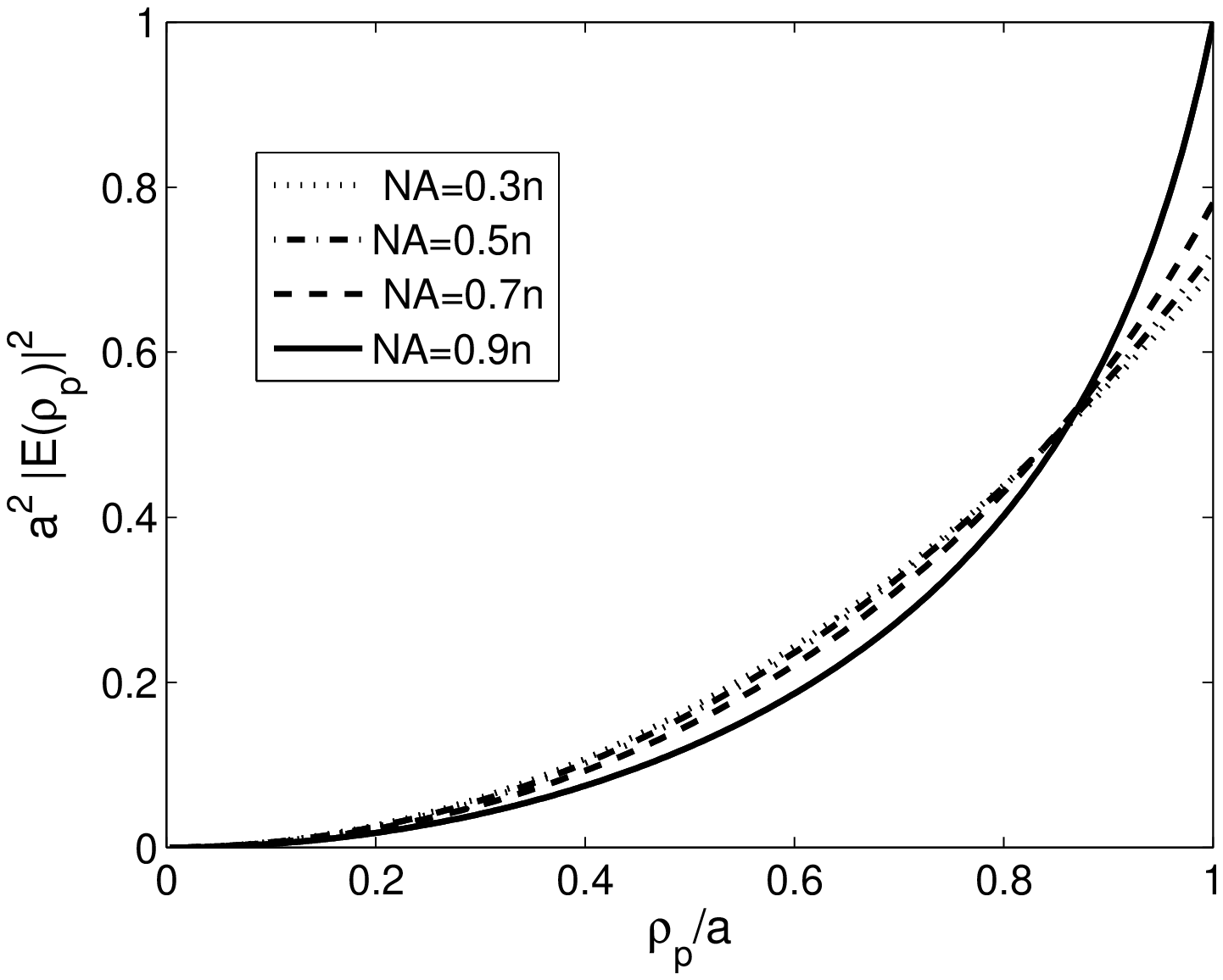}
   \caption{\label{Fig.E2pupil0} Electric energy density
   $a^2 |\mathbf{E}(\varrho_p|^2$ in the lens pupil as function of
    the  radial coordinate $\varrho_p$, corresponding to
    the maximum longitudinal component
in focus for several $\mbox{NA}/n$. All values are relative to the value for $\mbox{NA}/n=0.9$ at the rim of the pupil.}
\end{center}
\end{figure}

Suppose now that  the vector $\hat{\mathbf{v}}$ is in the $(x,z)$-plane and the
angle with the positive $z$-axis is $\vartheta_v$.
Then (\ref{eq.vtheta}) holds and
(\ref{eq.Epupil2}) becomes:
\begin{eqnarray}
\mathbf{E}^p(\varrho_p,\varphi_p) & = &
  \frac{i}{\sqrt{2}\Lambda \lambda_0 f} \left(\frac{\mu_0}{\epsilon_0}\right)^{1/2}
\left[  \left(\frac{\sqrt{f^2 - \varrho_p^2}}{f} \, \cos \varphi_p \sin\vartheta_v +
\frac{\varrho_p}{f^{1/2} (f^2 - \varrho_p^2)^{1/4} } \cos\vartheta_v \right)\,
\hat{\bfrho}_p
\right. \nonumber \\
& & \left.
  -\frac{f^{1/2}}{(f^2 -\varrho_p^2)^{1/4}} \, \sin \varphi_p\,\sin\vartheta_v\,  \hat{\bfphi}_p
  \right],
  \label{eq.Epupil_interm}
  \end{eqnarray}
 The pupil field is always
 linearly   polarized, with direction of polarization that varies with the pupil point.
 The phase of the field is the same in all points of the pupil.
In Fig. \ref{Fig.quiverpupil} we show snapshots of the electric field in the pupil for
$\vartheta_v=0$ (i.e. the longitudinal component in focus is maximum), $\vartheta_v=90^o$
(i.e. the $x$-component in focus is maximum) and the intermediate cases
$\vartheta_v=30^o$ and $\vartheta_v=60^o$. In all cases $\mbox{NA}/n=0.9$.
 For $\vartheta_v=0$ the electric field is radially polarized, for
$\vartheta_v=90^o$ it is the field of a plane wave polarized parallel to the $x$-axis.
For $\vartheta_v=30^o$ and $\vartheta_v=60^o$ intermediate cases occur.
In Fig. \ref{Fig.E2pupil0} the squared modulus of the electric pupil field is shown as function of the normalized
pupil coordinate $\bar{\varrho}_p=\varrho_p/a$, with $a=f\mbox{NA}/n$, for several values of $\mbox{NA}/n$ and
for $\vartheta_v=0^o$ (i.e. the pupil fields correspond to fields with maximum longitudinal component).
The square modulus shown has been rescaled by the factor $a^2$, hence the function
\begin{equation}
  \bar{\varrho}_p \mapsto a^2 |\mathbf{E}(a\bar{\varrho}_p|^2,
  \label{eq.E2rescaled}
  \end{equation}
is shown. The reason is that the power in the pupil is proportional to the following integral over the normalized radial
pupil coordinate:
\begin{equation}
  \int_0^1 a^2 |\mathbf{E}(a\bar{\varrho}_p|^2 \, \bar{\varrho}_p\, d \bar{\varrho}_p.
  \label{eq.power}
  \end{equation}
  It is seen that for higher $\mbox{NA}/n$ the electric energy density is more concentrated near the rim of the pupil.
But even for $\mbox{NA}/n=0.9$,    the energy density is smoothly varying inside the pupil.
Hence it seems quite possible to realize these pupil fields using
appropriately programmed spatial light modulators. It may be important in some applications to
realize the radially increasing amplitude without absorbing a substantial part of the light.  This
could be done with optical elements which  refract  energy close to the axis towards the
rim  of the pupil.

\section{Discussion}
We have studied fields which, with respect to some cartesian coordinate system $(x,y,z)$, consist
of plane waves that all propagate in the positive $z$-direction in a homogeneous isotropic lossless medium
with refractive index $n$. The sinus of the maximum angle between  the plane wave vectors and the $z$-axis is the
numerical aperture of the plane wave superposition.
Let $\hat{\mathbf{v}}$ be a unit vector.
For given $\hat{\mathbf{v}}$, given  $\mbox{NA}$ and given mean power flow in the positive $z$-direction,
 we have determined closed formulas for the plane wave amplitudes
for which the projection of the electric field in the direction of $\hat{\mathbf{v}}$ in the origin is larger or equal
than that of all fields with the same $\mbox{NA}$ and the same maximum power flow.
Closed formulae have been derived for the plane wave amplitudes. By choosing $\hat{\mathbf{v}}$ parallel to the
$z$-axis, we get the field for which the amplitude of the longitudinal component is maximum.
The width of this longitudinal component is considerably smaller than that of the Airy spot for the
same power and the same $\mbox{NA}$.
By choosing $\mathbf{v}$ perpendicular to the $z$-axis, along the $x$-axis say, we obtain the field for which
the amplitude of the $x$-component of the electric field in the origin is maximum.
This field is similar to the field of a focused linearly polarized plane, but it is not identical to it.
Also for directions of $\hat{\mathbf{v}}$ that are intermediate between longitudinal and transverse, the
optimum fields were studied. In particular the FWHM of the components in the plane $z=0$  where compared for
several values of the numerical aperture.

By using the vectorial diffraction theory of Ignatowsky and Richards and Wolf, the
field distributions in the pupil of a diffraction limited lens of given $\mbox{NA}$ was
derived such that the focused are the previously discussed fields for which the
projection of the electric field along the $\hat{\mathbf{v}}$-direction is maximum in the focal point.
Closed formulas were obtained. When $\hat{\mathbf{v}}$ is chosen equal to the unit vector along the
$z$ axis, the pupil field is radially polarized. When $\hat{\mathbf{v}}$ is parallel to the $x$-direction,
the pupil field is almost identical to that of a plane wave that is polarized parallel to the
$x$-direction. In general, the pupil fields that yield the optimum fields near the focal plane, are
linearly polarized and in phase in all points of the pupil. But the state of polarization and the
amplitude varies with position in the pupil.

  \appendix
\section{The focused field of a linearly polarized plane wave}
\label{app.plw}
  We recall here the expression for the electric field
  near the focal plane of a lineary polarized plane wave
in the pupil of a lens of numerical aperture $\mbox{NA}$, as derived
in the theory  of
Ignatowsky \cite{ignatowsky}, \cite{ignatowskyb}  and
 Richards and Wolf \cite{wolf}, \cite{richards}.
With respect the polar basis, the $x$-polarized pupil field can be written
  \begin{equation}
  \mathbf{E}^p(\varrho_p,\varphi_p) =   E_{\varrho}^p(\varrho_p,\varphi_p)\hat{\bfrho}_p + E_{\varphi}^p(\varrho_p,\varphi_p)\hat{\bfphi}_p,
    \label{eq.apppupil2}
  \end{equation}
  with
  \begin{eqnarray}
    \hat{\bfrho}_p & = & \cos\varphi_p \, \hat{\mathbf{x}} + \sin \varphi_p \, \hat{\mathbf{y}},
    \label{eq.appbfrho}\\
    \hat{\bfphi}_p & = & -\sin\varphi_p \, \hat{\mathbf{x}} + \cos \varphi_p \, \hat{\mathbf{y}},
    \label{eq.appbfphi}
    \end{eqnarray}
and
\begin{eqnarray}
    E_{\varrho}^p(\varrho_p,\varphi_p) & = & \cos \varphi_p, \label{eq.appErho}\\
    E_{\varphi}^p(\varrho_p,\varphi_p) & = & -\sin \varphi_p. \label{eq.appEphi}
    \end{eqnarray}
According to
 (\ref{eq.bfE}), (\ref{eq.bfH}),  the plane wave expansion of the electromagnetic field near the focal plane
 can be written as
 \begin{eqnarray}
\mathbf{E}(\mathbf{r}) & = & \frac{ n^2}{\lambda_0^2}
 \int_0^{\alpha_{\max}}\!\int_0^{2\pi}
   ( A_\alpha \hat{\bfalpha}+A_\beta \hat{\bfbeta}  ) \sin \alpha \cos \alpha \,
e^{i \mathbf{k}\cdot \mathbf{r}}\,
d \alpha\, d\beta, \label{eq.appbfE} \\
\mathbf{H}(\mathbf{r}) & = & \frac{ n^3}{\lambda_0^2} \left(\frac{\epsilon_0}{\mu_0}\right)^{1/2}\,
 \int_0^{\alpha_{\max}}\!\int_0^{2\pi}
   ( -A_\beta \hat{\bfalpha}+A_\alpha \hat{\bfbeta}  ) \sin \alpha \cos \alpha \,
e^{i \mathbf{k}\cdot \mathbf{r}}\,
d \alpha\, d\beta, \label{eq.appbfH}
\end{eqnarray}
where $\alpha$ and $\beta$ are spherical coordinates for the wave vectors
of the plane waves:
\begin{equation}
\mathbf{k} = k_0 n \left( \cos \beta \sin \alpha \hat{\mathbf{x}} +
                           \sin \beta \sin \alpha \hat{\mathbf{y}} +
                           \cos \alpha  \hat{\mathbf{z}}\right),
                                \label{eq.appkvec}
               \end{equation}
               and $\hat{\mathbf{\alpha}}$ and $\hat{\mathbf{\beta}}$ are given by
               (\ref{eq.hatalpha}), (\ref{eq.hatbeta}).
               With respect to cylindrical coordinates: $\mathbf{r}=\varrho \hat{\bfrho} + z \hat{\mathbf{z}}$, we have
               \begin{eqnarray}
               \mathbf{k}\cdot \mathbf{r}  =  k_0 n \left[ \varrho \sin \alpha \cos(\beta - \varphi) + z \cos \alpha \right].
               \label{eq.kdotr}
               \end{eqnarray}
               By substituting (\ref{eq.appErho}), (\ref{eq.appEphi}) into
(\ref{eq.richards11}) and using
  (\ref{eq.phi_pphi}): $\varphi_p=\beta+\pi$, we find
  \begin{eqnarray}
    A_\alpha(\alpha,\beta) & = & 2\pi i \frac{f}{k_0 n} \frac{\cos \beta}{\sqrt{\cos \alpha}},
    \label{eq.apAalpha}\\
    A_\beta(\alpha,\beta) & = & -2\pi i  \frac{f}{k_0 n} \frac{\sin \beta}{\sqrt{\cos \alpha}}
    \label{eq.apAbeta}
    \end{eqnarray}
    Hence,
   \begin{eqnarray}
      \mathbf{A}(\alpha, \beta)& =&
                2 \pi i \frac{f}{k_0 n \sqrt{\cos\alpha} }  \left[
       \cos \beta
\, \hat{\bfalpha}  - \sin \beta \,
\hat{\bfbeta}\right] \nonumber \\
& = &  2 \pi i \frac{f}{k_0 n  \cos^{1/2} \alpha}  \left\{
\left(\cos \alpha  \cos^2 \beta +  \sin^2\beta
\right)\,\hat{\mathbf{x}}
 +  \left(\cos  \alpha  - 1 \right)\cos \beta \sin
\beta \hat{\mathbf{y}} - \sin \alpha \cos \beta\,
\hat{\mathbf{z}}\right\}, \nonumber \\
\label{eq.appAE}
\end{eqnarray}
and
\begin{eqnarray}
-A_\beta \hat{\bfalpha} + A_\alpha \hat{\bfbeta} & = &
2 \pi i \frac{f}{k_0 n \sqrt{\cos\alpha} }  \left[ \sin \beta \hat{\bfalpha} + \cos \beta \hat{\bfbeta}\right]
\nonumber \\
& = & 2 \pi i \frac{f}{k_0 n \sqrt{\cos\alpha} } \left[ (\cos \alpha -1) \cos\beta \sin\beta \hat{\mathbf{x}}
 + ( \cos \alpha \sin^2 \beta + \cos^2\beta) \hat{\mathbf{y}} - \sin\alpha \cos\beta \hat{\mathbf{z}} \right],
 \nonumber \\
\label{eq.appAH}
\end{eqnarray}
where $f$ is the focal length.
Substitution into plane wave expansion
(\ref{eq.appbfE}) and  using   formulas
    (\ref{eq.cos}-\ref{eq.cossin}) and (\ref{eq.glnumu}) yields
  \begin{eqnarray}
E_x(\mathbf{r}) & = &  \frac{ i \pi n f }{\lambda_0} \left\{
g_0^{\frac{1}{2},1}(\varrho,z) +
g_0^{\frac{3}{2},1}(\varrho,z)
 + \left[g_2^{\frac{1}{2},1}(\varrho,z) - g_2^{\frac{3}{2},1}(\varrho,z) \right]
 \, \cos 2\varphi \right\}, \label{eq.Explw}\\
E_y(\mathbf{r})  & = &  \frac{ i \pi n f }{\lambda_0}
  \left[g_2^{\frac{1}{2},1}(\varrho,z) -
  g_2^{\frac{3}{2},1}(\varrho,z) \right]
 \, \sin 2\varphi, \label{eq.Eyplw}\\
E_z(\mathbf{r})  &  =  & \frac{ 2 \pi n f }{\lambda_0}
g_1^{\frac{1}{2},2}(\varrho,z) \, \cos
\varphi,
 \label{eq.Ezplw}
  \end{eqnarray}
  and
  \begin{eqnarray}
 H_x(\mathbf{r}) & = & \frac{i \pi n^2 f}{\lambda_0} \sqrt{\frac{\epsilon_0}{\mu_0}}   [  g_2^{\frac{1}{2},1}(\varrho,z) -g_2^{\frac{3}{2},1}(\varrho,z) ]\sin(2\varphi),
 \label{eq.Hxplw}\\
 H_y(\mathbf{r}) & = & \frac{i \pi n^2 f}{\lambda_0} \sqrt{\frac{\epsilon_0}{\mu_0}}\left\{ g_0^{\frac{1}{2},1}(\varrho,z) + g_0^{\frac{3}{2},1}(\varrho,z) - [ g_2^{\frac{1}{2},1}(\varrho,z) - g_2^{\frac{3}{2},1}(\varrho,z)] \cos(2\varphi)
  \right\}, \nonumber \\ \label{eq.Hyplw} \\
  H_z(\mathbf{r}) & = & \frac{2 \pi n^2 f}{\lambda_0} \sqrt{\frac{\epsilon_0}{\mu_0}}
  \, g_1^{\frac{1}{2},2}(\varrho,z)\sin\varphi.
  \label{eq.Hzplw}
  \end{eqnarray}
 The cylindrical field components are
 \begin{eqnarray}
 E_\varrho(\mathbf{r}) & = & E_x(\mathbf{r})\cos\varphi + E_y(\mathbf{r}) \sin \varphi \nonumber \\
 & = & \frac{i \pi n f}{\lambda_0} \left[ g_0^{\frac{1}{2},1}(\varrho,z) + g_0^{\frac{3}{2},1}(\varrho,z)
 + g_2^{\frac{1}{2},1}(\varrho,z) - g_2^{\frac{3}{2},1}(\varrho,z)\right] \cos\varphi,
 \label{eq.Erho} \\
  E_\varphi(\mathbf{r}) & = & -E_x(\mathbf{r})\sin\varphi + E_y(\mathbf{r}) \cos \varphi \nonumber \\
 & = & -\frac{i \pi n f}{\lambda_0} \left[ g_0^{\frac{1}{2},1}(\varrho,z) + g_0^{\frac{3}{2},1}(\varrho,z)
 - g_2^{\frac{1}{2},1}(\varrho,z) + g_2^{\frac{3}{2},1}(\varrho,z)\right] \sin\varphi,
 \label{eq.Ephi}
 \end{eqnarray}
 and
 \begin{eqnarray}
 H_\varrho(\mathbf{r}) & = & H_x(\mathbf{r})\cos\varphi + H_y(\mathbf{r}) \sin \varphi \nonumber \\
 & = & \frac{i \pi n^2 f}{\lambda_0} \sqrt{\frac{\epsilon_0}{\mu_0}}\left[ g_0^{\frac{1}{2},1}(\varrho,z) + g_0^{\frac{3}{2},1}(\varrho,z)
 + g_2^{\frac{1}{2},1}(\varrho,z) - g_2^{\frac{3}{2},1}(\varrho,z)\right] \sin\varphi,
 \label{eq.Hrho} \\
  H_\varphi(\mathbf{r}) & = & -H_x(\mathbf{r})\sin\varphi + H_y(\mathbf{r}) \cos \varphi \nonumber \\
 & = & \frac{i \pi n^2 f}{\lambda_0} \sqrt{\frac{\epsilon_0}{\mu_0}}
 \left[ g_0^{\frac{1}{2},1}(\varrho,z) + g_0^{\frac{3}{2},1}(\varrho,z)
 - g_2^{\frac{1}{2},1}(\varrho,z) + g_2^{\frac{3}{2},1}(\varrho,z)\right] \cos\varphi.
 \label{eq.Hphi}
 \end{eqnarray}
 In the plane $z=0$ all functions $g^{\nu,\mu}_\ell$ are real. Therefore, the squared modulus of the electric field
 in the $z=0$-plane is
 \begin{eqnarray}
 |\mathbf{E}(\varrho,\varphi,0)|^2 & = & |E_\varrho(\varrho,\varphi,0)|^2 + |E_\varphi(\varrho,\varphi,0)|^2
 + |E_z(\varrho,\varphi,0)|^2 \nonumber \\
 & = & \frac{\pi^2 n^2 f^2}{\lambda_0^2} \left\{
  \left[g_0^{\frac{1}{2},1}(\varrho,0) + g_0^{\frac{3}{2},1}(\varrho,0)
 + g_2^{\frac{1}{2},1}(\varrho,0) - g_2^{\frac{3}{2},1}(\varrho,0)\right]^2 \cos^2\varphi \right. \nonumber\\
 & & \left.+ \left[
  g_0^{\frac{1}{2},1}(\varrho,0) + g_0^{\frac{3}{2},1}(\varrho,0)
 - g_2^{\frac{1}{2},1}(\varrho,0) + g_2^{\frac{3}{2},1}(\varrho,0)\right]^2 \sin^2\varphi  \right. \nonumber \\
 & & \left.+ \left[g_1^{\frac{1}{2},2}(\varrho,0)\right]^2 \, \cos^2\varphi \right\}.
 \label{eq.Eintens}
 \end{eqnarray}
 The spot shape is elliptical. To obtain a measure of the spot size we  average the squared modulus
 over $0 < \varphi < 2\pi$:
 \begin{eqnarray}
 \frac{1}{2\pi}\int_0^{2\pi} |\mathbf{E}(\varrho,\varphi,0)|^2 \, d\varphi & = &
 \frac{\pi^2 n^2 f^2}{2\lambda_0^2} \left\{
   \left[g_0^{\frac{1}{2},1}(\varrho,0) + g_0^{\frac{3}{2},1}(\varrho,0)
 + g_2^{\frac{1}{2},1}(\varrho,0) - g_2^{\frac{3}{2},1}(\varrho,0)\right]^2 \right. \nonumber\\
 & & \left. +\left[
  g_0^{\frac{1}{2},1}(\varrho,0) + g_0^{\frac{3}{2},1}(\varrho,0)
 - g_2^{\frac{1}{2},1}(\varrho,0) + g_2^{\frac{3}{2},1}(\varrho,0)\right]^2 \right. \nonumber \\
 & & \left. + \left[g_1^{\frac{1}{2},2}(\varrho,0)\right]^2 \right\} \nonumber \\
 & = &  \frac{\pi^2 n^2 f^2}{\lambda_0^2}
  \left\{ \left[ g^{\frac{1}{2},1}_0(\varrho,0) + g^{\frac{3}{2},1}_0(\varrho,0)\right]^2
  \right.  \nonumber \\
    & & \left. + \left[ g^{\frac{1}{2},1}_2(\varrho,0) - g^{\frac{3}{2},1}_2(\varrho,0)\right]^2
    + \frac{1}{2} \left[ g^{\frac{1}{2},2}_1(\varrho,0)\right]^2 \right\}.
 \label{eq.aveEintens}
 \end{eqnarray}
 The FWHM in the $z=0$-plane is then defined by $2\varrho_0$ with $\varrho_0$ such that
 \begin{equation}
 \frac{1}{2\pi}\int_0^{2\pi} |\mathbf{E}(\varrho_0,\varphi,0)|^2 \, d\varphi =
 \frac{1}{2} |\mathbf{E}(\mathbf{0})|^2.
 \label{eq.FHMAiry}
 \end{equation}

 The real part of the complex Poynting vector in cylindrical coordinates is:
 \begin{eqnarray}
 \mbox{Re} \, S_\varrho(\mathbf{r}) & = & \frac{1}{2} \mbox{Re} \left[ E_\varphi H_z^*- E_z H_\varphi^*\right] \nonumber \\
 & = & -\frac{\pi^2 n^3 f^2}{\lambda_0^2} \sqrt{\frac{\epsilon_0}{\mu_0}} \,\mbox{Im}
 \left[ (g_0^{\frac{1}{2},1} + g_0^{\frac{3}{2},1} - g_2^{\frac{1}{2},1} + g_2^{\frac{3}{2},1} ) (g_1^{\frac{1}{2},2})^* \right],
  \label{eq.appSrho}
 \end{eqnarray}
 \begin{eqnarray}
 \mbox{Re} \, S_\varphi(\mathbf{r}) & = & \frac{1}{2} \mbox{Re} \left[ E_z H_\varrho^*- E_\varrho H_z^*\right] =0,
   \label{eq.appSphi}
 \end{eqnarray}
 \begin{eqnarray}
 \mbox{Re} \, S_z(\mathbf{r}) & = & \frac{1}{2} \mbox{Re}
 \left[ E_\varrho H_\varphi^*- E_\varphi H_\varrho^*\right] \nonumber \\
 & = & \frac{\pi^2 n^3 f^2}{2 \lambda_0^2} \sqrt{\frac{\epsilon_0}{\mu_0}} \,
 \left( |g_0^{\frac{1}{2},1} +g_0^{\frac{3}{2},1}|^2- |g_2^{\frac{1}{2},1} - g_2^{\frac{3}{2},1}|^2 \right).
 \label{eq.appSz}
 \end{eqnarray}
 The total power flow $P$ through a plane $z=\mbox{constant}$ is most easily calculated by substituting
 (\ref{eq.apAalpha}) and (\ref{eq.apAbeta}) into
 (\ref{eq.defP}). One finds
 \begin{equation}
 P = \frac{\pi}{2} n f^2 \sqrt{\frac{\epsilon_0}{\mu_0}} \, \sin^2\alpha_{\max}.
 \label{eq.P0plw}
 \end{equation}
   We have along the $z$-axis:
 \begin{eqnarray}
    g_1^{\nu,\mu}(0,z)=g_2^{\nu,\mu}(0,z)=0,
    \nonumber
    \end{eqnarray}
    for all $\nu$ and $\mu$, and
    \begin{eqnarray}
    g_0^{\frac{1}{2},1}(0,z) & = & \int_0^{\alpha_{\max}} e^{ik_0n z\cos \alpha}
    \, \cos^{1/2} \alpha \sin \alpha \, d \alpha \nonumber \\
    & = &   \frac{e^{ik_0 n z} - 1}{i k_0 n z }-
    \frac{ e^{i k_0 n z\cos\alpha_{\max}} - 1 }{
    i k_0 n z } \, \cos^{1/2}\alpha_{\max} \nonumber \\
    & &  -i \sqrt{\frac{\pi}{2}} \frac{
      {\cal C}(\sqrt{ (2/\pi) k_0 n z \cos \alpha_{\max}})
  - \sqrt{ (2/\pi)k_0 n z \cos \alpha_{\max}} +   i
  {\cal S}( \sqrt{ (2/\pi)k_0 n z \cos \alpha_{\max}})  } { (k_0 n z)^{3/2} }  \nonumber \\
   & &  + i \sqrt{\frac{\pi}{2}} \frac{
      {\cal C}(\sqrt{ (2/\pi)k_0 n z })
  - \sqrt{ (2/\pi) k_0 n z } +   i
  {\cal S}( \sqrt{ (2/\pi) k_0 n z })  } { (k_0 n z)^{3/2}  },
\nonumber \\
     \label{eq.g0121} \\
    g_0^{\frac{1}{2},1}(\varrho,-z) & = &
    g_0^{\frac{1}{2},1}(\varrho,z)^*,
     \label{eq.g0121pi}
     \end{eqnarray}
     where
     \begin{eqnarray}
       {\cal C}(z) & = & \int_0^z \cos\left(\frac{\pi}{2} t^2\right) \, d
       t,  \label{eq.C}  \\
      {\cal S}(z) & = & \int_0^z \sin\left(\frac{\pi}{2} t^2\right) \, d
       t. \label{eq.S}
       \end{eqnarray}
       Furthermore,
   \begin{eqnarray}
    g_0^{\frac{3}{2},1}(0,z) & = & \int_0^{\alpha_{\max}} e^{ik_0n z \cos \alpha}
    \, \cos^{3/2} \alpha \sin \alpha \, d \alpha \nonumber \\
    & = &
    \frac{e^{ik_0 n z}- 1}{ik_0n z}  -
   \frac{ e^{i k_0 n z\cos\alpha_{\max}} -1}{i k_0 n z}
   \cos^{3/2}\alpha_{\max}  \nonumber \\
   & & + \frac{3}{2} \frac{e^{ik_0 n z}-1 - i k_0 n  z}{(k_0n z)^2}
   - \frac{3}{2}
    \frac{e^{ik_0 n z \cos\alpha_{\max}}-1 - i k_0 n z\cos\alpha_{\max}}
    {(k_0n z)^2}\, \cos^{1/2}\alpha_{\max}  \nonumber \\
    & & + \frac{3}{2}\sqrt{\frac{\pi}{2}} \frac{ {\cal C}(\sqrt{(2/\pi) k_0 n z
      \cos\alpha_{\max}}) - \sqrt{(2/\pi) k_0 n z\cos\alpha_{\max}}}
      {((2/\pi)k_0 n z)^{5/2} }
      \nonumber \\
      & &  +   i\frac{3}{2}\sqrt{\frac{\pi}{2}} \frac{
      {\cal S}(\sqrt{(2/\pi) k_0 n z\cos\alpha_{\max}}) - \frac{\pi}{6}
      ((2/\pi)k_0 n z\cos \alpha_{\max})^{3/2}  }
      {((2/\pi)k_0 n z)^{5/2} }
      \nonumber \\
      & & - \frac{3}{2}\sqrt{\frac{\pi}{2}} \frac{ {\cal C}(\sqrt{(2/\pi) k_0 n z
      }) - \sqrt{(2/\pi) k_0 n z} +
      i \left[ {\cal S}(\sqrt{(2/\pi) k_0 n z}) - \frac{\pi}{6}
      ((2/\pi)k_0 n z)^{3/2}\right] }
      {((2/\pi)k_0 n z)^{5/2} },
         \nonumber \\
     \label{eq.g0321} \\
    g_0^{\frac{3}{2},1}(0,-z) & = &
    g_0^{\frac{3}{2},1}(0,z)^*.
     \label{eq.g0321pi}
     \end{eqnarray}
     In particular,
  \begin{eqnarray}
    g_0^{\frac{1}{2},1}(0,0) & = &
     \frac{2}{3} ( 1-\cos^{3/2}\alpha_{\max}),
    \label{eq.g0121zero} \\
    g_0^{\frac{3}{2},1}(0,0) & = &
    \frac{2}{5} ( 1-\cos^{5/2}\alpha_{\max}).
    \label{eq.g0321zero}
    \end{eqnarray}
    Hence,  along the optical axis, the electric field is
\begin{equation}
   \mathbf{E}(0,0,z) = \frac{i \pi n f}{\lambda_0} \left[
   g_0^{\frac{1}{2},1}(0,z) +
   g_0^{\frac{3}{2},1}(0,z)\right]\, \hat{\mathbf{x}},
   \label{eq.Eopticaxis}
   \end{equation}
     and the electric energy density  is
    \begin{eqnarray}
 |\mathbf{E}(0,0,z)|^2 & = &
 |E_x(0,0,z)|^2  \nonumber \\
 & = &
  \frac{\pi^2 n^2 f^2}{\lambda_0^2} | g_0^{\frac{1}{2},1}( 0,z) + g_0^{\frac{3}{2},1}( 0,z)
  |^2.
  \label{eq.EEdens}
  \end{eqnarray}
  In the focal point:
  \begin{equation}
 |\mathbf{E}(\mathbf{0})|^2  =
  \frac{ \pi^2 n^2 f^2}{\lambda_0^2}  \left[
  \frac{16}{15} - \frac{2}{3} \cos^{3/2}\alpha_{\max}
  - \frac{2}{5} \cos^{5/2} \alpha_{\max}
  \right]^2.
  \label{eq.EEdens0}
  \end{equation}
Now
\begin{eqnarray}
g_0^{\frac{1}{2},1}( 0,z) + g_0^{\frac{3}{2},1}(
0,z) & = &  \int_0^{\alpha_{\max}} e^{i k_0 n z\cos
\alpha} ( \cos^{1/2}\alpha + \cos^{3/2}\alpha)
\sin\alpha \, d\alpha \nonumber \\
& = & 2\int_0^{\alpha_{\max}} e^{i k_0 n z\cos \alpha}
 \sin\alpha \, d\alpha + {\cal I}(z),
 \label{eq.approx1}
 \end{eqnarray}
 where
 \begin{equation}
 {\cal I}(z) = \int_0^{\alpha_{\max}} e^{i k_0 n z\cos
 \alpha} (\cos^{1/2}\alpha + \cos^{3/2}\alpha -2) \, \sin
 \alpha \, d\alpha.
 \label{eq.defI}
 \end{equation}
 For small numerical aperture:
 \begin{eqnarray}
 \int_0^{\alpha_{\max}} e^{i k_0 n z\cos \alpha}
 \sin\alpha \, d\alpha & = &  e^{i \frac{k_0 n z}{2}
 (1+\cos\alpha_{\max})} \frac{\sin\left[ \frac{k_0n z}{2}
  (1-\cos\alpha_{\max})\right]}{\frac{k_0 n z}{2}
 }
 \nonumber \\
 & = &e^{i \frac{k_0 n z}{2}
  (1+\cos\alpha_{\max})} \frac{\sin\left( \frac{k_0 n z}{4}
  \alpha_{max}^2\right)}{\frac{k_0 n z}{4}
    \alpha_{\max}^2}\, \frac{\alpha_{\max}^2}{2} + O(\alpha_{\max}^4),
 \nonumber \\
 \label{eq.result1}
 \end{eqnarray}
 where $O(\alpha_{\max}^4)$ denotes a term  which modulus is smaller than
 $C \alpha_{\max}^4$, where $C$ is a positive number that is independent of $z$.
 Furthermore, ${\cal I}(z)$ can be
 estimated as follows
 \begin{eqnarray}
 |{\cal I}(z)| & \leq  &\int_0^{\alpha_{\max}} ( 2-\cos^{1/2}\alpha
 - \cos^{3/2}\alpha) \, \sin \alpha \, d\alpha  \nonumber
 \\
 & = & 2(1-\cos\alpha_{\max}) -
 \frac{2}{3}(1-\cos^{3/2}\alpha_{\max}) -
 \frac{2}{5}(1-\cos^{5/2}\alpha_{\max}) = O(\alpha_{\max}^4).
 \label{eq.estimation}
 \end{eqnarray}
 Hence,
\begin{equation}
|g_0^{\frac{1}{2},1}(z, 0) + g_0^{\frac{3}{2},1}(z,
0)|^2  =  \left[\frac{\sin\left(
\frac{k_0 n z }{4}\alpha_{\max}^2\right)}{\frac{k_0 n z }{4}
 \alpha_{\max}^2} \right]^2  \alpha_{\max}^4 + O(\alpha_{\max}^6).
 \label{eq.result2}
 \end{equation}
 The first term is identical to the intensity distribution  along
 the optical axis in the scalar paraxial theory. This shows that
 when the numerical aperture is small, the focal depth of the
 intensity in the vectorial theory is the same as in the scalar
 paraxial theory.


\end{document}